\documentclass{jfm}

\usepackage{graphicx}
\usepackage{natbib}
\usepackage{amsmath,amssymb}
\usepackage{epstopdf}
\usepackage{color}
\usepackage{bm}
\usepackage[normalem]{ulem}
\usepackage{mathtools}
\usepackage{tabularx}

\def\bfR      {\mathbf{R}}
\def\bfL      {\mathbf{L}}
\def\bfS      {\mathbf{S}}
\def\bfP      {\mathbf{P}}
\def\bfA      {\mathbf{A}}
\def\bfB      {\mathbf{B}}
\def\bfC      {\mathbf{C}}
\def\bfD      {\mathbf{D}}
\def\bfI      {\mathbf{I}}
\def\bfu       {\mathbf{u}}
\def\bff        {\mathbf{f}}
\def\bfphi    {\bm{\phi}}
\def\bfpsi    {\bm{\psi}}
\def\bfxi      {\bm{\xi}}
\def\bfzeta  {\bm{\zeta}}
\def\bfs {\mathbf{s}}
\def\bfp {\mathbf{p}}
\def\bfe {\mathbf{e}^{kin}}
\def\gnu {\gamma_{\nu}}
\def\gnut {\gamma_{\nu_t}}
\def\ret  {Re_{\tau}}

\def\diag {\mathrm{diag}}

\title[The colour of forcing statistics in resolvent analyses of turbulent channel flows]
{The colour of forcing statistics in resolvent analyses of turbulent channel flows}

\author[P. Morra, P. A. S. Nogueira, A. V. G. Cavalieri, D. S. Henningson]%
{Pierluigi Morra$^{1}$,\ns
Petr\^{o}nio A. S. Nogueira$^{2}$,\ns
Andr\'e V. G. Cavalieri$^2$\break
and Dan S. Henningson$^1$}
\affiliation{$^1$KTH Royal Institute of Technology, Dept. of Engineering Mechanics,  FLOW, SE-10044, Stockholm, Sweden\\[\affilskip]
$^2$Instituto Tecnol\'ogico de Aeron\'autica, Aerodynamics Department, S\~ao Jos\'e dos Campos, 12228900, Brazil}

\volume{???}
\pagerange{???--???}
\date{?; revised ?; accepted ?. - To be entered by editorial office}
\begin{document}
\maketitle
\begin{abstract}
The cross-spectral density (CSD) of the non-linear forcing term arising as input in resolvent analyses of turbulent flows is here explicitly quantified for the first time for turbulent channel flows. The resolvent is built upon the mean flow, and the non-linear forcing term is associated to the velocity fluctuations only. The computation of the non-linear forcing is based on snapshots of direct numerical simulations (DNS) of turbulent channel flows at friction Reynolds numbers $\ret=179$ and $\ret=543$. The non-linear forcing is computed at fixed time instants and the realizations are used with the Welch method. The CSDs are computed for highly energetic structures typical of buffer-layer and large-scale motions, for different temporal frequencies. The CSD of the non-linear forcing term is shown not to be uncorrelated (or white) in space, which implies the forcing is structured. Since the non-linear forcing is non-solenoidal by construction and the velocity field of the incompressible Navier--Stokes is affected only by the solenoidal part of the forcing, the CSD associated to the solenoidal part of the non-linear forcing is evaluated. It is shown that the solenoidal part of the non-linear forcing term consists in the combination of oblique streamwise vortices and a streamwise component which counteract each other, as in a destructive interference. It is seen that a low-rank approximation of the forcing, which includes only the pair of most energetic symmetric and antisymmetric SPOD (spectral proper orthogonal decomposition) modes, leads to the bulk of the response for all the cases presented. The projections of the non-linear forcing term onto the right-singular vectors of the resolvent are evaluated. It is seen that the left-singular vectors of the resolvent associated with very low-magnitude singular values are non-negligible because the non-linear forcing term has a non-negligible projection onto the linear sub-optimals of the resolvent analysis. The same projections are computed when the stochastic forcing is modelled with an eddy-viscosity approach. It is here clarified that this modelling leads to an improvement in the accuracy of the prediction of the response since the resulting projection coefficients are closer to those associated with the non-linear forcing term evaluated from DNS data.
\end{abstract}
\section{Introduction}
\label{sec:intro}
Elongated streaky structures of spanwise alternating high and low streamwise velocity contain most of the fluctuating energy of wall-bounded turbulent shear flows. The presence of these structures in wall-bounded turbulent shear flows was observed for the first time with the visualizations of \cite{kline1967a}, which revealed that turbulent boundary layers contain streaks with an average spanwise spacing of $\lambda^{+}_z \approx 100$ in the buffer layer and with a higher average spacing farther from the wall. Later, larger streaky structures were observed in the logarithmic and outer region of the turbulent boundary layer. There two types of structures were characterized: \textit{large-scale} motions (LSM) with average spanwise spacings $\lambda_z \approx \delta - 1.5\delta$ (with $\delta$ the outer length scale) and streamwise length $\lambda_x \approx 2\lambda_z$ \citep{corrsin1954a,kovasznay1970a}, and \textit{very large-scale} motions (VLSM) with average spanwise spacings $\lambda_z \approx \delta - 1.5\delta$ and streamwise length $\lambda_x \simeq O(10\lambda_z)$ \citep{komminaho1996a,kim1999a,hutchins2007a}. \\ \indent
The process responsible for the occurrence of streaks in transitional and turbulent flows is the lift-up mechanism in which low-energy quasi-streamwise vortices immersed in a shear flow induce high-energy streamwise streaks \citep{moffatt1967a,ellingsen1975a,landahl1980a} with algebraic growth in time \citep{ellingsen1975a,gustavsson1991a}. The energy amplification obtained by the lift-up mechanism is due to the non-normality of the linearized Navier--Stokes operator (e.g. \cite{boberg1988a,reddy1993a, trefethen1993a}). \cite{gustavsson1991a}, \cite{butler1992a}, \cite{reddy1993a}, \cite{schmid2001a} present the energy amplification capabilities of some laminar baseflows. \cite{boberg1988a} propose that the subcritical onset of turbulence in wall-bounded flows may be caused by a process in which the low-energy quasi-streamwise vortices, which lead to the amplification of high-energy streamwise streaks, are regenerated by nonlinear effects associated to the breakdown of the streaks. A similar self-sustained process is introduced by \cite{hamilton1995a} to explain the dynamics of buffer-layer streaks in the turbulent regime. \cite{schoppa2002a} associate the breakdown of the streaks to a secondary non-modal energy amplification while \cite{waleffe1995a} and \cite{reddy1998a} to a modal secondary instability. \cite{hwang2010a,hwang2011a} provide evidence that similar coherent self-sustained processes maintain every streaky structure from buffer-layer to large-scale motions.\\ \indent
Despite general agreement on the existence of self-sustained processes in wall-bounded turbulence, there are different understandings of the mechanisms involved. A common technique to tackle this problem consists in writing the Navier--Stokes as a linear evolution equation with non-linear feedback; the non-linear feedback being a forcing which includes the instantaneous non-linear advection terms associated to the velocity fluctuations. The flow is decomposed into a time-invariant reference state and a fluctuation component, the reference state is usually assumed to be known, and the fluctuations become the unknown variable. Thus, the system dynamics is clearly described as the combination of the energy amplification and the energy redistribution mechanisms, which are represented by the linear operator and the non-linear forcing.\\ \indent
The choice of the reference state and the treatment of the non-linear forcing in the mentioned framework is key. A set of studies bypass the explicit computation of the non-linear term and deals solely with a linear operator. An approach consists in taking the time-averaged field as the reference state and introducing assumptions to avoid the computation of the non-linear forcing \citep{malkus1956a,butler1993a,farrell1993a,mckeon2010a}. Another approach, firstly proposed in \cite{reynolds1967a}, \cite{reynolds1972a} and later revived by \cite{bottaro2006a}, \cite{delalamo2006a}, \cite{cossu2009a}, \cite{pujals2009a}, \cite{hwang2010a}, \cite{hwang2010b}, consists in modelling part of the non-linear forcing with an eddy viscosity and introducing assumptions to avoid the computation of the remainder of the non-linear forcing. This leads to a modified linear operator. A review of these two approaches with the benefits and limitations of using an eddy-viscosity model can be found in \cite{mckeon2017a} or \cite{cossu2017a}. 
A set of studies takes the time-averaged field as reference state and mimics the effects of the non-linear feedback via the introduction of a correction term in the linearized dynamics. This correction term is a stochastic forcing which aims at minimizing the errors introduced by employing a linearized system, and needs to be designed. The correction term is usually assumed to be linear in the fluctuation quantities, so it is given by a ``to-be-designed'' linear operator applied to the fluctuation quantities. Therefore, a design problem which aims at providing this linear operator is introduced. In some studies the design results in solving an optimization problem which aims at minimizing the error between the linearized and the non-linear Navier--Stokes \citep{jovanovic2001a,chevalier2006a,zare2017a,illingworth2018a}, as in a Kalman filter estimation framework. Another possibility consists in computing this ``to-be-designed'' linear operator as a least-squares approximation in a resolvent-based framework \citep{towne2020a}. It is noteworthy that the forcing is always designed to be a linear function of the fluctuation velocities only, so it is implicitly set to be solenoidal. Nevertheless, the non-linear forcing term of the Navier--Stokes is non-solenoidal, although its solenoidal part is the only one affecting the velocity field \citep{chorin1993a}.\\ \indent
The general idea of the aforementioned works is to provide a method based on a linear system such that the prediction of the velocity field with a linear system is as close as possible to direct numerical simulations (DNS). Therefore, quantifying the non-linear forcing term is not the objective. However, having access to a quantitative characterization of the non-linear forcing term can be helpful for addressing the domain of validity of modelling techniques which want to mimic its effects on the dynamics. \cite{beneddine2016a} show that such quantification is convenient also to assess the validity of resolvent analyses. Moreover, even though there is general agreement on the existence of self-sustained coherent motions and visualizations of such motions are presented, there is no documentation of the structure of the non-linear forcing terms resulting from the fluctuation velocities. Since these non-linear forcing terms cause the feedback mechanism in the linearized system dynamics, its quantification is of interest to further understand the ``recycling'' of the amplified outputs in the input from the non-linear terms. Although it is understood that the linear optimal forcing resulting from resolvent analyses similar to \cite{mckeon2010a} is not parallel to the non-linear forcing term in a turbulent channel flow, and it is inferred that this non-linear forcing term has significant projection onto to the linear sub-optimals from such analysis, there is no such verification.\\ \indent
When dealing with coherent motions in the framework of stochastic forcing and response, the resolvent operator and the spectral proper orthogonal decomposition (SPOD) prove to be useful tools. SPOD assures the resulting modes to be coherent in space and time \citep{towne2018a}, while the resolvent operator can describe the input-output relationship between the cross-spectral densities (CSD) of the input and the output, on which the SPOD is based. The usage of the CSD allows to isolate the dominant energetic structures, so it avoids blurring the interpretation of the results. It is known that a relation exists between the SPOD modes and the singular modes of the resolvent operator \citep{towne2018a,lesshafft2019a,cavalieri2019a}, but investigations have been based on assumptions, without quantifying the effects of the non-linear forcing term as input.\\ \indent
In this work a quantification of the non-linear forcing term, usually treated as an input in resolvent analyses, is accomplished for turbulent channel flows. The investigated flows have friction Reynolds numbers $\ret=179$ and $\ret=543$. The resolvent framework is employed. The reference state upon which the resolvent is built is the time-averaged field, so the non-linear forcing term consists of the advection term with the fluctuation velocities. The non-linear forcing is quantified through its CSD for the most energetic near-wall and large-scale structures, such that the input-output relation of the most energetic motions is highlighted. The CSD of the non-linear forcing term is computed directly from snapshots of DNS data by means of the Welch method, with the same technique discussed by \cite{nogueira2020a}. The complete non-solenoidal forcing and its solenoidal part are presented, and its effects on the output discussed. The expected coherence of the forcing is here quantified. Moreover, inspired by the discussion in \cite{beneddine2016a}, a quantification of the key parameters to assess the validity of resolvent analyses is here possible and performed. Thus, an evaluation of the errors introduced by neglecting or modeling the non-linear forcing is possible and presented. The aim of this work is to compensate the lack of an explicit characterization of the non-linear forcing term in the literature, and provide a foundation for all the studies which choose to include assumptions about this non-linear forcing term in order to facilitate the mathematical treatment.
\\ \indent
The paper is structured as follows. In \S~2 the governing equations of the problem addressed are summarized. In \S~3 the results of the direct numerical simulations are presented, the CSD of the forcing is shown and discussed. In \S~4 the effects of the non-linear forcing term are analyzed and the low-rank property of the associated CSD demonstrated. In \S~5 an assessment of the errors introduced when resorting to modeling the non-linear forcing term are discussed by comparison with the non-linear forcing term computed from DNS data. The results are summarized and discussed in \S~6. Further details about the operators involved in the analysis and the Welch method are provided in Appendix~\ref{app:OSS}, \ref{app:divfree}, and \ref{app:welch}.\par
%
%
\section{Governing equations}
%
%
\subsection{Evolution equations for the fluctuation quantities}
This work focuses on the dynamics of perturbations about the time-averaged fields in a channel flow. The flow is incompressible and the density constant. The quantities treated here are non-dimentionalized, and the Reynolds number $Re = (3/2) U_{bulk} h/\nu$ is based on the channel half-height $h$, the constant mass-averaged streamwise velocity $U_{bulk}$, and the fluid molecular viscosity $\nu$. The domain is described with Cartesian coordinates $\bm{x} = (x,y,z)^T$, which correspond to the streamwise, wall-normal and spanwise directions. The total velocity and pressure fields can be described as the superposition of the time-averaged fields and the fluctuations, $\bm{u}_{tot} = \bm{U} + \bm{u}$ and $p_{tot} = P + p$, as in the Reynolds decomposition. Here, $\bm{U} = (U(y),0,0)^{T}$ is the mean flow in the channel, $P=P(x)$ the time-averaged pressure field, $\bm{u} = (u(\bm{x},t),v(\bm{x},t),w(\bm{x},t))^{T}$ the perturbation velocity, and $p = p(\bm{x},t)$ the perturbation pressure; $t$ being the non-dimensional time. Both the mean flow and the perturbation velocity are subject to the incompressibility condition $\nabla \cdot \bm{U} = 0$ and $\nabla \cdot \bm{u} = 0$, with $\nabla=(\partial_{x},\partial_{y},\partial_{z})^{T}$ such that $\nabla^2 = \nabla \cdot \nabla$. Assuming $\bm{U}$ and ${P}$ to be known, the momentum equations
\begin{equation}
\label{eq:momeq}
\frac{\partial \bm{u}}{\partial t} + (\bm{U} \cdot \nabla) \bm{u} + (\bm{u} \cdot \nabla) \bm{U}  = -\nabla p + \frac{1}{Re} \nabla^2 \bm{u} +\bm{b} + \bm{f},
\end{equation}
are the evolution equations for the fluctuations, where the density is included in the pressure term, $\bm{b} = -\nabla P + {Re}^{-1} \nabla^2 \bm{U} - (\bm{U} \cdot \nabla) \bm{U}$ includes the contribution of time-averaged quantities only, and $\bm{f} = - (\bm{u} \cdot \nabla) \bm{u}$ the instantaneous Reynolds stresses from the fluctuations; it is assumed that there is no external body force. It is noticeable that equation~\eqref{eq:momeq} has the same structure of the perturbation equation of a flow linearized about an equilibrium solution of the N-S equations, in which case $\bm{b}=0$ by construction.
%
%
\subsection{Harmonic and stochastic forcing analysis}
Since the mean flow is homogeneous in the wall-parallel directions, the Fourier transform can be applied along those directions and the same manipulations to obtain the Orr-Sommerfeld and Squire equations can be performed. Thus, by introducing $\bm{\hat{q}} = (\hat{v}(\alpha,y,\beta,t),\hat{\omega}_y(\alpha,y,\beta,t))$ as the vector containing the wall-normal velocity and vorticity Fourier modes, equation~\eqref{eq:momeq} can be written in terms of $\bm{\hat{q}}(\alpha,y,\beta,t)e^{i(\alpha x + \beta z)}$ and $\bm{\hat{f}}(\alpha,y,\beta,t)e^{i(\alpha x + \beta z)}$ Fourier modes. Moreover, by discretizing the wall-normal direction with $N_y$ points, and by introducing the vectors $\bf{\hat{q}}$, $2N_y\times 1$, and $\bf{\hat{f}}$, $3N_y\times 1$, as the discrete counterparts of $\bm{\hat{q}}$ and $\bm{\hat{f}}$, equation~\ref{eq:momeq} reduces to the system 
\begin{equation}
\label{eq:OSS}
\frac{\partial \bf{\hat{q}}}{\partial t} = \bf{A} \bf{\hat{q}} +\bf{B}\bf{\hat{f}},
\end{equation}
where $\bm{b}$ is not included because the focus of this study is on $\alpha \neq 0$ and $\beta\neq0$, and $\bm{b}$ is constant along the wall-parallel directions. In the discretized domain the Fourier modes of the fluctuation velocity $\bm{\hat{u}}(\alpha,y,\beta,t)e^{i(\alpha x + \beta z)}$ correspond to the vector $\hat{\bf{u}}$, $3N_y\times 1$, and can be computed as $\hat{\bf{u}} = \bf{C}\hat{\bf{q}}$, whereas $\hat{\bf{q}} = \bf{D}\hat{\bf{u}}$. The expressions for the matrices $\bf{A}$, $\bf{B}$, $\bf{C}$, $\bf{D}$ are given in Appendix~\ref{app:OSS}. For the sake of readability from now on the dependency on the wave-numbers $\alpha$ and $\beta$ is no more written explicitly.\\ \indent
Since $\bm{U}$ and $P$ are the time-averaged fields of a turbulent channel flow, the system described by equation~\eqref{eq:OSS} is linearly stable \citep{reynolds1967a}, so a finite amplitude forcing can be studied by performing the Fourier transform in time on equation~\eqref{eq:OSS}. Then, the harmonic forcing $\hat{\bf{f}} = \tilde{\bf{f}}(\omega)e^{-i \omega t}$ and the harmonic response $\hat{\bf{u}} = \tilde{\bf{u}}(\omega)e^{-i \omega t}$ are related by $\tilde{\bf{u}}(\omega) = \bfR(\omega) \tilde{\bf{f}}(\omega)$, with
\begin{equation}\label{eq:R}
\bf{R}(\omega) = -\bf{C}(i \omega \bf{I} + \bf{A})^{-1} \bf{B},
\end{equation}
the $3N_y \times 3N_y$ matrix form of the resolvent operator (with boundary conditions $\tilde{v} = \partial \tilde{v}/\partial y = \tilde{\omega}_y = 0$, or equivalently $\tilde{u} = \tilde{v} = \tilde{w} = 0$, at $y = \pm 1$). Resorting to a singular value decomposition (SVD) allows to express the resolvent matrix $\bf{R}(\omega)$ in terms of its left-singular vectors $\bfphi_i(\omega)$, its singular values $\sigma_i(\omega)$, and its right-singular vectors $\bfpsi_i(\omega)$, such that
\begin{subequations}
\begin{align}
\bfpsi_{i}^{H} \mathbf{W} \bfpsi_j &= \delta_{ij},\label{eq:U}\\
\bfphi_{i}^{H} \mathbf{W} \bfphi_j &= \delta_{ij} ,\label{eq:V}
\end{align}
\end{subequations}
with $^H$ the complex conjugate transpose, $\delta_{ij}$ the Kronecker delta, and $\mathbf{W}$ the positive definite hermitian matrix, $3N_y\times 3N_y$, of quadrature weights necessary to compute the energy norm on the discrete grid of the the wall-normal direction. This decomposition explicitly shows if low rank approximations based on the singular values $\sigma_{i}$ are applicable.\\ \indent
If instead of harmonic excitation a stochastic and statistically stationary forcing is considered, the response will also be stochastic and statistically stationary. In this case the Fourier transform in time cannot be applied because $\int_{-\infty}^{\infty}| \hat{\bfu}|^2 \ \mathrm{d}t< \infty$ (or $\int_{-\infty}^{\infty}| \hat{\mathbf{q}}|^2 \ \mathrm{d}t< \infty$) and $\int_{-\infty}^{\infty}| \hat{\bff}|^2 \ \mathrm{d}t < \infty$ do not hold \citep{chibbaro2014a}. A quantity that exists and can be computed for a statistically stationary process is the cross-spectral density (CSD) \citep{stark1986a}, which is defined for the vectors $\hat{\bfu}$ and $\hat{\bff}$ as
\begin{subequations}
\begin{align}
\bfS(\omega)& = \lim_{T \to \infty}  \frac{ \ \mathbb{E}\left[ \left( \frac{1}{2\pi}\int_{-T}^{T}\hat{\bfu}(t) e^{-i\omega t} \ \mathrm{d}t \right)\left(\frac{1}{2\pi}\int_{-T}^{T}\hat{\bfu}(t)^{H} e^{i\omega t}\ \mathrm{d}t \right) \right]}{2T},\label{eq:CSDa} \\
\bfP(\omega)& = \lim_{T \to \infty}  \frac{ \ \mathbb{E}\left[ \left(\frac{1}{2\pi} \int_{-T}^{T}\hat{\bff}(t) e^{-i\omega t} \ \mathrm{d}t \right)\left(\frac{1}{2\pi}\int_{-T}^{T}\hat{\bff}(t)^{H} e^{i\omega t}\ \mathrm{d}t \right) \right]}{2T},\label{eq:CSDb}
\end{align}
\end{subequations}
where $T$ is the total time, the expectation $\mathbb{E} \left[ \cdot \right]$ is the ensemble average over different stochastic realizations, and the cross-spectral densities are the $3N_y \times 3N_y$ matrices $\bfS(\omega)$ and $\bfP(\omega)$. The diagonals of $\bfS(\omega)$ and $\bfP(\omega)$ contain the power-spectral density (PSD) of the three velocity and forcing components at the discrete points of the wall-normal direction for a given angular frequency $\omega$ (and the omitted $\alpha$ and $\beta$). For the sake of readability the streamwise, wall-normal and spanwise components on the diagonal of $\bfS(\omega)$ and $\bfP(\omega)$ are from now on referred to with the $N_y\times 1$ vectors $\bfs_{uu}, \ \bfs_{vv}, \ \bfs_{ww}$, and $\bfp_{uu}, \ \bfp_{vv}, \ \bfp_{ww}$. Then, the premultiplied streamwise kinetic energy spectra can be computed as
\begin{equation}
\alpha \beta  \bfe_{uu} = \alpha \beta \int_{-\infty}^{\infty} \bfs_{uu}(\omega) \ \mathrm{d}\omega.
\end{equation}
Since the system in equation~\eqref{eq:OSS} is stable, it also holds \citep{stark1986a}
\begin{equation}\label{eq:Rinout}
\bfS(\omega) = \bfR(\omega)\bfP(\omega)\bfR(\omega)^{H},
\end{equation}
which is the input-output relation of the CSDs. For the sake of readability from now on the dependency on the angular frequency $\omega$ is no more written explicitly.\\ \indent
Since $\bfS$ and $\bfP$ are CSDs, the Karhunen--Lo\`eve decomposition can be performed, which for quantities in the frequency domain is referred to as spectral proper orthogonal decomposition (SPOD) \citep{picard2000a,towne2018a}. The SPOD modes of the matrices $\bfS$ and $\bfP$ correspond to the $3N_y \times 1$ eigenvectors $\bfxi_i$ and $\bfzeta_i$ of the matrix eigenvalue problems $\mathbf{SW}\bfxi_i=\mu_{i}\bfxi_i$ and $\mathbf{PW}\bfzeta_i=\eta_i \bfzeta_i$. The eigenvectors are orthogonal in $\bfxi_i^H\mathbf{W}\bfxi_j=\delta_{ij}$ and $\bfzeta_i^H\mathbf{W}\bfzeta_j=\delta_{ij}$. Thus, it holds the expansion $\bfS = \sum_i \mu_i \bfxi_i\bfxi^{H}_i$ and $\bfP = \sum_i \eta_i \bfzeta_i\bfzeta^{H}_i$, such that the diagonal of $\bfS$ and $\bfP$ can be written as
\begin{subequations}
\begin{align}
\diag(\bfS) = \sum_{i} \mu_i |\bfxi_i |^2,\label{eq:Sspod} \\
\diag(\bfP) = \sum_{i} \eta_i |\bfzeta_i |^2, \label{eq:Pspod}
\end{align}
\end{subequations}
with $| \cdot |$ the absolute value of each entry of the vector.
It is noticeable that since the SPOD modes are orthogonal in the inner product associated to the energy norm, the ratios $\mu_{i}/\sum_{i}\mu_i$ and $\eta_{i}/\sum_{i}\eta_i$ represent the fraction of power associated with the $i$-th SPOD mode.\\ \indent
The computation of the left- and right-singular vectors of $\bfR$ allows to split the input-output relation $\bfS=\bfR\bfP\bfR^H$ into three steps: (i) the projection of $\bfP$ onto the right-singular vectors $\bfpsi_i$, which results in a scalar
\begin{equation}\label{eq:bi}
b_i^2 = \diag(\bfpsi_i^H\mathbf{W}\bfP\mathbf{W}\bfpsi_i)
\end{equation}
for each $\bfpsi_i$, (ii) the amplification or damping of the associated singular values $\sigma_i$ by $b_i$, which results in a scalar $a_i = \sigma_i b_i$, and (iii) the linear combination of the left-singular vectors $\bfphi_i$ weighted with $a_i$ such that
\begin{equation}\label{eq:bcoeff}
\diag(\bfS) =  \sum_i |\bfphi_i|^2 a_i^2.
\end{equation}
Since $a_i^2 = \sigma_i^2 b_i^2$ and $\sigma_i$ do not depend on $\bfP$, it is the coefficients $b_i$ which quantify the contribution of the forcing $\bfP$ to the output $\bfS$. Note that if $b_i = 1$ then $\bfP \equiv \bfI$.\\ \indent
It is noticeable that $\bfS$ is built upon a solenoidal vector field while $\bfP$ is built upon a non-solenoidal vector field; in fact, the divergence of the non-linear forcing term $\bm{f}$ is non-zero. Moreover, if the flow is incompressible only the solenoidal part of the forcing affects the velocity field \citep{chorin1993a}. Therefore, if the forcing is written as the sum of a solenoidal and an irrotational vector field, the irrotational part gives a null response in equation~\eqref{eq:Rinout}. This implies that $\bfR$ is singular. Since the irrotational part of the input results in a null output, the solenoidal part of $\bfP$ can be retrieved from the velocity field, and it coincides with $\bfL\bfS\bfL^{H}$, where
\begin{equation}
\bfL = -\bfC(i \omega \bfI + \bfA)\bfD.
\end{equation}
It should be noted that $\bfL$ is not exactly the inverse of $\bfR$ in equation~\eqref{eq:Rinout} because $\bfR$ is singular. If the forcing $\hat{\bff}$ is known, its solenoidal part can be computed also as $\bfC \bfB\hat{\bff}$, which can be employed to evaluate the solenoidal part of $\bfP$. Resorting to $\bfL$ or $\bfC \bfB$ to compute the solenoidal part of the forcing is equivalent. A more detailed discussion about $\bfL$ and $\bfC \bfB$ is presented in Appendix~\ref{app:divfree}.\\ \indent
\subsection{Modeling the non-linear forcing terms}
The input-output relationship described by equation~\eqref{eq:Rinout} includes the contribution of the non-linear terms, those responsible for the Reynolds stresses, in the input $\bfP$. The non-linear terms are usually unknown, and the input $\bfP$ is modeled. The lack of knowledge about the non-linear terms implies that the accuracy of these modeling techniques cannot be based on a direct comparison with them, instead the error in the prediction of the velocity field or its statistics is evaluated. Since this work aims at presenting the actual $\bfP$ which appears in the NS, its direct comparison with a modeled $\bfP$ can be evaluated. Moreover, $\bfP$ has never been quantified from instantaneous realizations $\hat{\bff}$ of a turbulent channel flow. Thus, a comparison with the results from often used modeling methods is clearly of interest.\\ \indent
The two modeling approaches discussed in this work are: (i) the assumption that the non-linear terms are uncorrelated in space $\bfP = \gnu \bfI$ (with $\gnu$ a normalization scalar, and $\bfI$ the identity), and (ii) the introduction of an eddy-viscosity $\nu_t$ to model a part of the non-linear terms via the Boussinesq expression; in (ii) the unmodeled part of the non-linear terms is treated as uncorrelated in space. The two approaches lead to the predictions
\begin{subequations}
\begin{align}
\bfS_{\bfR_{\nu}} &=  \gnu \bfR \bfR^H, \label{eq:Rnu} \\
\bfS_{\bfR_{\nu_t}} &=  \gnut \bfR_{\nu_t} \bfR_{\nu_t}^H, \label{eq:Rnut}
\end{align}
\end{subequations}
where $\bfR_{\nu_t}$ is a modified resolvent which includes the eddy-viscosity modeling (details about the operator are given in Appendix~\ref{app:OSS}), and $\gnut$ a normalization scalar.\\ \indent
The forcing necessary to obtain the prediction $\bfS_{\bfR_{\nu_t}}$ by means of $\bfR$, such that $\bfS_{\bfR_{\nu_t}} = \bfR\bfP_{\nu_t}\bfR^H$, can be computed as
\begin{equation}\label{eq:Pnut}
\bfP_{\nu_t} = \bfL \bfS_{\bfR_{\nu_t}} \bfL^{H},
\end{equation}
which quantifies how the eddy-viscosity approach models $\bfP$. The effects of the modeling with $\bfP = \gnu \bfI$, of the eddy-viscosity approach with $\bfP = \bfP_{\nu_t}$, are compared to the $\bfP$ computed from instantaneous realizations of $\hat{\bff}$ via the coefficients $b_i$ in equation~\eqref{eq:bi}.\\ \indent
The normalization scalars $\gnu$ and $\gnut$ are function of the wave-numbers $\alpha, \ \beta$ and the angular frequency $\omega$. They are computed here from the power-spectral densities as
\begin{subequations}
\begin{align}
\gnu(\alpha,\beta,\omega)    &= \frac{|| \diag(\bfS) ||_{\infty}} {||  \diag(\bfR_{\nu} \bfR^H_{\nu}) ||_{\infty}}, \\
\gnut(\alpha,\beta,\omega) &= \frac{||  \diag(\bfS) ||_{\infty}} {||  \diag(\bfR_{\nu_t}\bfR^H_{\nu_t}) ||_{\infty}},
\end{align}
\end{subequations}
and represent a rescaling factor to compensate for the lack of knowledge on the amplitude of the modeled forcing term such that the responses match the DNS $\diag(\bfS)$ in the $\infty$-norm. The scalars also give an indication about the offset of the prediction of the amplitude; in fact, if the modeled forcing were to coincide with the one computed from the DNS data, $\gnu = 1$ or $\gnut=1$. For the sake of readability, from now on the explicit dependency of the scalars $\gnu, \ \gnut$ on $\alpha, \ \beta, \ \omega$ is dropped. The $\infty$-norm of a vector $\mathbf{g}$ is intended as
$|| \mathbf{g} ||_{\infty} = \max_i |g_i|$ with $i$ the $i$-th term of the vector. 
%
%
\section{Direct numerical simulations}\label{sec:DNSresults}
\begin{figure}%
\begin{center}
{\includegraphics[width=0.329\textwidth,trim={0 0 0 0},clip]{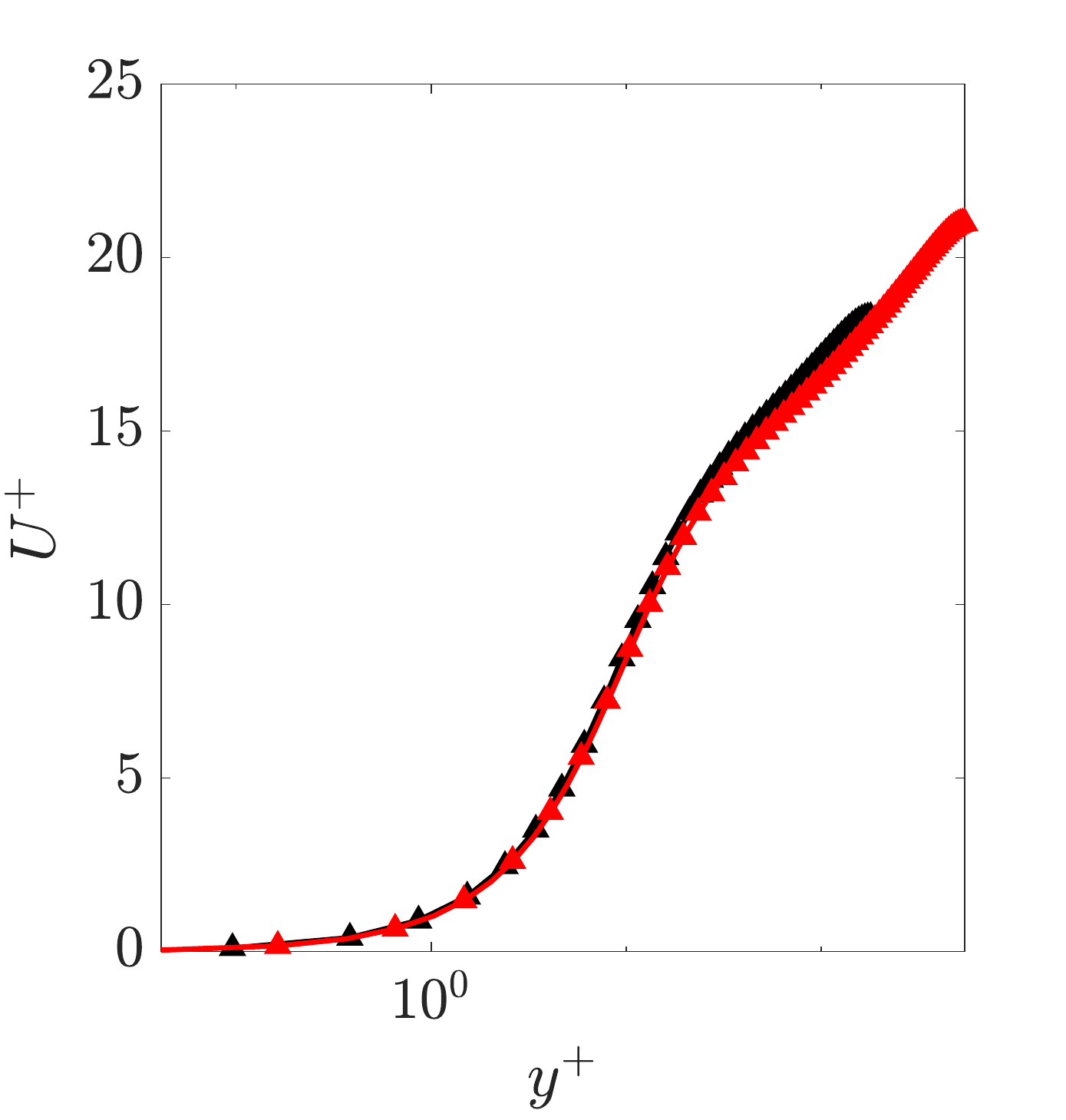}}\put(-130,120){$a)$}\hspace{0.2cm}
{\includegraphics[width=0.28\textwidth,trim={0 0 0 0},clip]{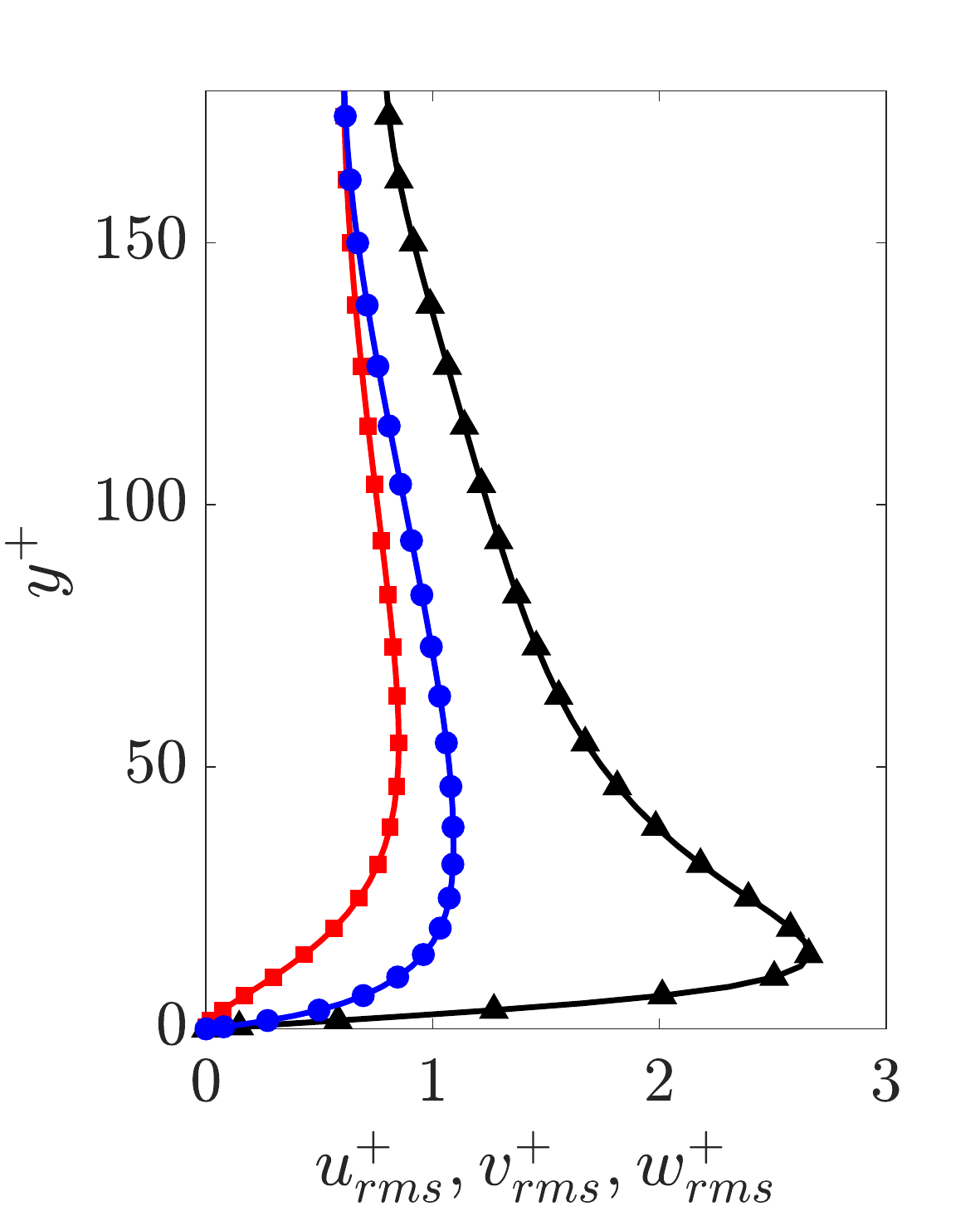}}\put(-110,120){$b)$}\hspace{0.2cm}
{\includegraphics[width=0.28\textwidth,trim={0 0 0 0},clip]{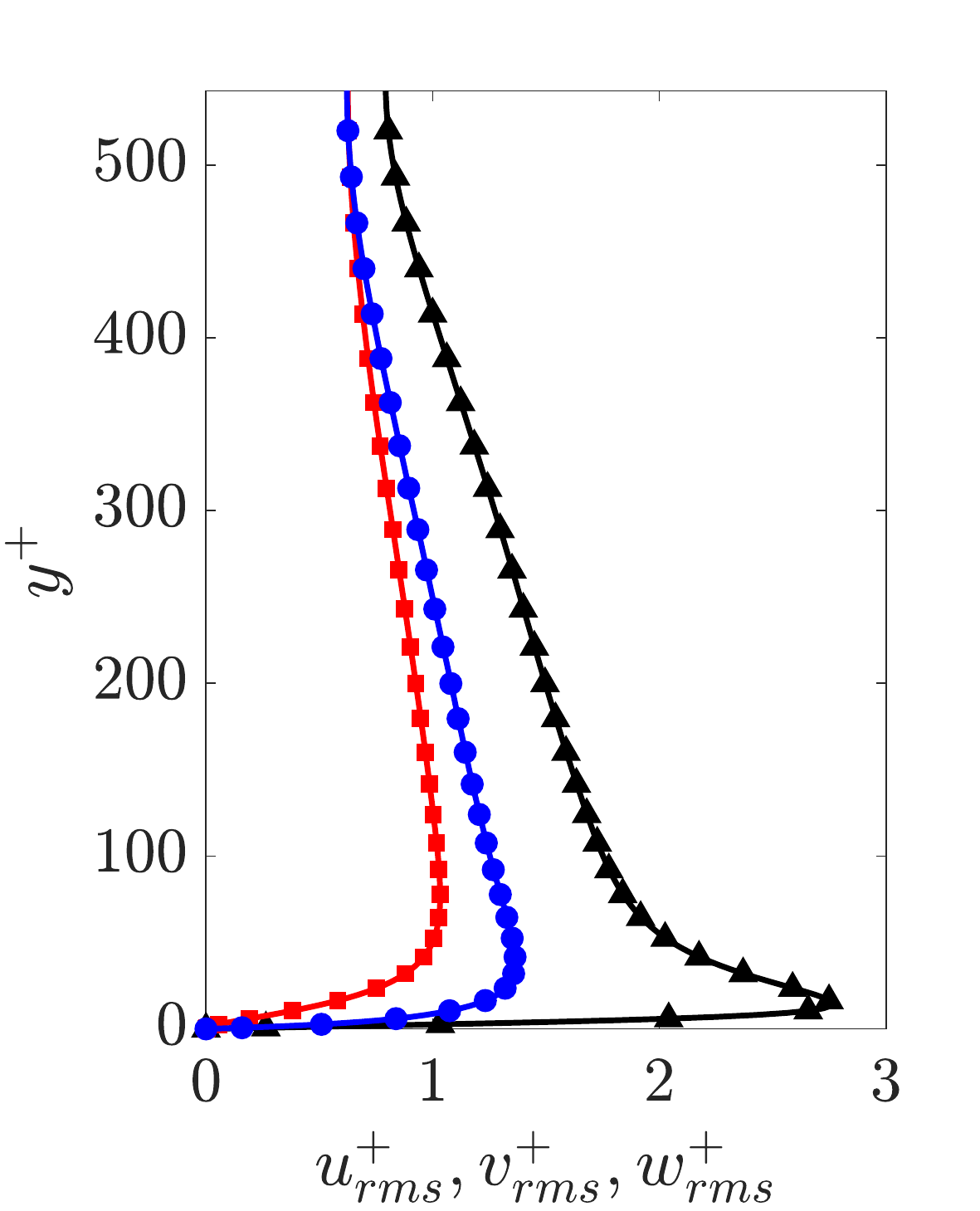}}\put(-110,120){$c)$}
\end{center}
\caption{Comparison of mean velocity $U^+$ and \textit{rms} values from DNS data and the results presented in \cite{delalamo2003a}. a) Mean velocity $U^+$ in inner units; symbols: present DNS; lines: reference. b) $u_{rms}^+$,$v_{rms}^+$,$w_{rms}^+$ in inner units for $\ret=179$; symbols: present DNS; lines: reference; black is $u_{rms}^+$, red is $v_{rms}^+$, blue is $w_{rms}^+$. c) $u_{rms}^+$,$v_{rms}^+$,$w_{rms}^+$ in inner units for $\ret=543$; symbols: present DNS; lines: reference; black is $u_{rms}^+$, red is $v_{rms}^+$, blue is $w_{rms}^+$.}
\label{fig:Urms}
\end{figure}

The flow cases analyzed are at $\ret = 179$ and $\ret = 543$ with the box details presented in table~\ref{tab:cases}. The mean velocity profile and the \textit{rms} of the three velocity components are presented in figure~\ref{fig:Urms}, where it is shown that the profiles are in agreement with the results of \cite{delalamo2003a}. The CSD are computed with the Welch's method as suggested by  \cite{martini2019a} for a dynamical system and tested in \cite{nogueira2020a} for the minimal turbulent unit of a Couette flow. More details in Appendix~\ref{app:welch}.
\begin{table}
\centering
\begin{tabular}{rrrrrrrr}
\multicolumn{1}{c}{$\ret$}  & \multicolumn{1}{c}{$Re_{bulk}$}  & \multicolumn{1}{c}{$L_x$}  & \multicolumn{1}{c}{$L_z$}            & \multicolumn{1}{c}{$\Delta x ^+$} & \multicolumn{1}{c}{$\Delta z^+$}  & \multicolumn{1}{c}{$\Delta y_{min} ^+$} & \multicolumn{1}{c}{$\Delta y_{max}^+$}\\ 
\\
179 & 2800 & $4\pi$ & $2\pi$ & 11.78 & 5.89  & $5.42 \times 10^{-2}$ & 4.42 \\
543 & 10000 &$2\pi$ & $\pi$ & 8.89 & 4.44 & $4.09 \times 10^{-2}$ & 6.67 \\
\end{tabular}
\caption{Reynolds number, box dimensions, and details about the resolution used in the DNS.}
\label{tab:cases}
\end{table}
%
%
\begin{figure}%
\begin{center}
{\includegraphics[width=0.48\textwidth,trim={5 15 15 60},clip]{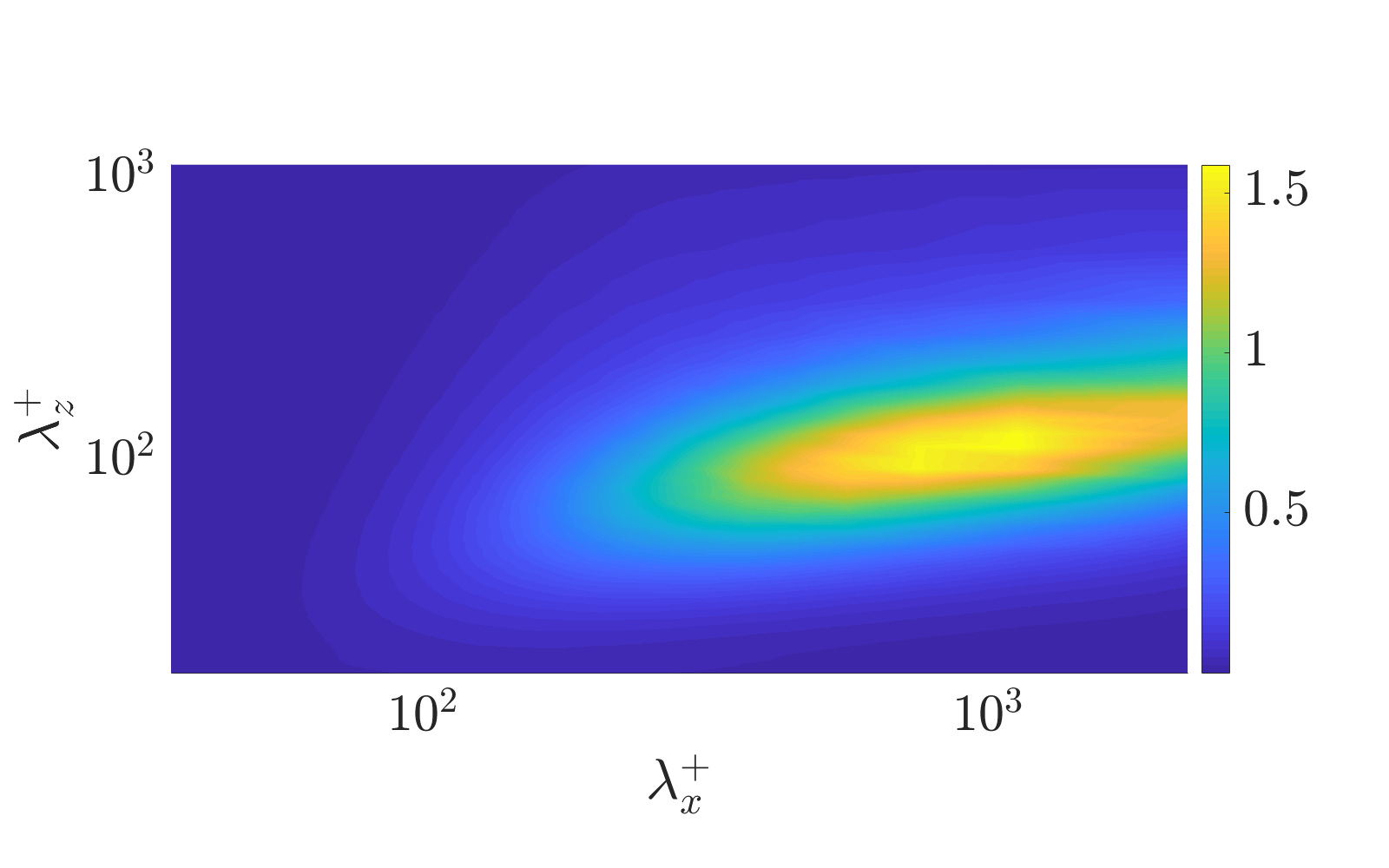}}
{\includegraphics[width=0.48\textwidth,trim={5 15 15 60},clip]{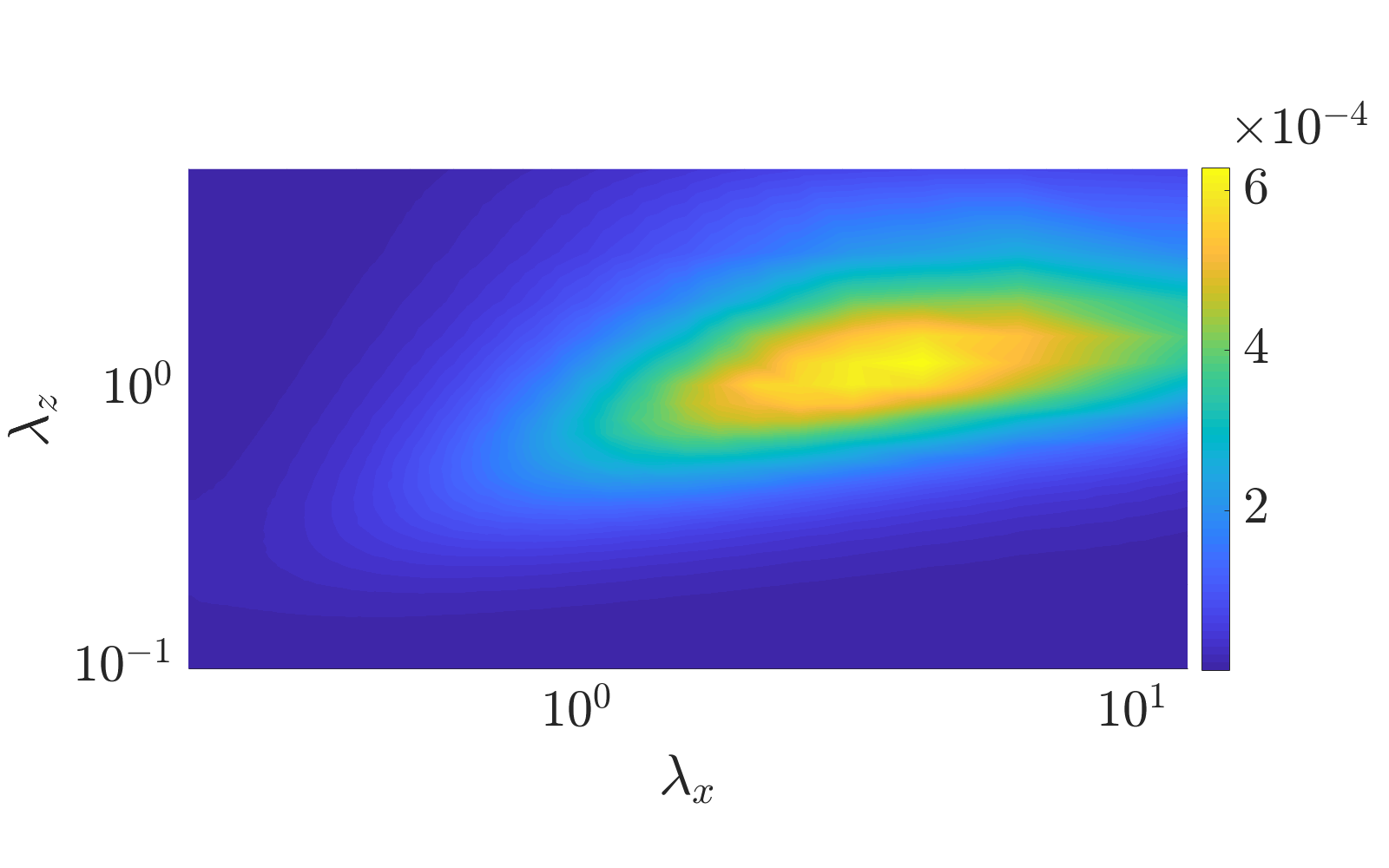}}
\end{center}
\caption{$\ret = 179$. Premultiplied streamwise energy spectra $\alpha \beta \bfe_{uu}$. Left: $y^+ = 15$ plane, spectra in inner units. Right: $y = 0.5$ plane, spectra in outer units.}
\label{fig:abEuuRet180}
\end{figure}%
\begin{figure}%
\begin{center}
{\includegraphics[width=0.4\textwidth,trim={0 0 0 0},clip]{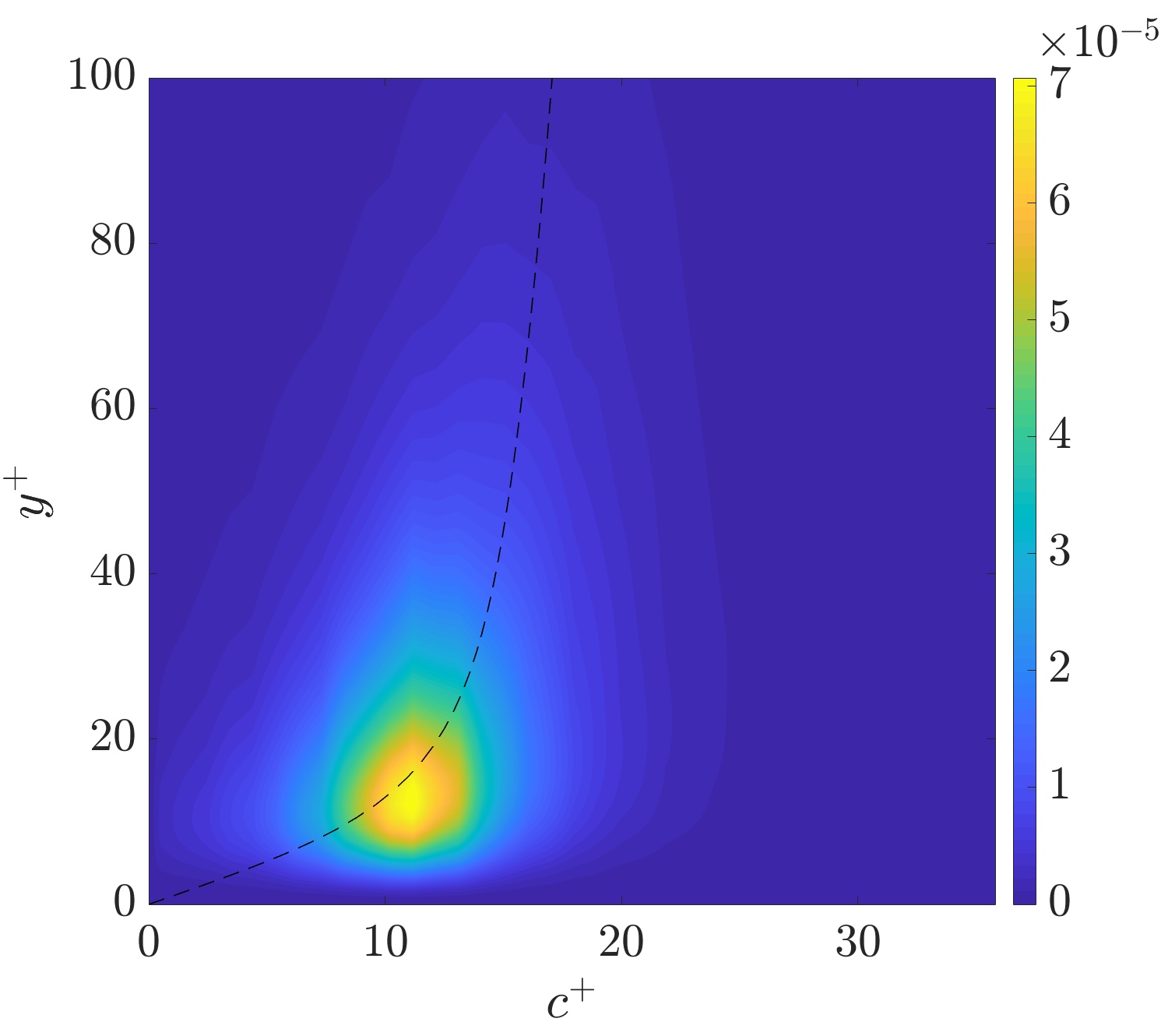}}\put(-155,125){$a)$}\hspace{0.8cm}
{\includegraphics[width=0.4\textwidth,trim={0 0 0 0},clip]{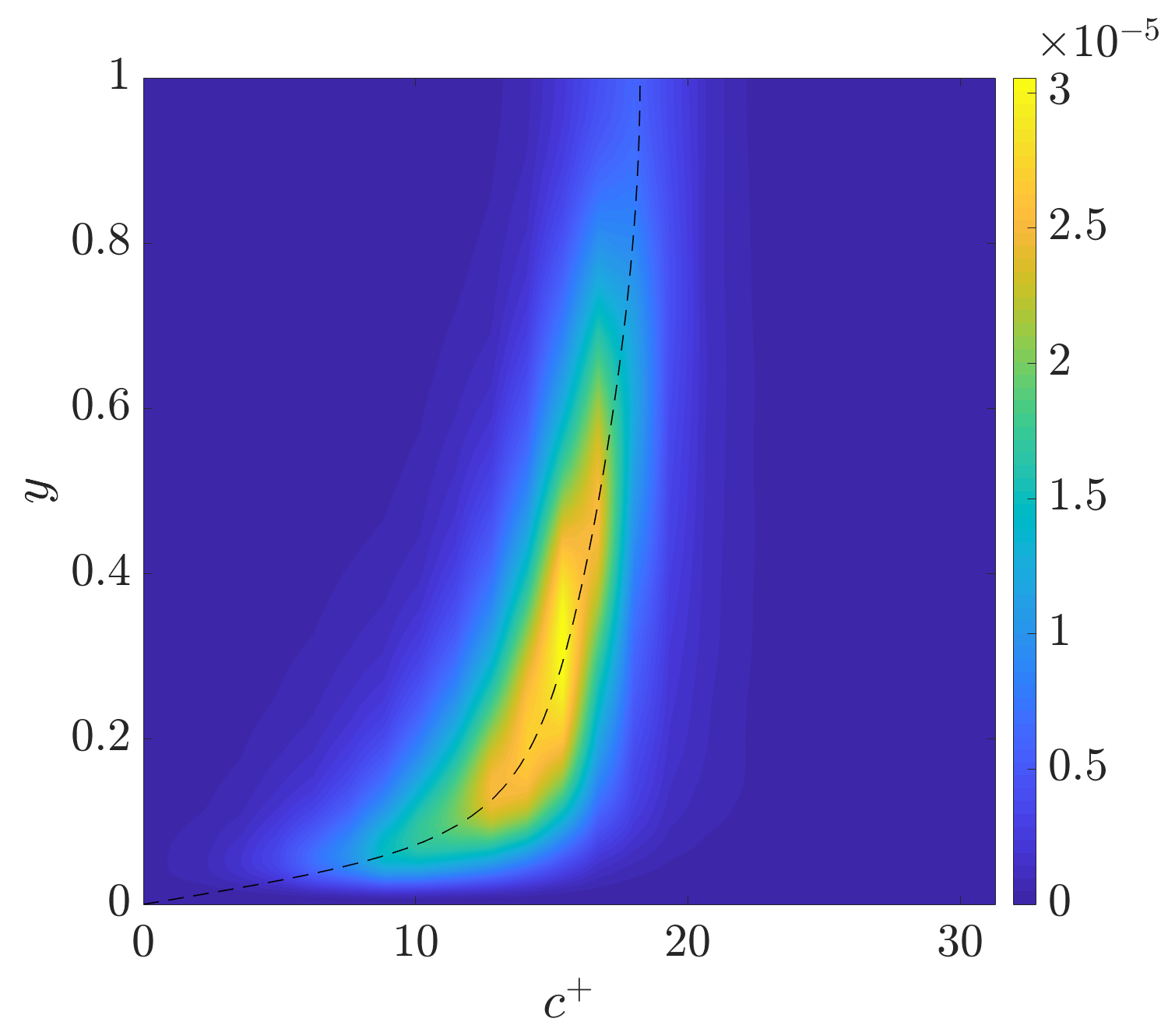}}\put(-150,125){$b)$}\\
{\includegraphics[width=0.4\textwidth,trim={0 0 0 0},clip]{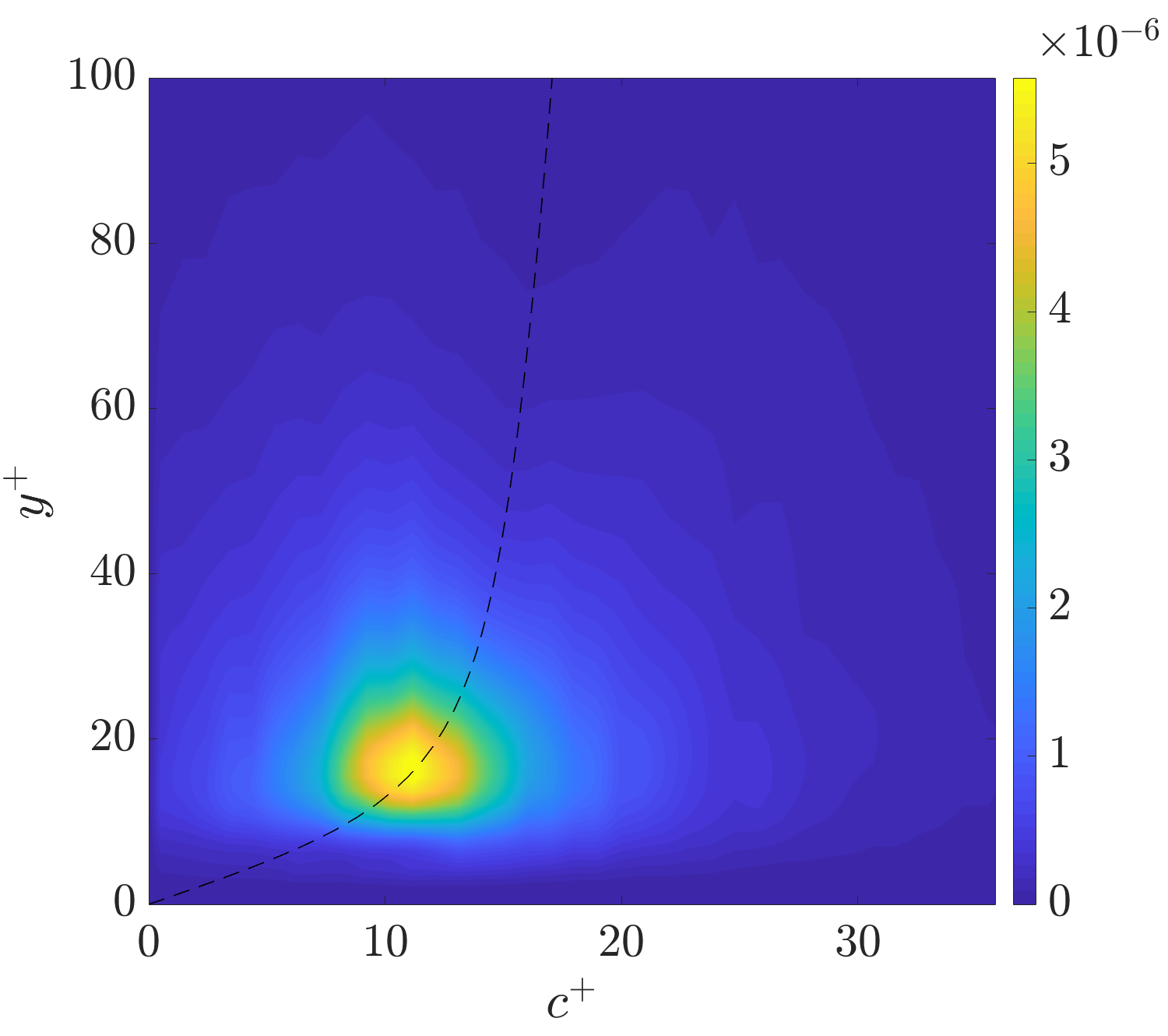}}\put(-155,125){$c)$}\hspace{0.8cm}
{\includegraphics[width=0.4\textwidth,trim={0 0 0 0},clip]{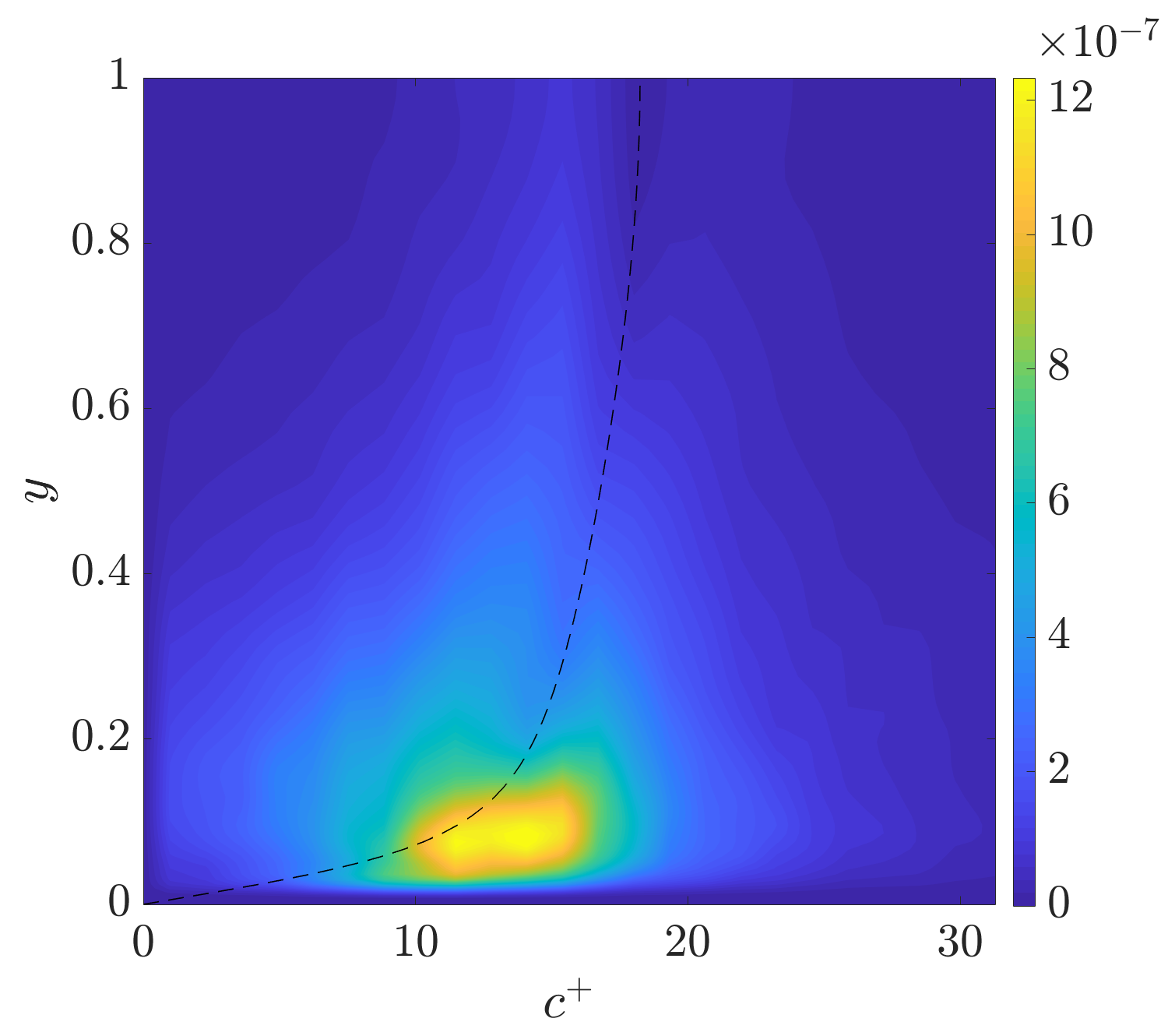}}\put(-150,125){$d)$}\\
\end{center}
\caption{$\ret = 179$. Streamwise velocity power spectral densities (PSD) $\bfs_{uu}$ and $\bfp_{uu}$ versus phase speed $c^+=\omega^+/\alpha^+$. Dashed black line: mean velocity $U^+$ in wall units. a) $\bfs_{uu}$, near-wall structures, $(\lambda_x^+,\lambda_z^+) = (1130,113)$; b) $\bfs_{uu}$, large-scale structures, $(\lambda_x,\lambda_z) = (4.19,1.26)$; c) $\bfp_{uu}$, near-wall structures, $(\lambda_x^+,\lambda_z^+) = (1130,113)$; d) $\bfp_{uu}$, large-scale structures, $(\lambda_x,\lambda_z) = (4.19,1.26)$.}
\label{fig:SPRet180}
\end{figure}%
\begin{figure}%
\begin{center}
{\includegraphics[width=0.44\textwidth,trim={10 0 50 0},clip]{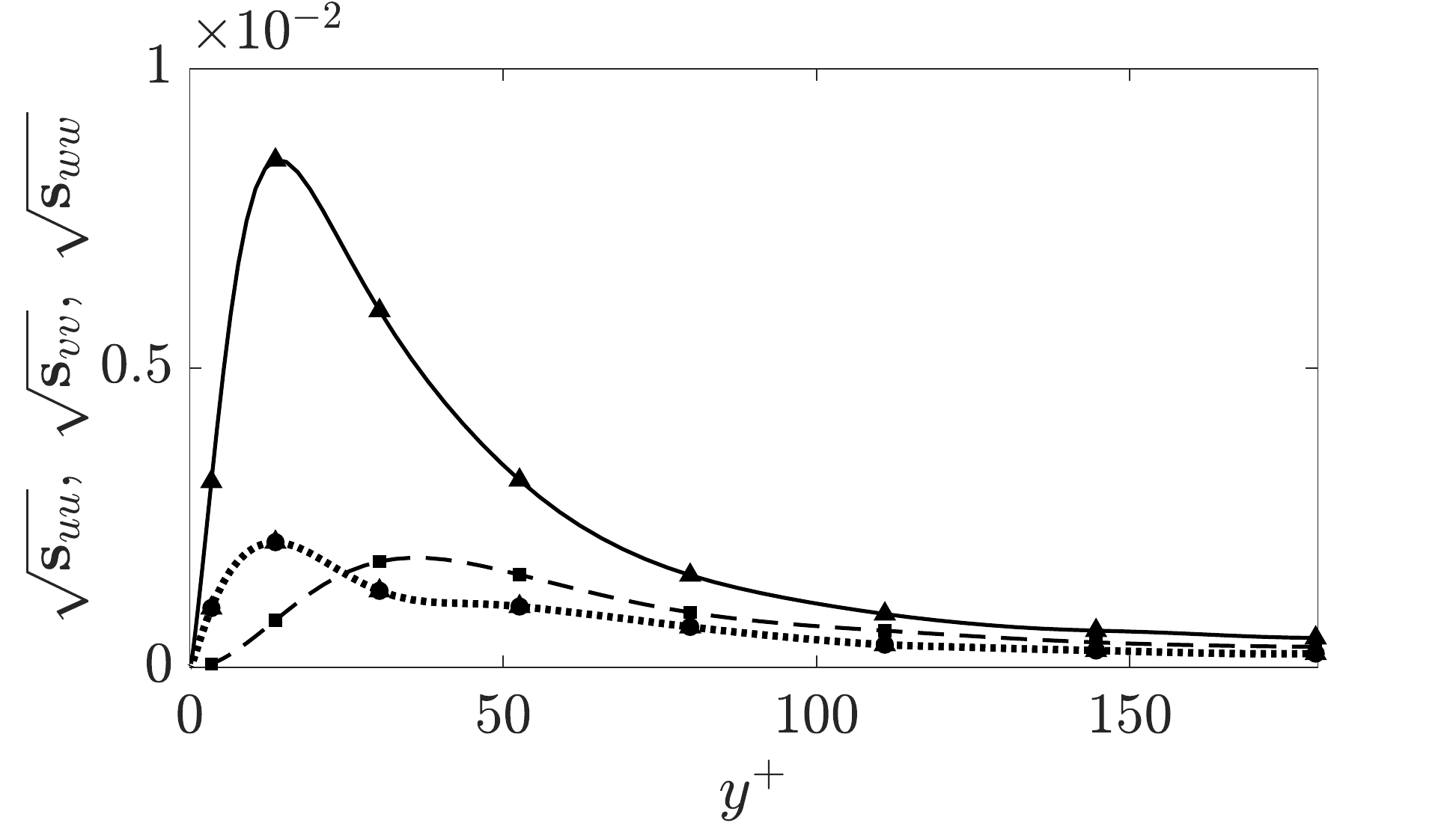}}\put(-180,95){$a)$}\hspace{0.8cm}
{\includegraphics[width=0.43\textwidth,trim={0 0 0 0},clip]{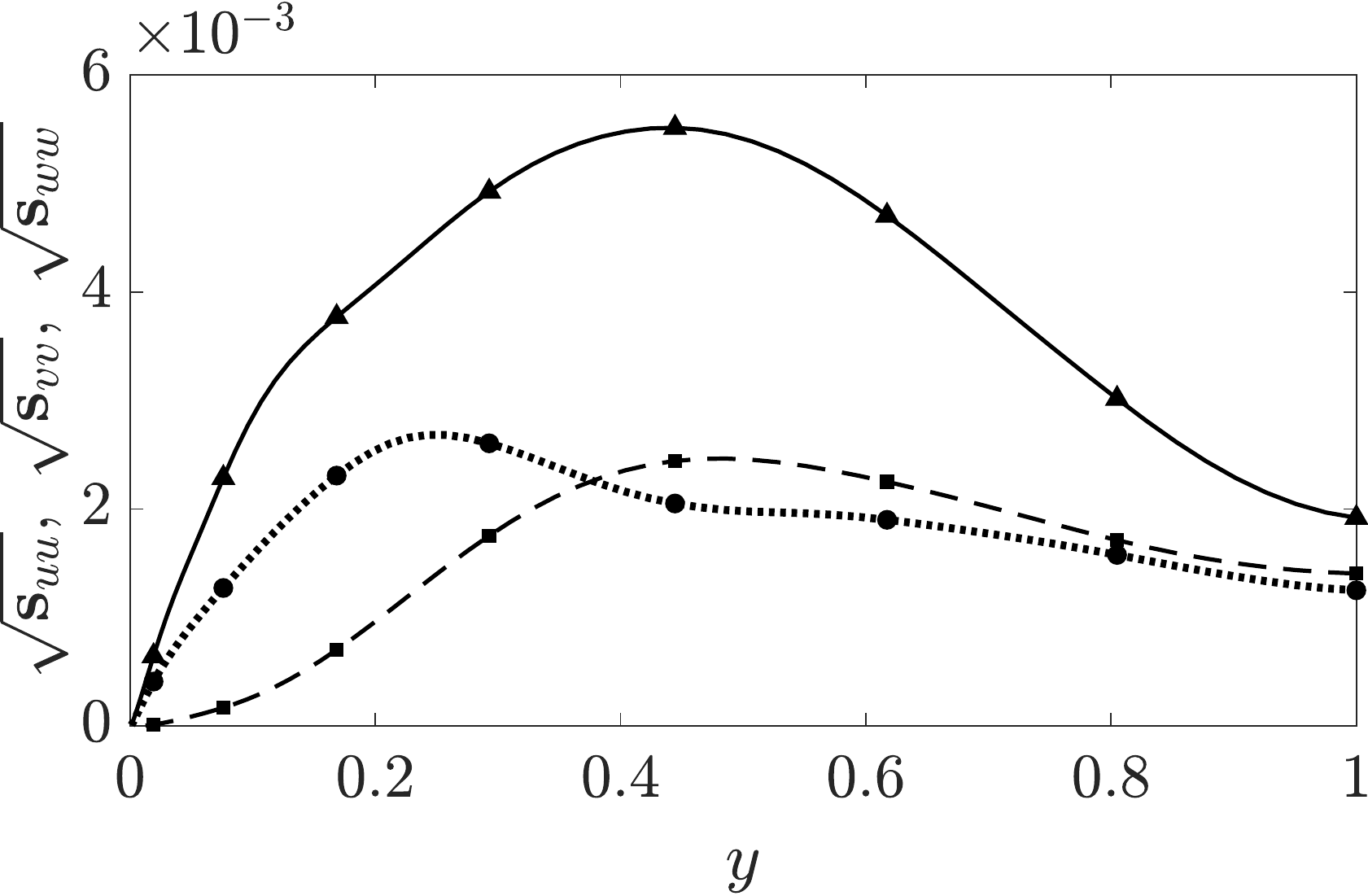}}\put(-175,95){$b)$}\\
{\includegraphics[width=0.44\textwidth,trim={0 0 0 0},clip]{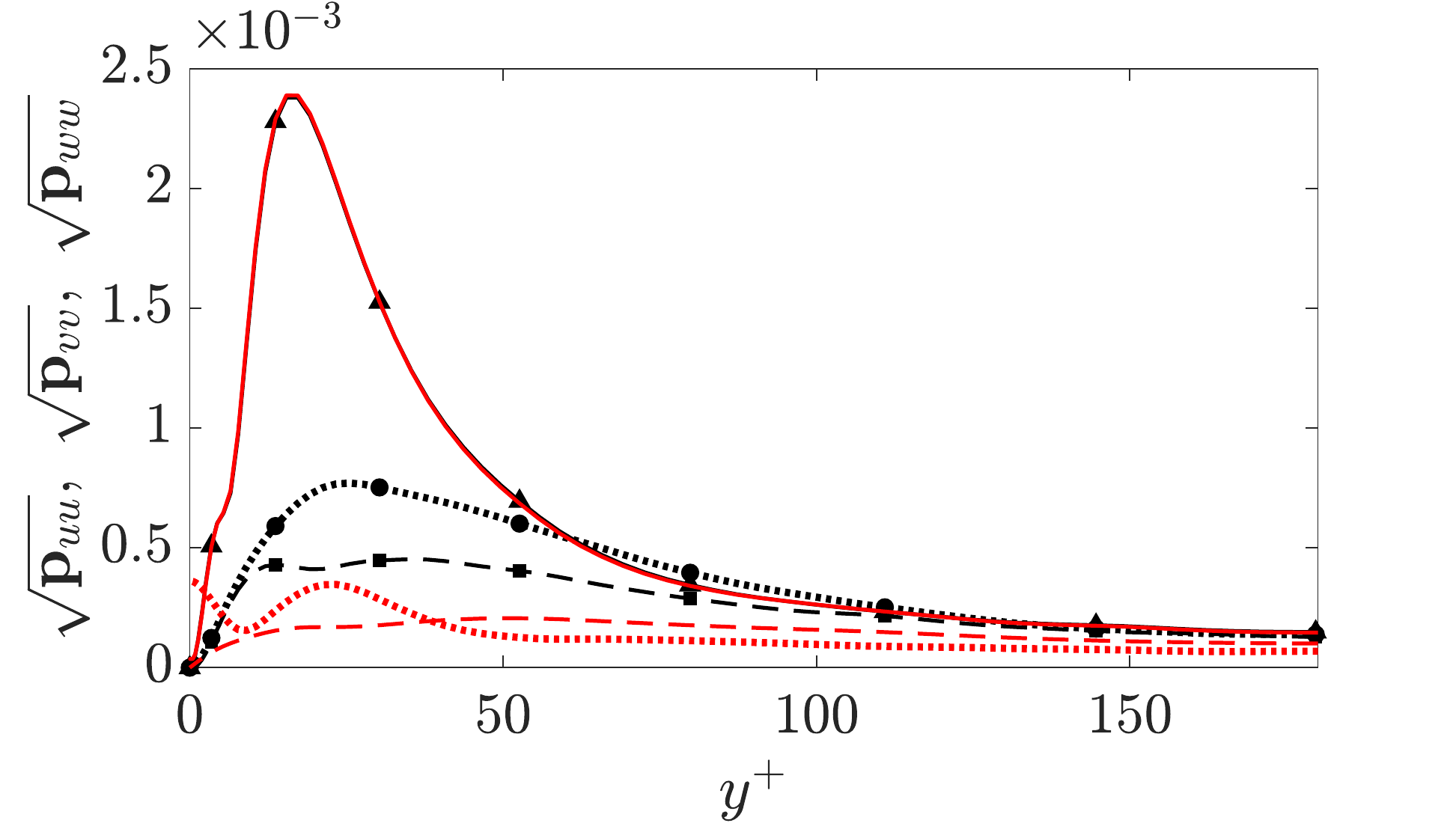}}\put(-180,95){$c)$}\hspace{0.8cm}
{\includegraphics[width=0.44\textwidth,trim={0 0 0 0},clip]{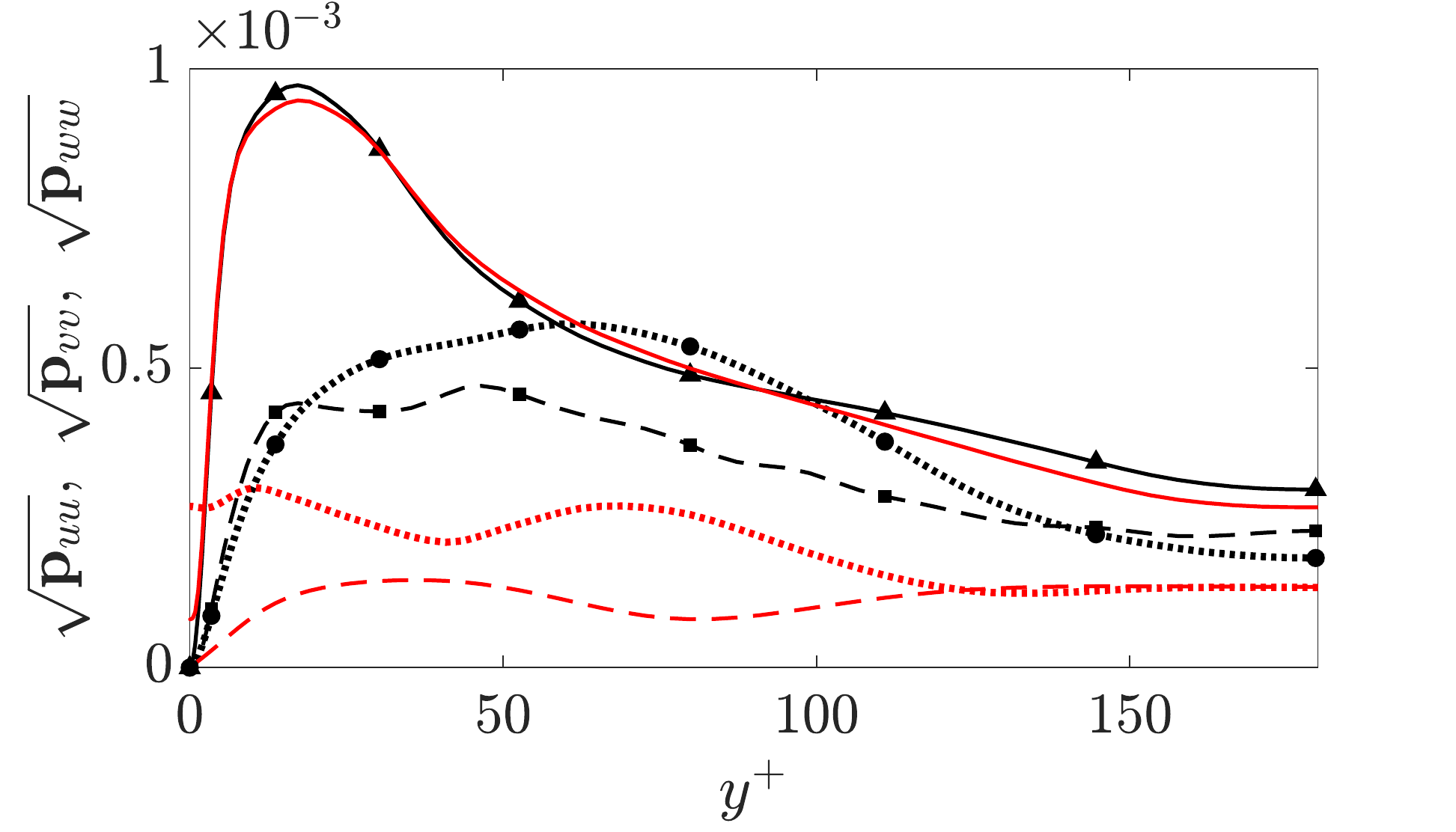}}\put(-180,95){$d)$}\\
\end{center}
\caption{$\ret = 179$. PSD. Symbols: $\bfS = \bfR\bfP\bfR^H$; triangle: streamwise component; circle: spanwise component; square: wall-normal component. Black lines: DNS. Red lines: solenoidal part of the forcing from DNS. Solid line: streamwise component; dotted line: spanwise component; dashed line: wall-normal component. a) $\bfs_{uu}$, $\bfs_{vv}$, $\bfs_{ww}$ near-wall, $\omega_{max}^+ = 0.065$; b) $\bfs_{uu}$, $\bfs_{vv}$, $\bfs_{ww}$ large-scale, $\omega_{max} = 1.05$; c) $\bfp_{uu}$, $\bfp_{vv}$, $\bfp_{ww}$ near-wall, $\omega_{max}^+ = 0.065$; d) $\bfp_{uu}$, $\bfp_{vv}$, $\bfp_{ww}$ large-scale, $\omega_{max} = 1.05$.}
\label{fig:SPioRet180}
\end{figure}%
\subsection{$\ret = 179$}
The focus is on the near-wall and large-scale processes only, so the premultiplied streamwise kinetic energy spectra $\alpha \beta \bfe_{uu}$ at $y^+ = 15$ and $y = 0.5$ is shown in figure~\ref{fig:abEuuRet180}. The maxima of the premultiplied spectral densities are at $(\lambda_x^+,\lambda_z^+) = (1130,113)$ for the near-wall structures and $(\lambda_x,\lambda_z) = (4.19,1.26)$ for the large-scale structures; where $\lambda_x$ and $\lambda_z$ are streamwise and spanwise wavelengths normalized by the outer scale $h$, and the $^+$ superscript is used when quantities are normalized by the viscous scale. These wave numbers are chosen for the following analysis because there is the highest energetic activity, and because they represent the self-sustained dynamics of the near-wall and large-scale motions. Figures~\ref{fig:SPRet180}a,b show the PSD of the streamwise velocity fluctuation $\bfs_{uu}$ for the near-wall structures $(\lambda_x^+,\lambda_z^+) = (1130,113)$ (panel (a)) and the large-scale structures $(\lambda_x,\lambda_z) = (4.19,1.26)$ (panel b). It appears that the near-wall structures are localized close to the wall, with a peak at $y^+ \approx 15$ with $\omega_{max}^+ = 0.065$ which corresponds to a time in viscous scale $\lambda_t^+ \approx 100$, while the large-scales have a peak around $y= 0.35$ with $\omega_{max} = 1.05$. Moreover, figures~\ref{fig:SPRet180}c,d show the PSD of the streamwise forcing $\bfp_{uu}$ for the same near-wall and large-scale structures. Both the forcing terms have the peak at the $\omega_{max}$ of the corresponding $\bfS_{11}$. Moreover, it is noticeable that both the near-wall and the large-scale structures are forced by a near-wall forcing. This phenomenon can be appreciated also in figure~\ref{fig:SPioRet180}, where the forcing $\bfP$ computed from the DNS data is used to predict $\bfS = \bfR\bfP\bfR^H$. The fact that the curves and the symbols in figures~\ref{fig:SPioRet180}a,b are on top of each other is evidence of the accuracy of the computed $\bfP$. Figure~\ref{fig:SPioRet180}c,d show both the forcing $\bfP$ based on the non-linear terms and its solenoidal part. The streamwise component is nearly unchanged, whereas the wall-parallel components are different. In particular, the amplitude of the wall-parallel components of the solenoidal part of the forcing is lower than it is for the total forcing and the solenoidal forcing is non-zero on the wall, but it is parallel to it.\\ \indent
%
%
\begin{figure}%
\begin{center}
{\includegraphics[width=0.48\textwidth,trim={5 15 15 60},clip]{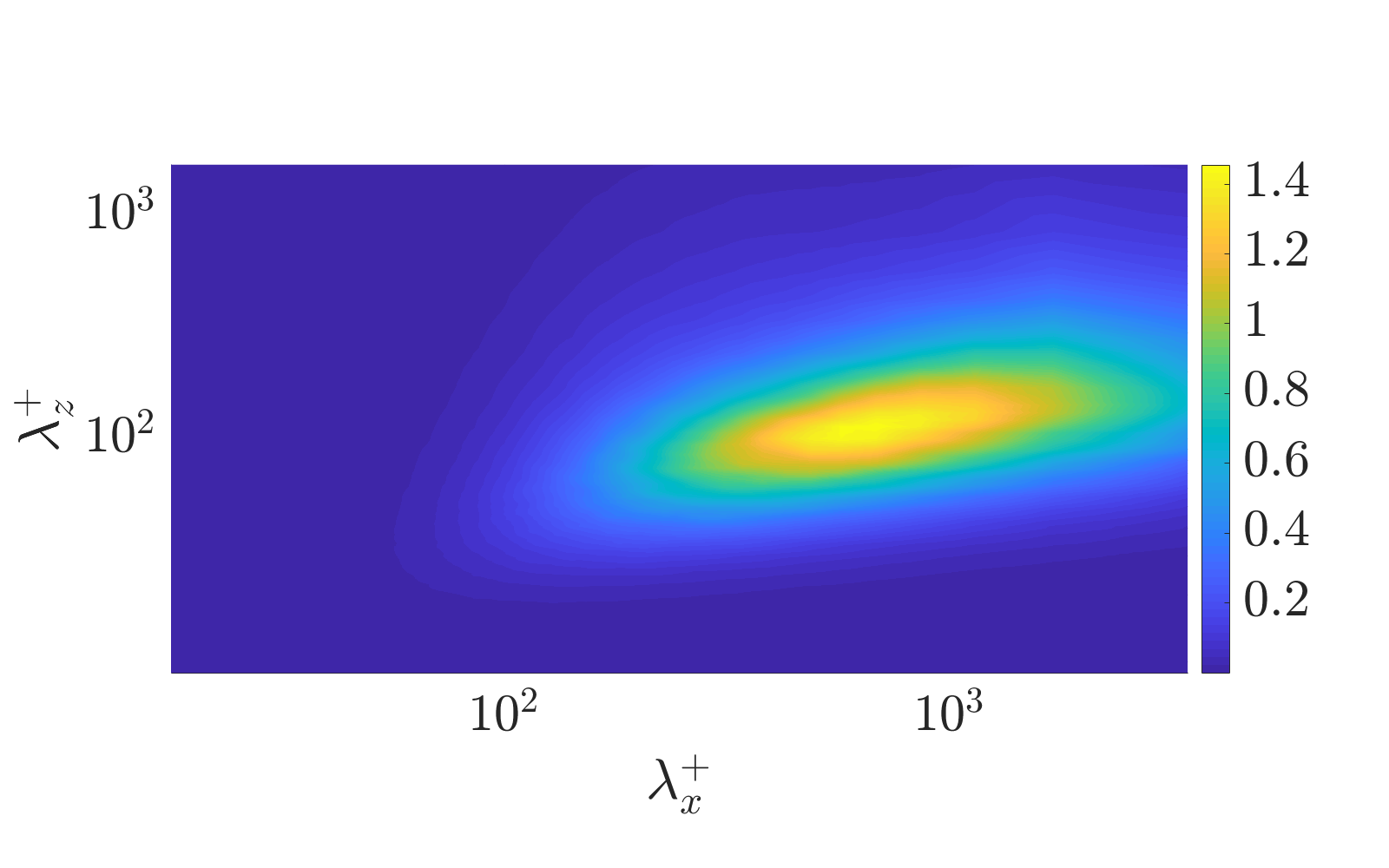}}
{\includegraphics[width=0.48\textwidth,trim={5 15 15 60},clip]{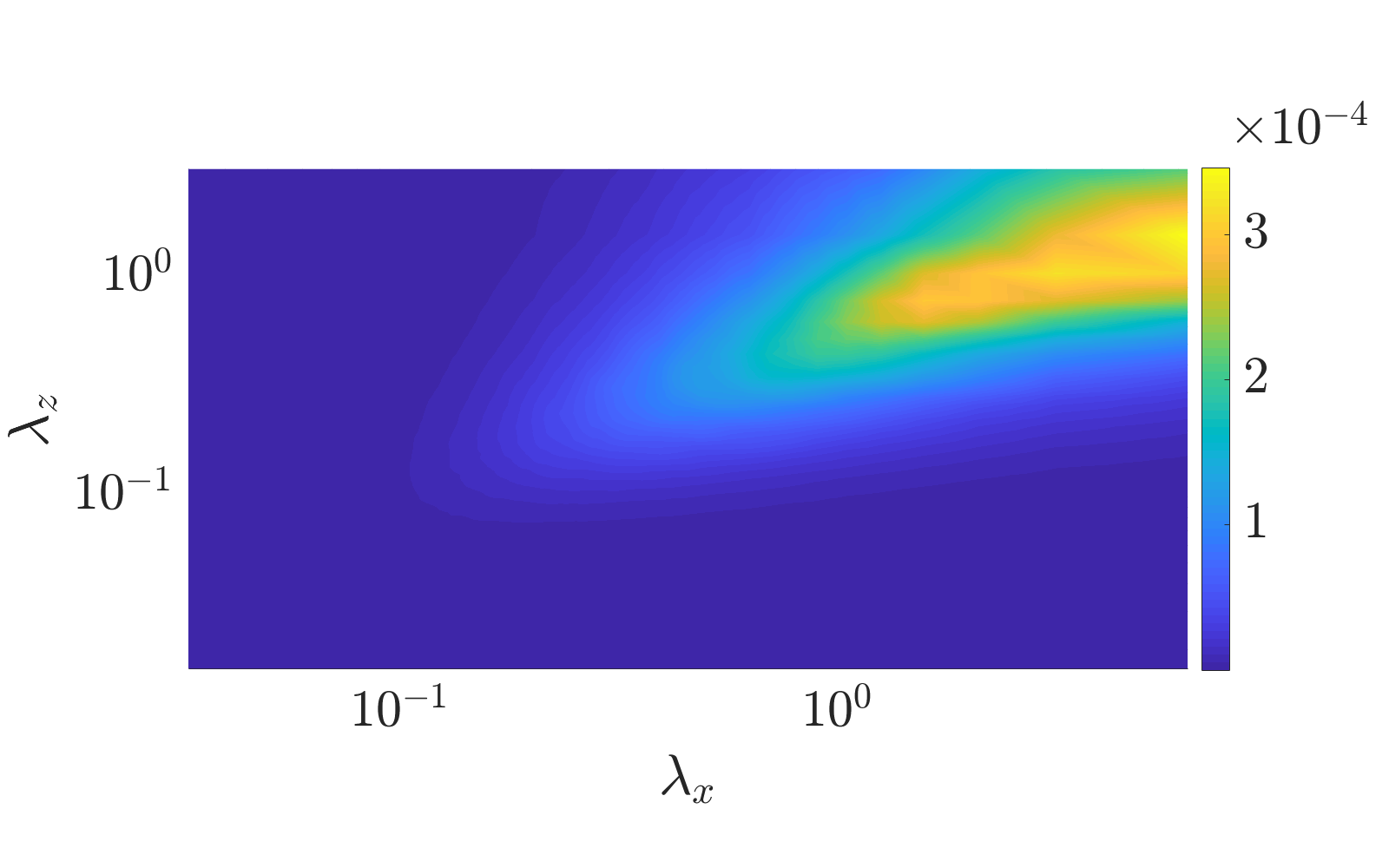}}
\end{center}
\caption{$\ret = 543$. Premultiplied streamwise energy spectra $\alpha \beta \bfe_{uu}$. Left: $y^+ = 15$ plane, spectra in inner units. Right: $y = 0.5$ plane, spectra in outer units.}
\label{fig:abEuuRet550}
\end{figure}%
\begin{figure}%
\begin{center}
{\includegraphics[width=0.4\textwidth,trim={0 0 0 0},clip]{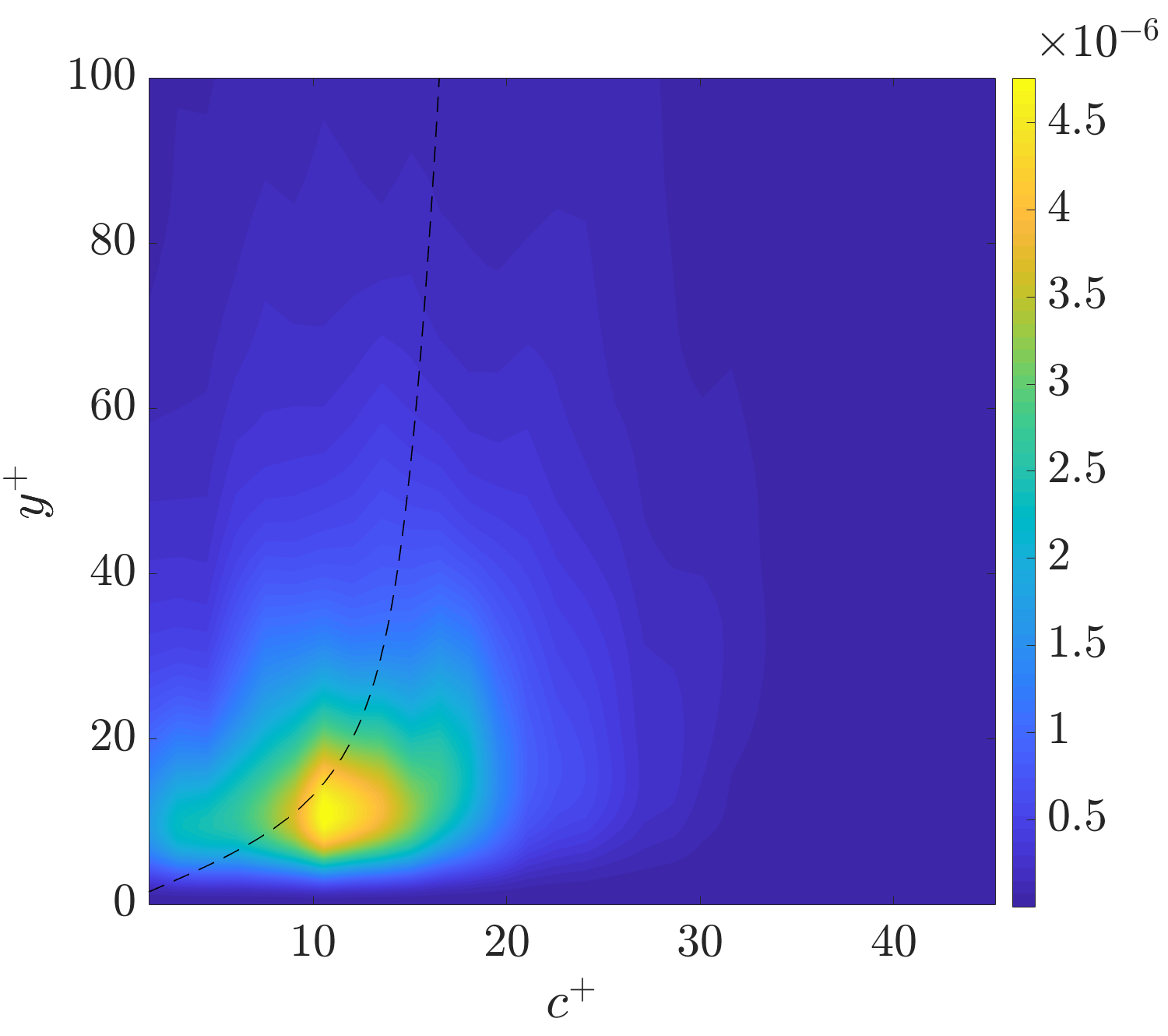}}\put(-155,125){$a)$}\hspace{0.8cm}
{\includegraphics[width=0.4\textwidth,trim={0 0 0 0},clip]{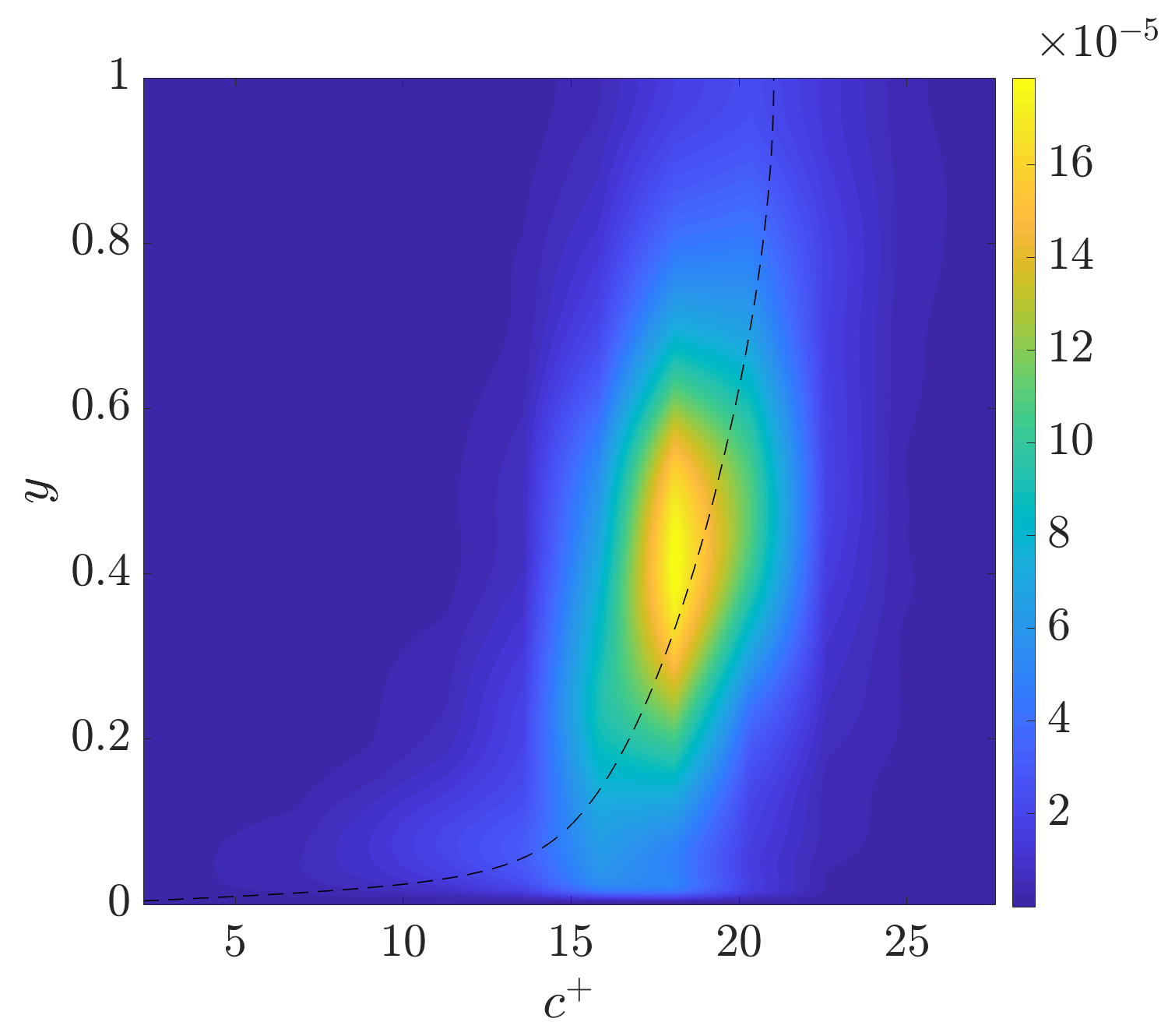}}\put(-150,125){$b)$}\\
{\includegraphics[width=0.4\textwidth,trim={0 0 0 0},clip]{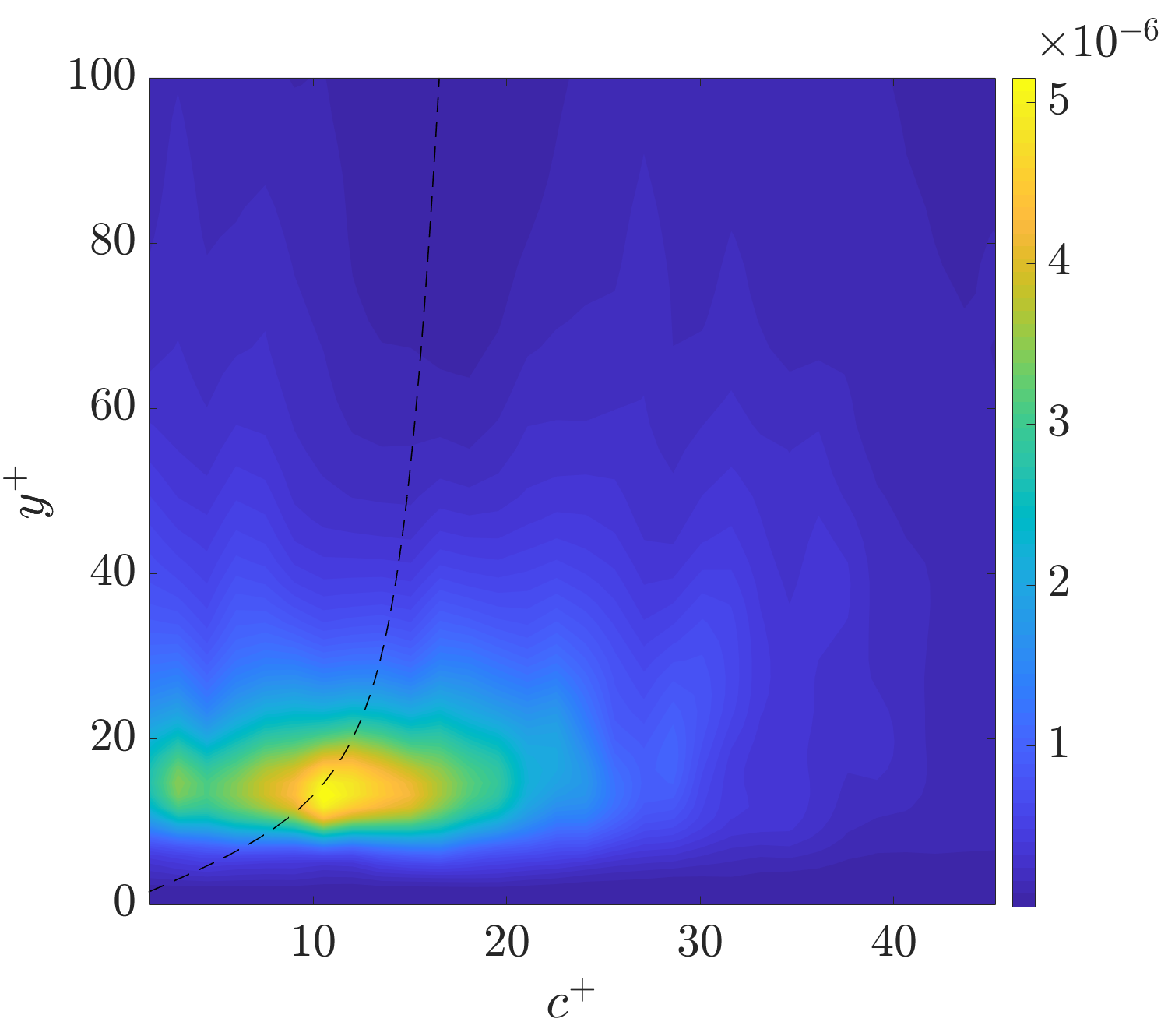}}\put(-155,125){$c)$}\hspace{0.8cm}
{\includegraphics[width=0.4\textwidth,trim={0 0 0 0},clip]{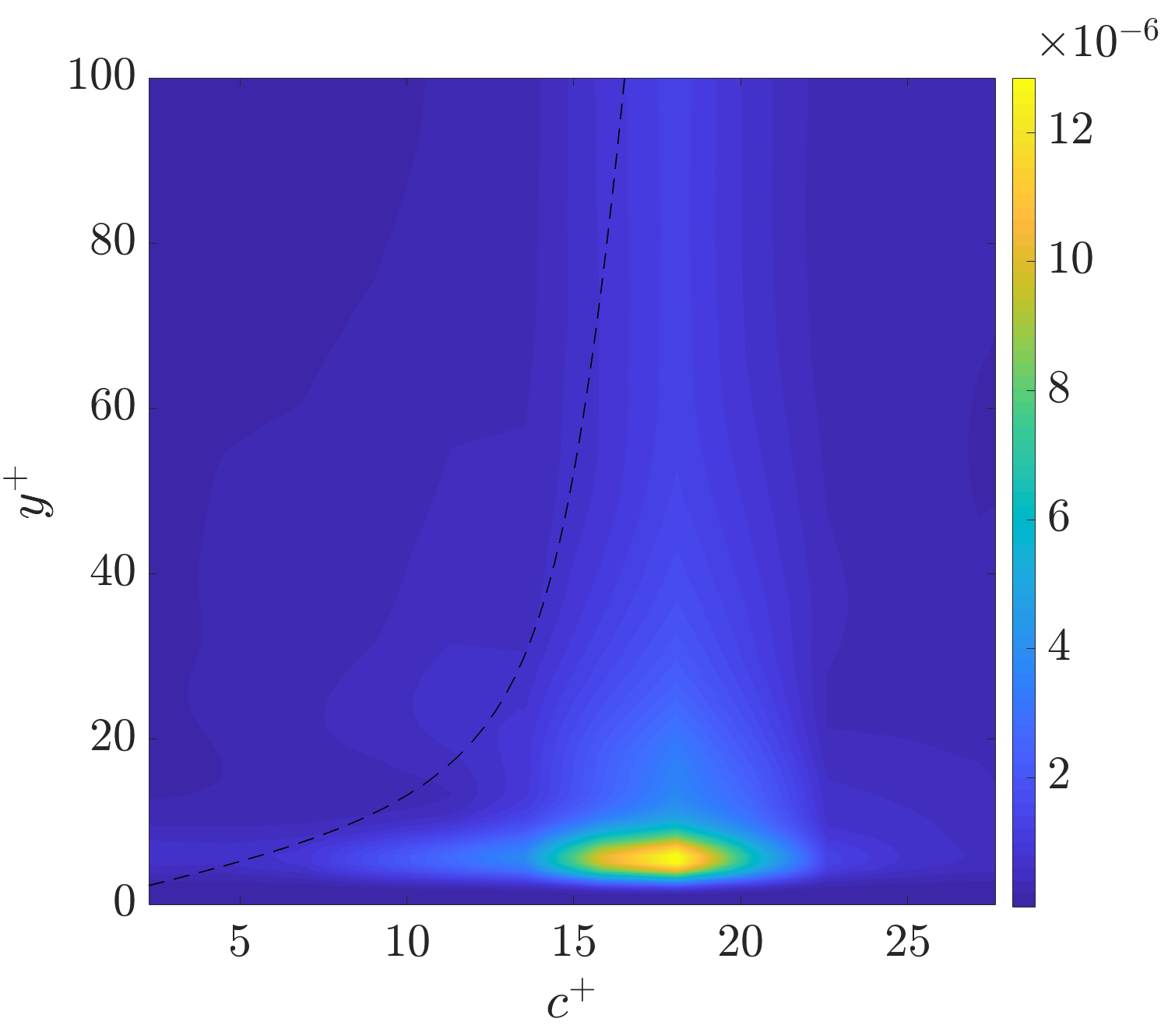}}\put(-155,125){$d)$}\\
\end{center}
\caption{$\ret = 543$. Streamwise velocity power spectral densities (PSD) $\bfs_{uu}$ and $\bfp_{uu}$ versus phase speed $c^+=\omega^+/\alpha^+$. Dashed black line: mean velocity $U^+$ in wall units. a) $\bfs_{uu}$, near-wall structures, $(\lambda_x^+,\lambda_z^+) = (1137,100)$; b) $\bfs_{uu}$, large-scale structures, $(\lambda_x,\lambda_z) = (6.28,1.57)$; c) $\bfp_{uu}$, near-wall structures, $(\lambda_x^+,\lambda_z^+) = (1137,100)$; d) $\bfp_{uu}$, large-scale structures, $(\lambda_x,\lambda_z) = (6.28,1.57)$.}
\label{fig:SPRet550}
\end{figure}%
\begin{figure}%
\begin{center}
{\includegraphics[width=0.44\textwidth,trim={0 0 0 0},clip]{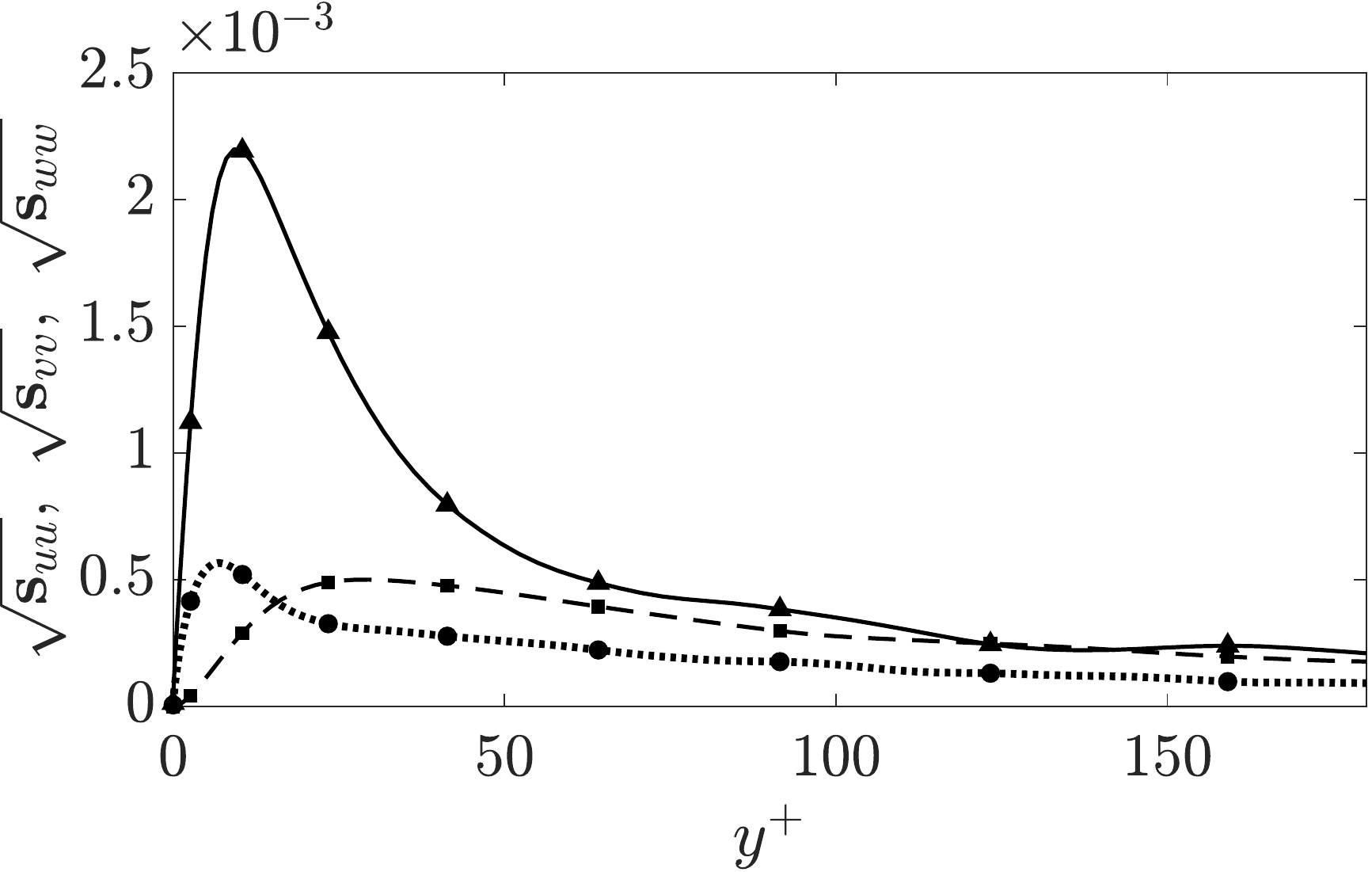}}\put(-180,95){$a)$}\hspace{0.7cm}
{\includegraphics[width=0.435\textwidth,trim={10 0 50 0},clip]{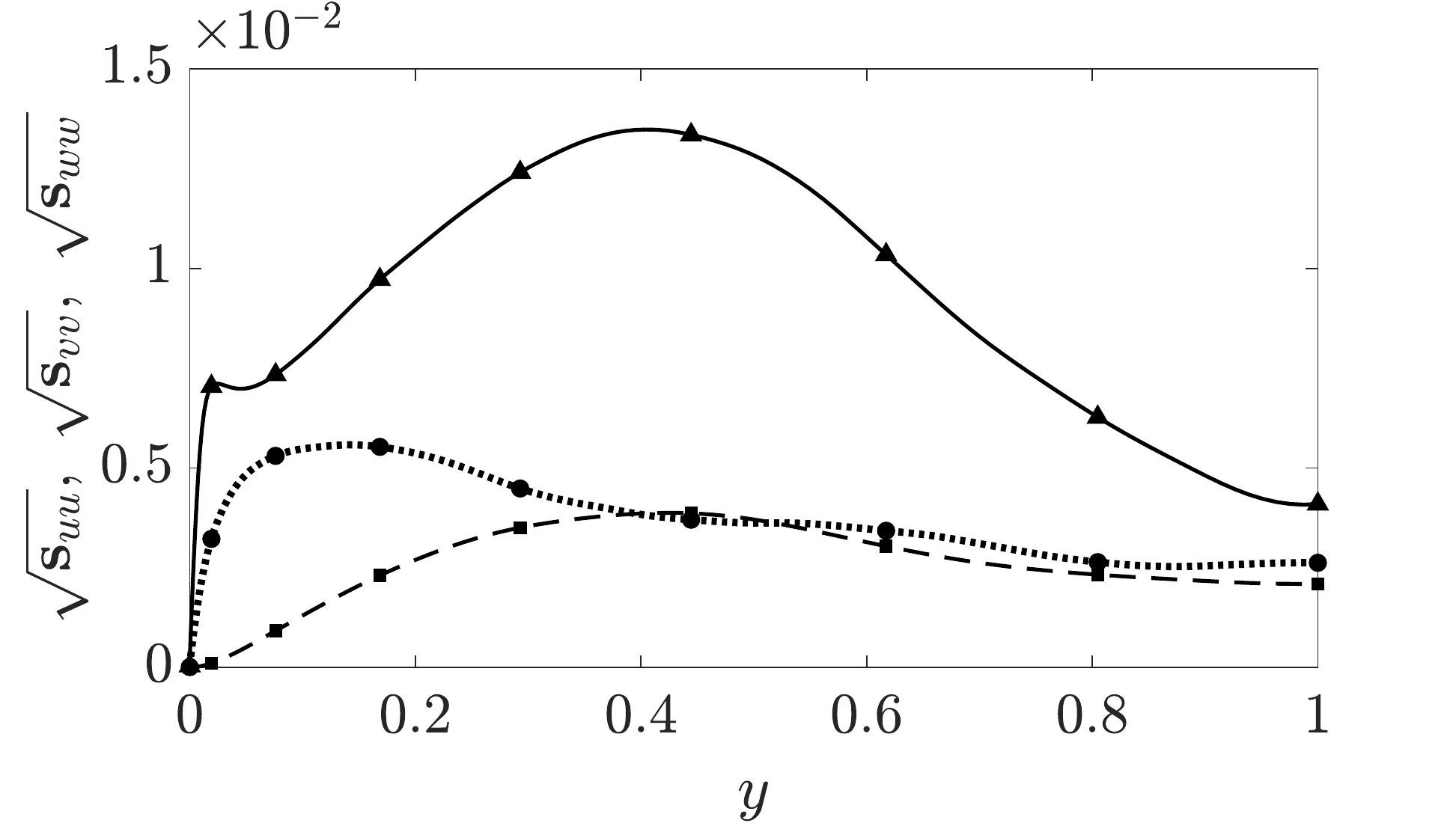}}\put(-180,95){$b)$}\\
{\includegraphics[width=0.444\textwidth,trim={0 0 0 0},clip]{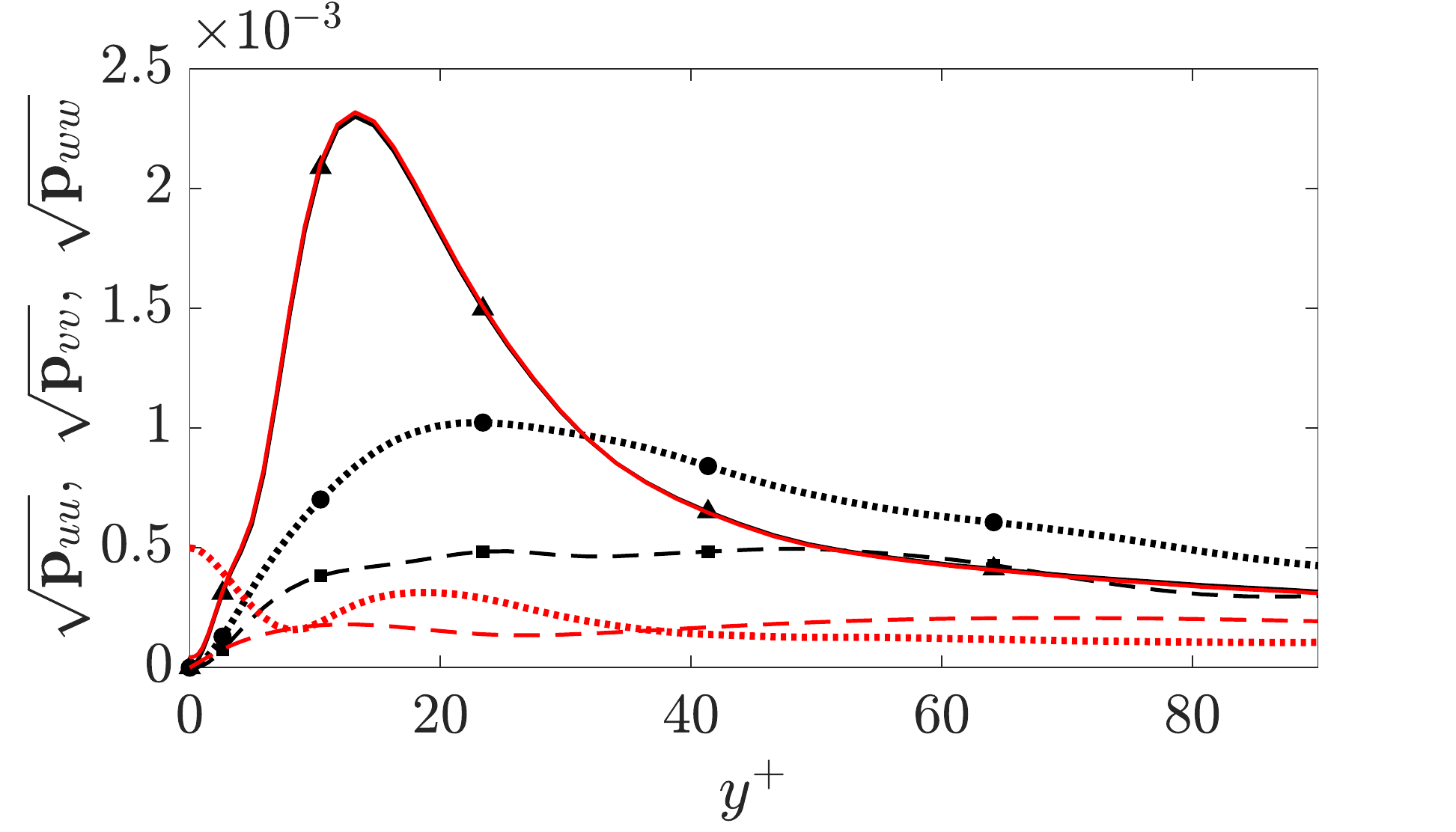}}\put(-180,95){$c)$}\hspace{0.7cm}
{\includegraphics[width=0.44\textwidth,trim={10 0 50 0},clip]{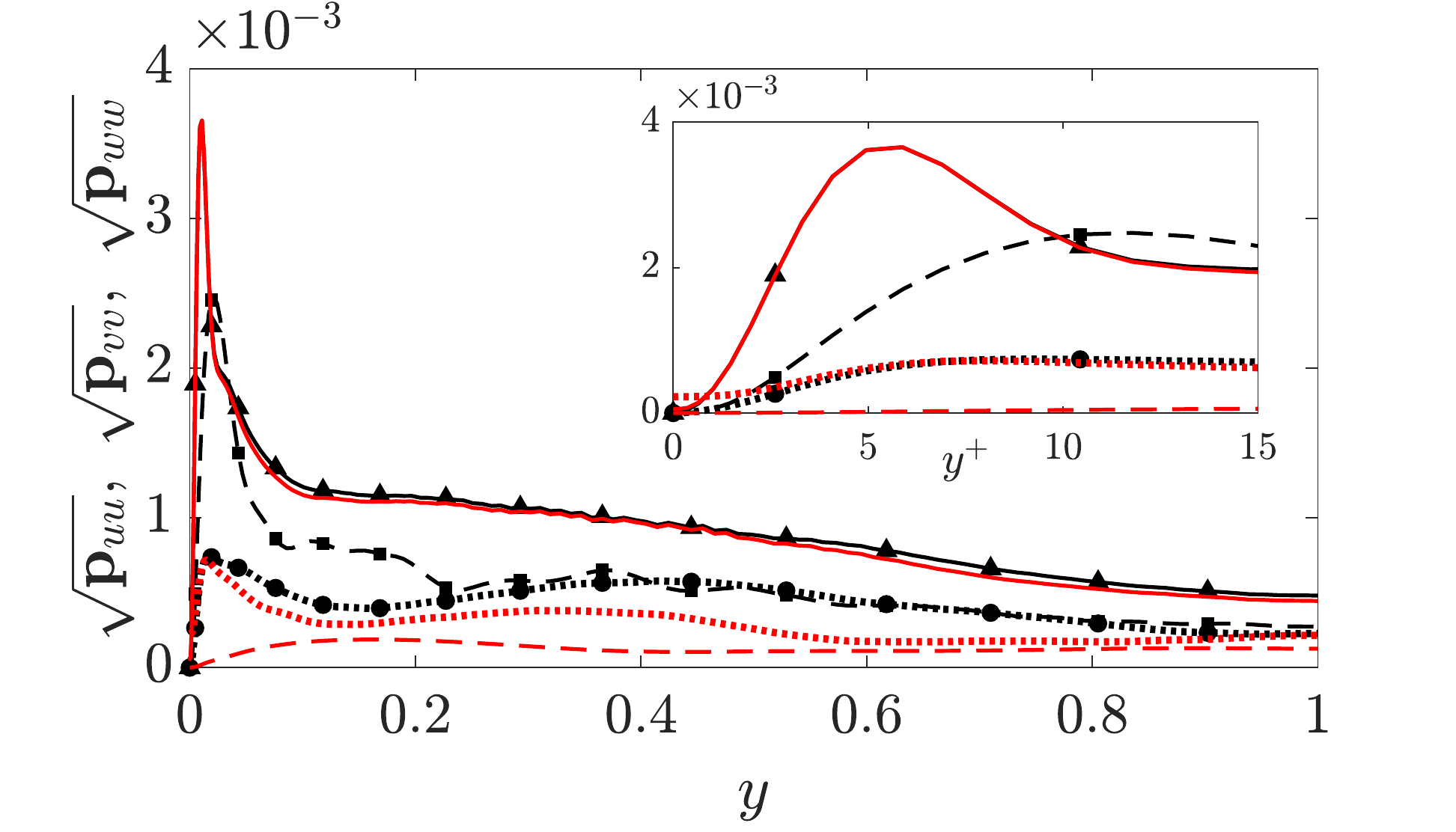}}\put(-180,95){$d)$}\\
\end{center}
\caption{$\ret = 543$. PSD. Symbols: $\bfS = \bfR\bfP\bfR^H$; triangle: streamwise component; circle: spanwise component; square: wall-normal component. Black lines: DNS. Red lines: solenoidal part of the forcing from DNS. Solid line: streamwise component; dotted line: spanwise component; dashed line: wall-normal component. a) $\bfs_{uu}$, $\bfs_{vv}$, $\bfs_{ww}$ near-wall, $\omega_{max}^+ = 0.065$ $(\lambda_t^+\approx 100)$; b) $\bfs_{uu}$, $\bfs_{vv}$, $\bfs_{ww}$ large-scale, $\omega_{max} = 0.65$; c) $\bfp_{uu}$, $\bfp_{vv}$, $\bfp_{ww}$ near-wall, $\omega_{max}^+ = 0.065$ $(\lambda_t^+\approx 100)$; d) $\bfp_{uu}$, $\bfp_{vv}$, $\bfp_{ww}$ large-scale, $\omega_{max} = 0.65$ (in-plot: zoom of near-wall region).}
\label{fig:SPioRet550}
\end{figure}%
\subsection{$\ret = 543$}
Figure~\ref{fig:abEuuRet550} shows the premultiplied streamwise kinetic energy spectra $\alpha \beta \bfe_{uu}$ at $y^+ = 15$ and $y=0.5$ for $\ret=543$. In this case the highest energetic activity is for $(\lambda_x^+,\lambda_z^+) = (1137,100)$ for the near-wall structures and $(\lambda_x,\lambda_z) = (6.28,1.57)$ for the large-scale structures, so these wave numbers are chosen for the following analysis. Figure~\ref{fig:SPRet550}a,b present the streamwise component $\bfS_{11}(\alpha,y,y,\beta,\omega)$ for both the near-wall and large-scale structures. The near-wall structures have a peak at $\omega_{max} = 0.065$ ($\lambda_t^+ \approx 100$) and at $y^+ = 15$, whereas large-scale structures have a peak at $\omega_{max} = 0.65$ and $y = 0.45$. Thus, an appreciable amount of scale separation is present for this Reynolds number. Figure~\ref{fig:SPRet550}c,d present the streamwise component of the forcing $\bfP_{11}$ for both near-wall and large-scale structures. The peaks of the forcing of both type of structures are localized in the inner layer. The near-wall structures show a peak at $\omega_{max}^+ = 0.065$ ($\lambda_t^+ \approx 100$) and $y^+ = 15$, while the large-scale structures show a peak at $\omega_{max} = 0.65$ and $y^+ = 6$. The input $\bfP$ and the output $\bfS$ show a peak at the same $\omega$. The shape of all the three components of $\bfP$ and $\bfS$ at the respective $\omega_{max}$ is shown in figure~\ref{fig:SPioRet550}, where also the relationship $\bfS = \bfR\bfP\bfR^H$ (the symbols in figures~\ref{fig:SPioRet550}a,b) is presented to demonstrate the accuracy of the computed input $\bfP$. Figures~\ref{fig:SPioRet550}c,d show the input $\bfP$. The near-wall structures present a peak of the streamwise component at $y^+ = 17$ and at $y^+ = 20$ for the wall-normal and the spanwise components. The large-scale structures have a peak in $\bfP$ at $y^+ = 6$ for the streamwise component, at $y^+ = 12$ for the wall-normal component, and at $y^+ = 9$ for the spanwise component. However, besides this near-wall peak, the forcing of large-scale structures is spatially extended throughout the channel. The solenoidal part of the input is also presented in figures~\ref{fig:SPioRet550}c,d (red lines without symbols). The streamwise component is nearly unchanged, whereas the transverse components are different, as it occurs for $\ret = 179$. Moreover, the amplitude of the wall-parallel components of the solenoidal part of the forcing is lower than it is for the total forcing and the solenoidal forcing is non-zero on the wall, but parallel to it.\\ \indent
%
%
\section{Input-output analysis}\label{sec:Pij}
%
%
\subsection{Effect of the sub-blocks of the input on the output}\label{subsec:blocks}
The effects of sub-blocks of the input $\bfP$ on the output $\bfS$ can be analyzed by representing the contribution of the terms
\begin{equation}\label{eq:Pij}
\bfP_{ij} = \lim_{T \to \infty}  \frac{ \ \mathbb{E}\left[ \left( \frac{1}{2\pi}\int_{-T}^{T}\hat{\bff}_{i}(t) e^{-i\omega t} \ \mathrm{d}t \right)\left(\frac{1}{2\pi}\int_{-T}^{T}\hat{\bff}_{j}(t)^{H} e^{i\omega t}\ \mathrm{d}t \right) \right]}{2T}
\end{equation}
to the output $\bfS=\bfR\bfP_{ij}\bfR^H$. Here, $\hat{\bff_i}$ represents a $3N_y\times 1$ vector made solely of the $i$-th $N_y\times 1$ component of the $3N_y\times1$ vector $\hat{\bff}$ and the other components set to zero; with $i=1,2,3$ corresponding to the streamwise, wall-normal and parallel directions. Therefore, the sum of the outputs associated to each $\bfP_{ij}$ is equal to the output $\bfS$ from the DNS data. This analysis highlights the importance of the off-diagonal blocks, which contain the cross-correlations of different components $\hat{\bff}_{i}$. Note that from now on, with the output $\bfR \bfP_{ij}\bfR^{H}$ computed from the off-diagonal blocks with $i\neq j$, it is meant the output given by $\bfR (\bfP_{ij} + \bfP_{ji})\bfR^{H}$ because $\bfP$ is hermitian.\\ \indent
%
\begin{figure}%
\begin{center}
{\includegraphics[width=0.45\textwidth,trim={0 0 0 0},clip]{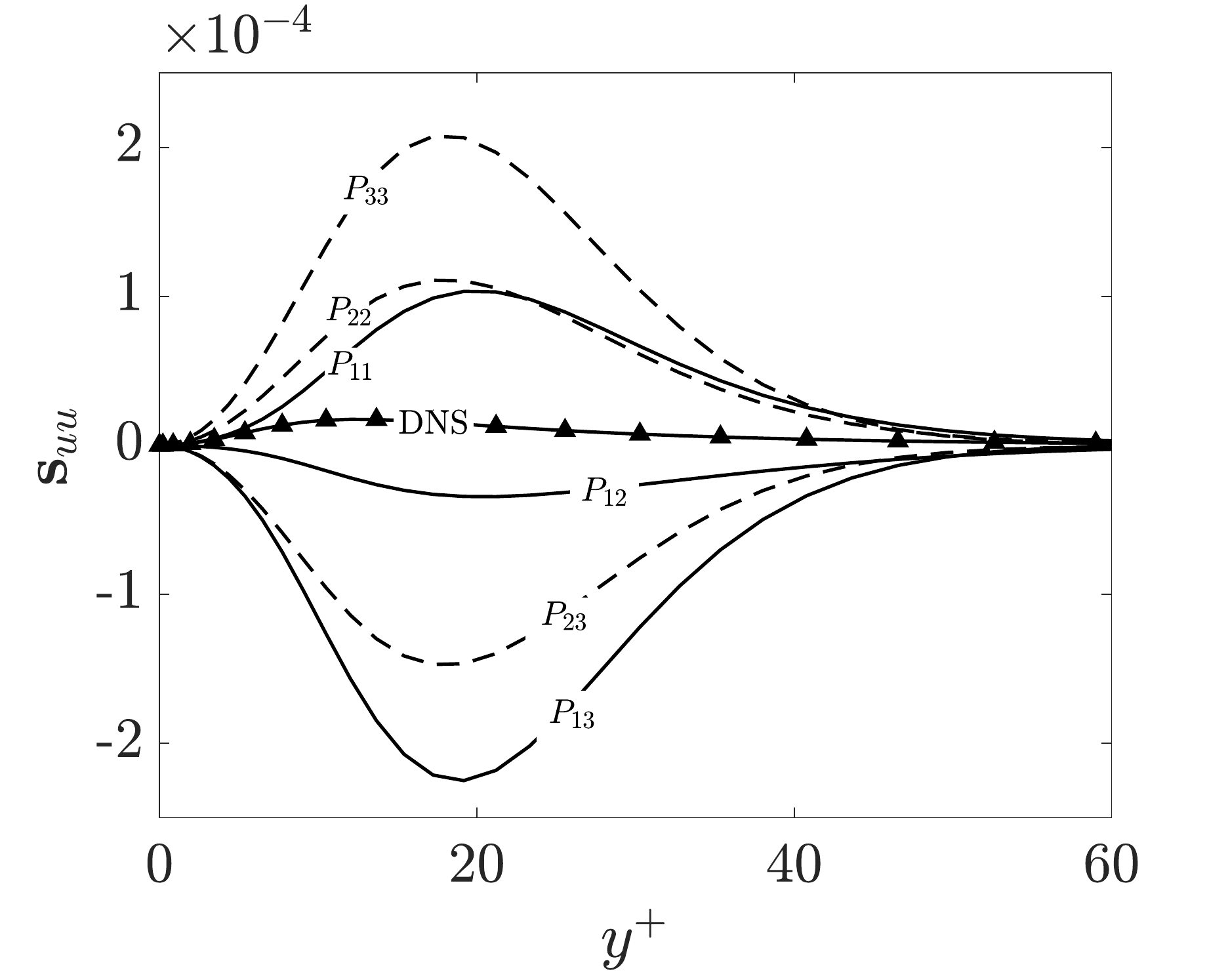}}\put(-171,125){$a)$}\hspace{0.5cm}
{\includegraphics[width=0.45\textwidth,trim={0 0 0 0},clip]{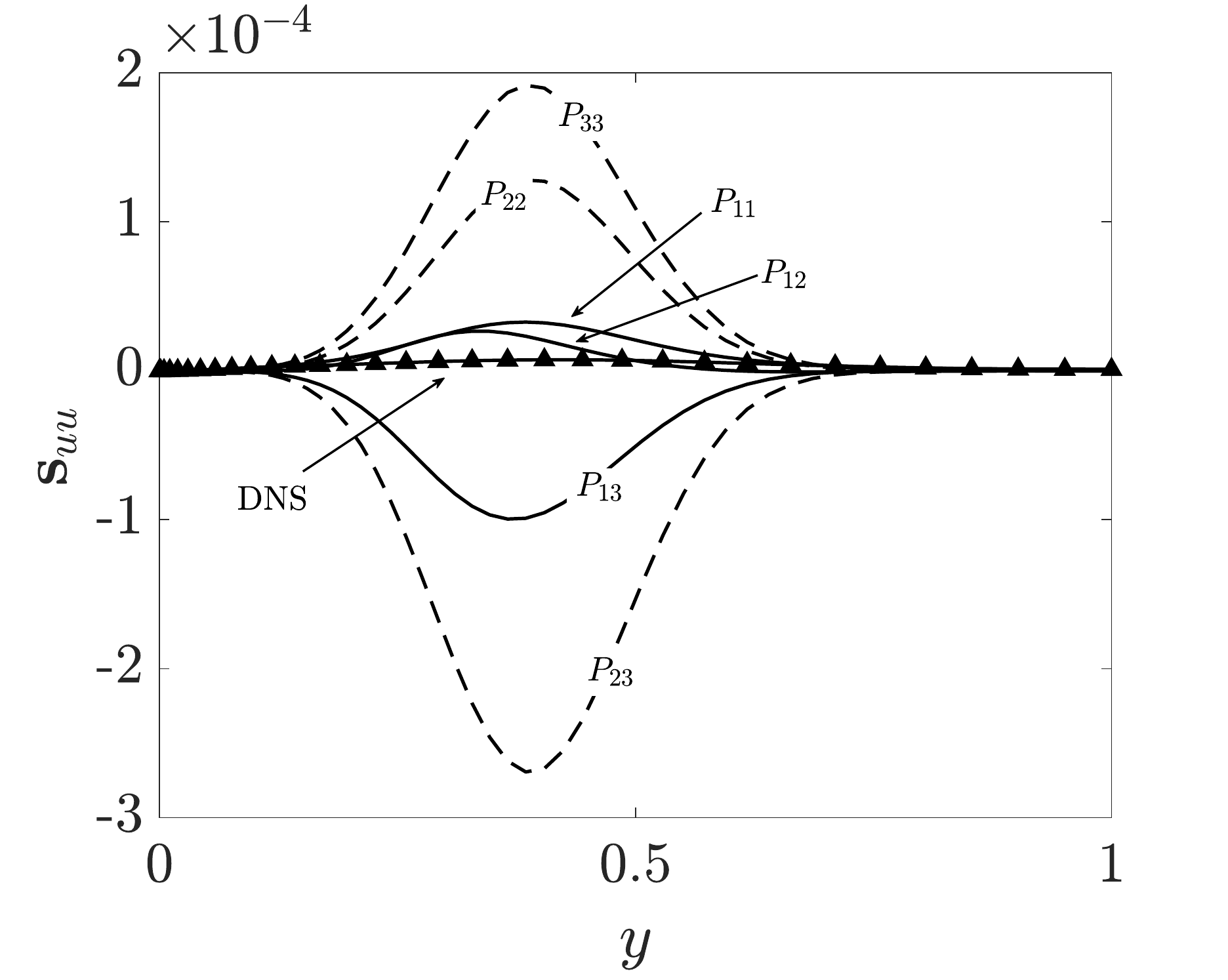}}\put(-171,125){$b)$}\\
\end{center}
\caption{$\ret = 179$. Response $\bfS = \bfR\bfP_{ij}\bfR^H$ from sub-blocks of $\bfP_{ij}$. Solid lines: response from $\bfP_{1j}$ (sub-blocks related to the streamwise component). Dashed lines: response from $\bfP_{ij}$ with $i\neq 1$. {\text{\sout{$\ \blacktriangle \ $}}}: DNS data. a) $\bfs_{uu}$ near-wall, $\omega_{max}^+ = 0.065$; b) $\bfs_{uu}$ large-scale, $\omega_{max} = 1.05$.}
\label{fig:SPcRet180}
\end{figure}%
For the case with $\ret=179$ the results are presented in figure~\ref{fig:SPcRet180} for $\bfS_{11}$. For both near-wall and large-scale structures the terms with $i \neq j$ give a negative contribution. In particular, for the large-scale structures in figure~\ref{fig:SPcRet180}b only the terms related to the spanwise direction ($j=3$) provide a negative contribution, whereas for the near-wall structures all the terms with $i\neq j$ give a negative contribution. Moreover, even though the magnitude of the streamwise component of the solenoidal part of the forcing is around one order of magnitude higher than the wall-normal and spanwise components, the forcing can not be approximated by the streamwise component only. In fact, the amplitudes of the outputs associated to the wall-normal and spanwise components has the same order of magnitude of the output associated to the streamwise component. This can be attributed to the lift-up effect, which greatly enhances the efficiency of forcing in the wall-normal or spanwise directions, leading to streamwise vortices that in turn form amplified streaks \citep{moffatt1967a,ellingsen1975a,landahl1980a,hwang2010a,hwang2011a,brandt2014a}. It is seen here that the wall-normal and spanwise components of the forcing, associated with lift-up, lead to high-amplitude outputs when considered in isolation (see curves associated with $\bfP_{22}$ and $\bfP_{33}$ in figures~\ref{fig:SPcRet180}a and b), but a complex phase relationship between the three forcing components leads to cancellations in such a way that the DNS output is of lower magnitude than what is predicted by considering a single forcing component. This is a first indication of relevance of forcing colour to the channel dynamics, as incoherent forcing components would lead to output PSDs that could simply be summed to form the full power spectral density. The high magnitude of cross terms shows that this is not the case: forcing components are coherent between them, and neglecting such coherence would lead to appreciable errors in the prediction of the output. Similar effects were observed for a minimal Couette flow unit by \cite{nogueira2020a}.\\ \indent
%
\begin{figure}%
\begin{center}
{\includegraphics[width=0.37\textwidth,trim={0 0 0 0},clip]{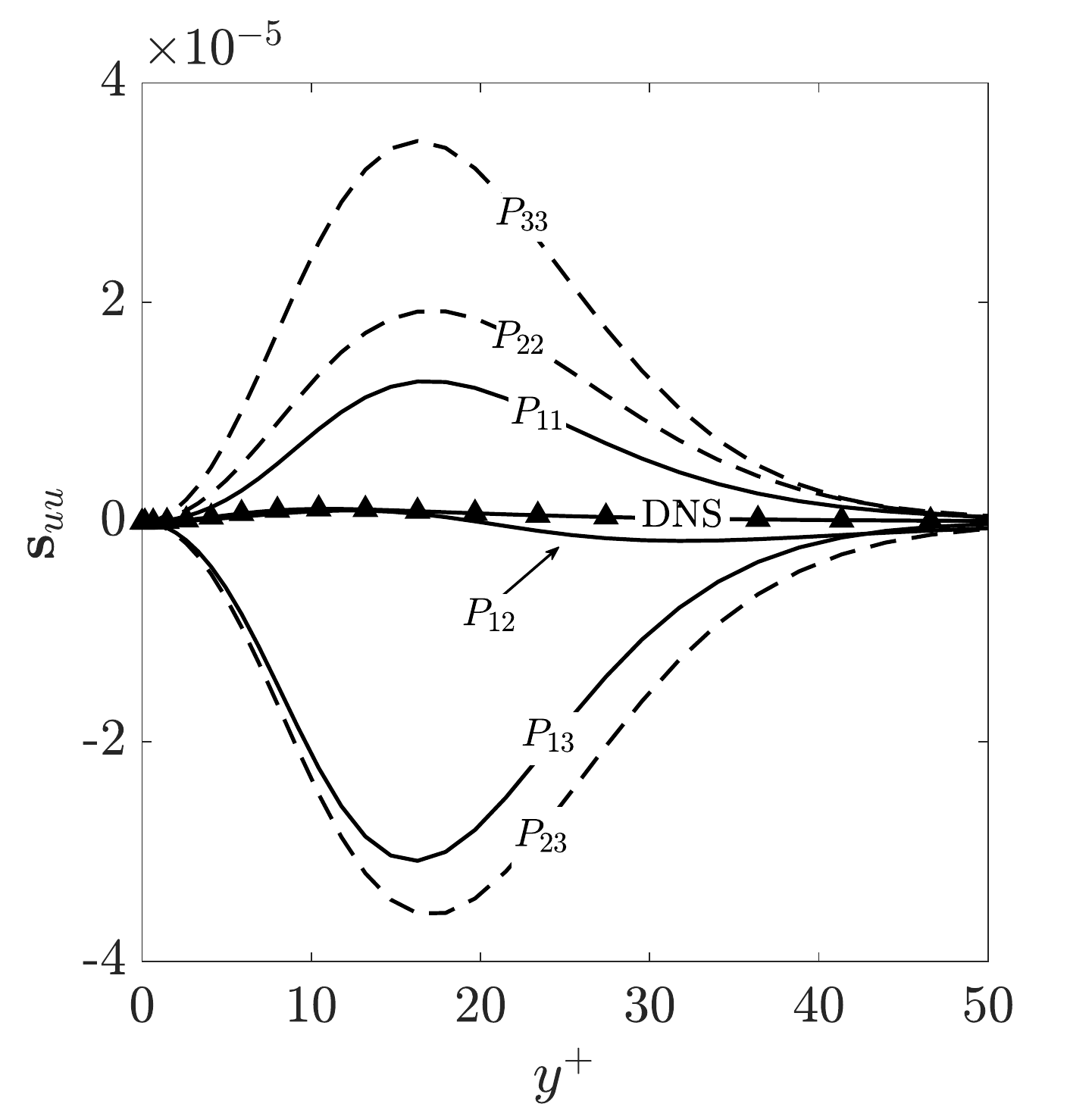}}\put(-140,130){$a)$}
{\includegraphics[width=0.37\textwidth,trim={0 0 0 0},clip]{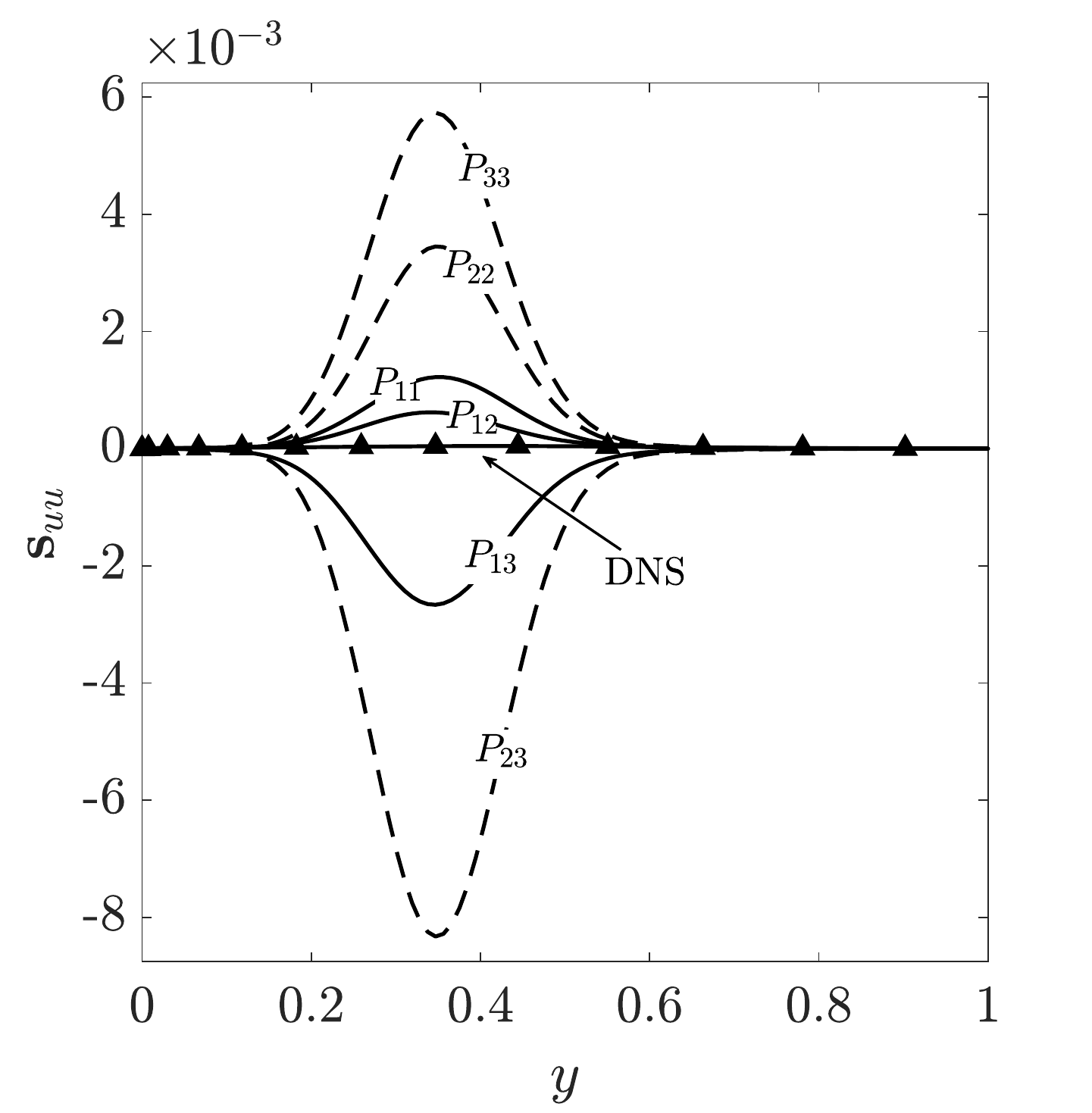}}\put(-140,130){$b)$}
{\includegraphics[width=0.268\textwidth,trim={0 0 0 0},clip]{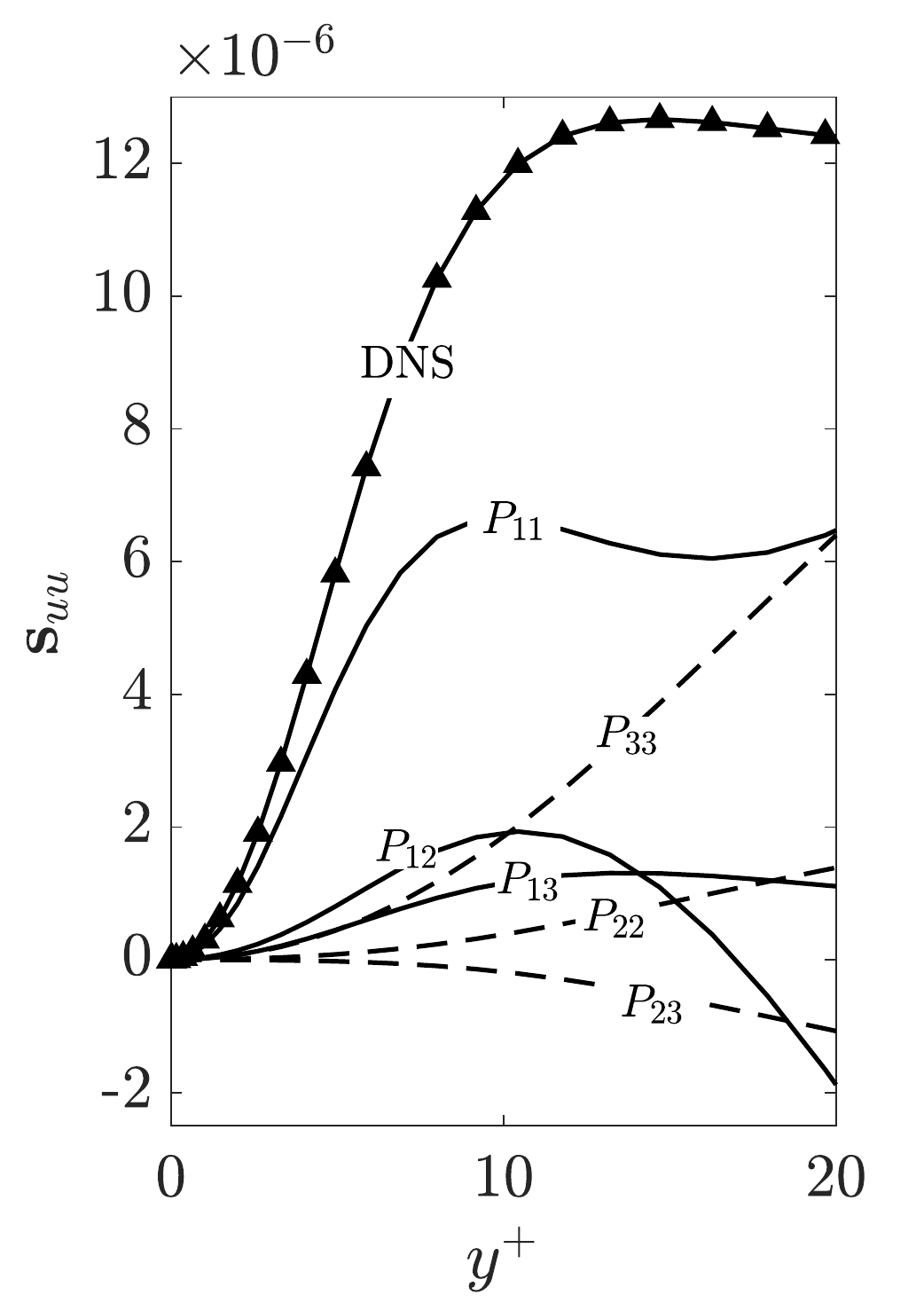}}\put(-102,130){$c)$}
\end{center}
\caption{$\ret = 543$. Response $\bfS = \bfR\bfP_{ij}\bfR^H$ from sub-blocks $\bfP_{ij}$. Solid lines: response from $\bfP_{1j}$ (sub-blocks related to the streamwise component). Dashed lines: response from $\bfP_{ij}$ with $i\neq 1$. {\text{\sout{$\ \blacktriangle \ $}}}: DNS data. a) $\bfs_{uu}$ near-wall, $\omega_{max}^+ = 0.065$ $(\lambda_t^+\approx 100)$; b) $\bfs_{uu}$ large-scale, $\omega_{max} = 1.05$; c) zoom of (b) in the near-wall region (note that $\bfP_{1j}$ are the only components contributing to the near-wall peak).}
\label{fig:SPcRet550}
\end{figure}%
For the case with $\ret=543$ the contribution of the terms $\bfP_{ij}$ on the output $\bfS$ are presented in figure~\ref{fig:SPcRet550}. As it occurs for $\ret = 179$, the off-diagonal terms ($i\neq j$) provide a negative contribution to the output, indicating that coherence between the forcing components, or forcing colour, is an important feature of the dynamics, as discussed above. Similarly to $\ret=179$, the forcing cannot be approximated by the streamwise component only, even though its peak amplitude is one order of magnitude higher than it is for the wall-normal and the spanwise components. In fact, as shown in figures~\ref{fig:SPcRet550}a,b, the amplitudes of the outputs associated to the wall-normal and spanwise components of the forcing are larger than the output associated to the streamwise component, due to the higher efficiency associated with the lift-up effect. Moreover, at $\ret = 543$ the large-scale structures show a low-amplitude near-wall peak in the PSD of the streamwise velocity component $\bfs_{uu}$ at $y^+ =12$. This low-amplitude near-wall peak is demonstrated to be generated by the streamwise component of the forcing $\bfP$. In fact, in figure~\ref{fig:SPcRet550}c it is clear that it is only the components of $\bfP$ associated with the streamwise forcing ($i = 1$) which are responsible for the peak around $y^+ = 12$ (compare the solid and the dashed lines in figure~\ref{fig:SPcRet550}c). Thus, the low-amplitude near-wall peak at $y^+=12$ in $\bfs_{uu}$ must be related to the near-wall peak at $y^+=6$ in $\bfp_{uu}$.
%
%
\subsection{Low-rank approximation}
\begin{figure}
\begin{center}
{\includegraphics[width=0.22\textwidth,trim={0 0 0 0},clip]{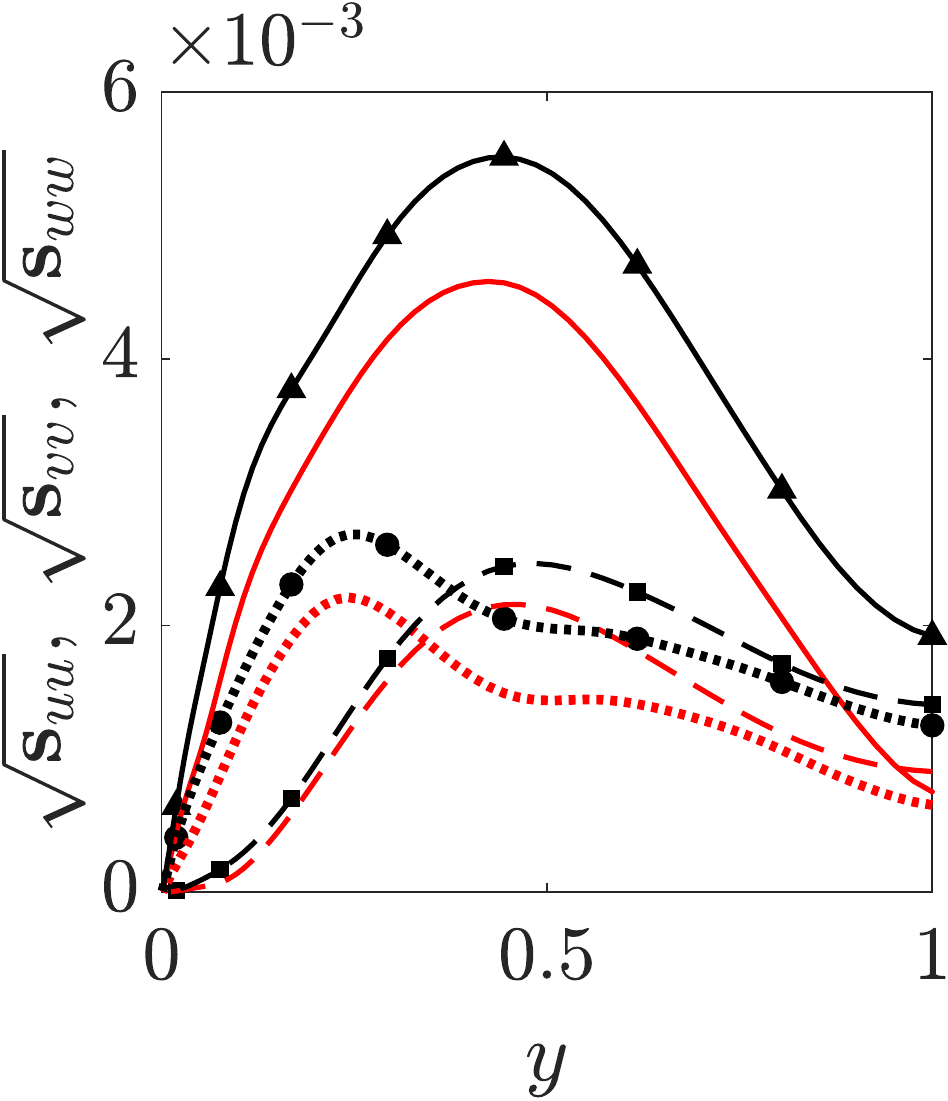}}\put(-88,90){$a)$}\hspace{0.2cm}
{\includegraphics[width=0.22\textwidth,trim={0 0 0 0},clip]{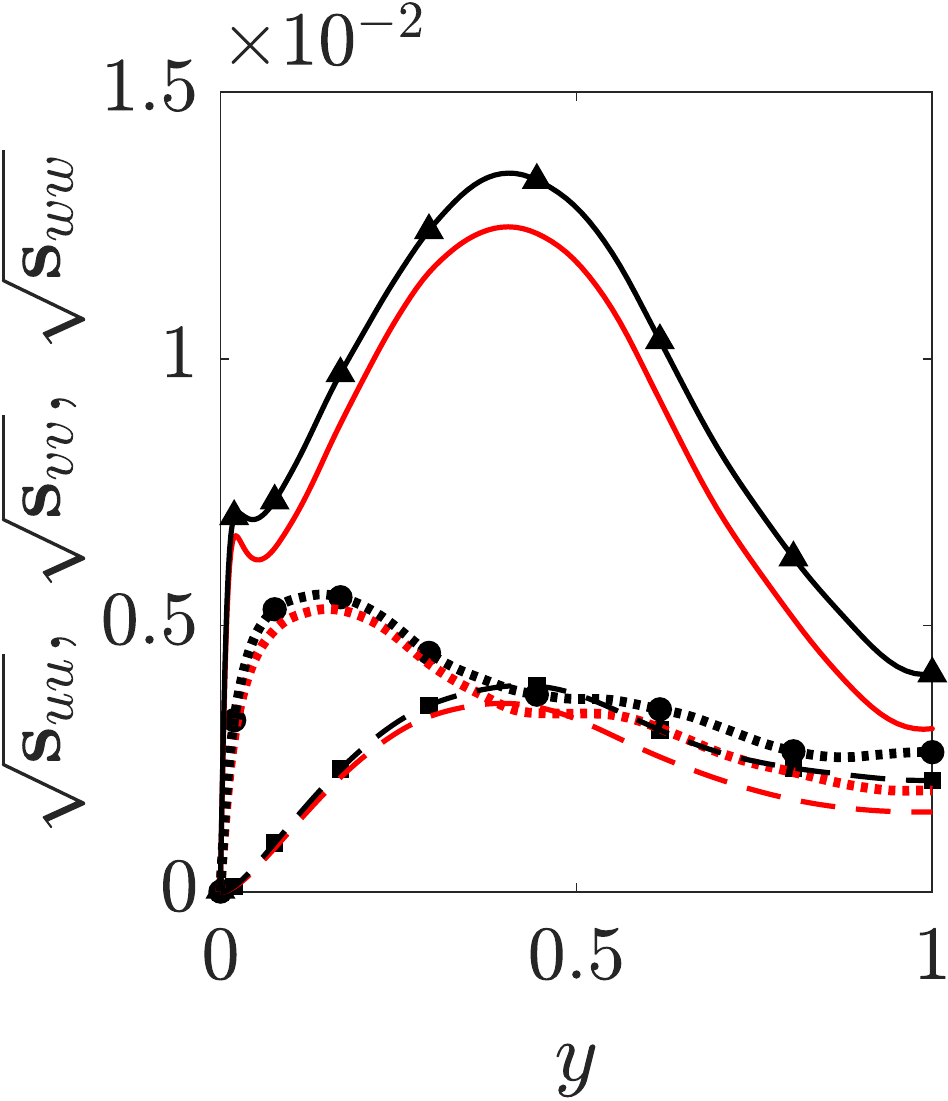}}\put(-88,90){$b)$}\hspace{0.2cm}
{\includegraphics[width=0.222\textwidth,trim={0 0 10 0},clip]{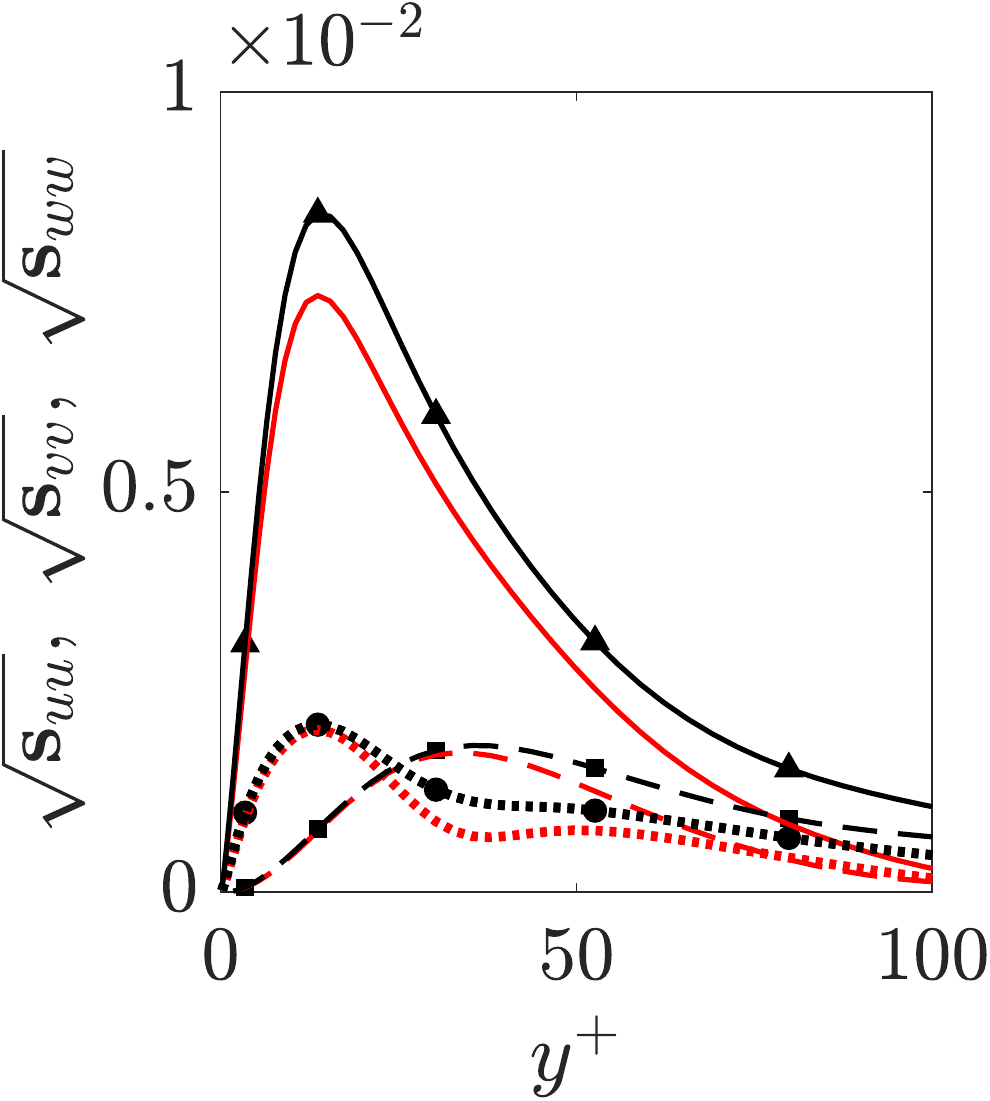}}\put(-88,90){$c)$}\hspace{0.2cm}
{\includegraphics[width=0.222\textwidth,trim={0 0 10 0},clip]{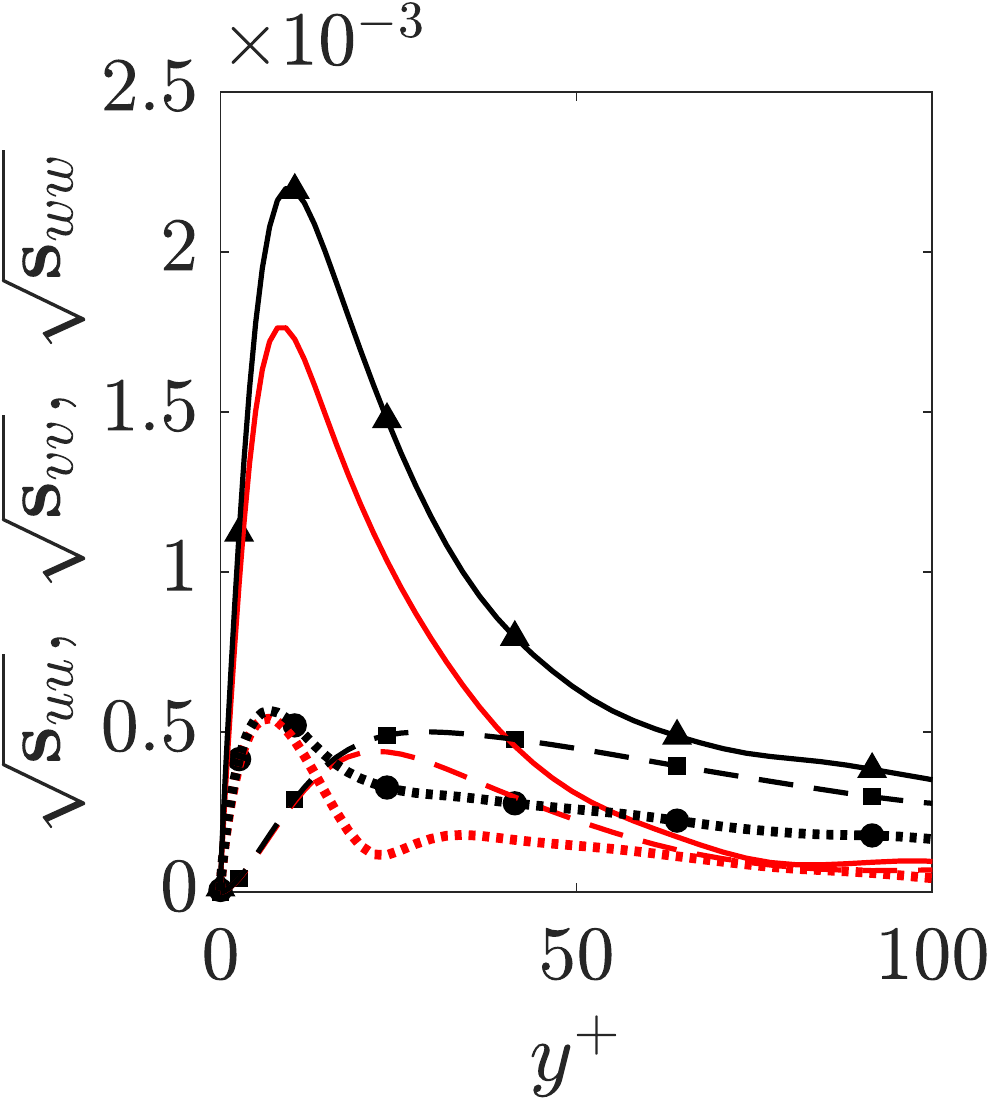}}\put(-88,90){$d)$}\\
\end{center}
\caption{PSD of the velocity fluctuations. Symbols: $\bfS = \bfR\bfP\bfR^H$; triangle: streamwise component; circle: spanwise component; square: wall-normal component. Black lines: DNS. Red lines: $\bfS_{lr}$ from low-rank approximation of $\bfP_{lr}$. Solid lines: streamwise component. Dashed lines: wall-normal component. Dotted lines: spanwise component. a,c) $\ret = 179$. b,d) $\ret = 543$. a,b) Large scales. c,d) Small scale.}
\label{fig:Slr}
\end{figure}%
\begin{figure}
\begin{center}
{\includegraphics[width=0.79\textwidth,trim={70 50 80 40},clip]{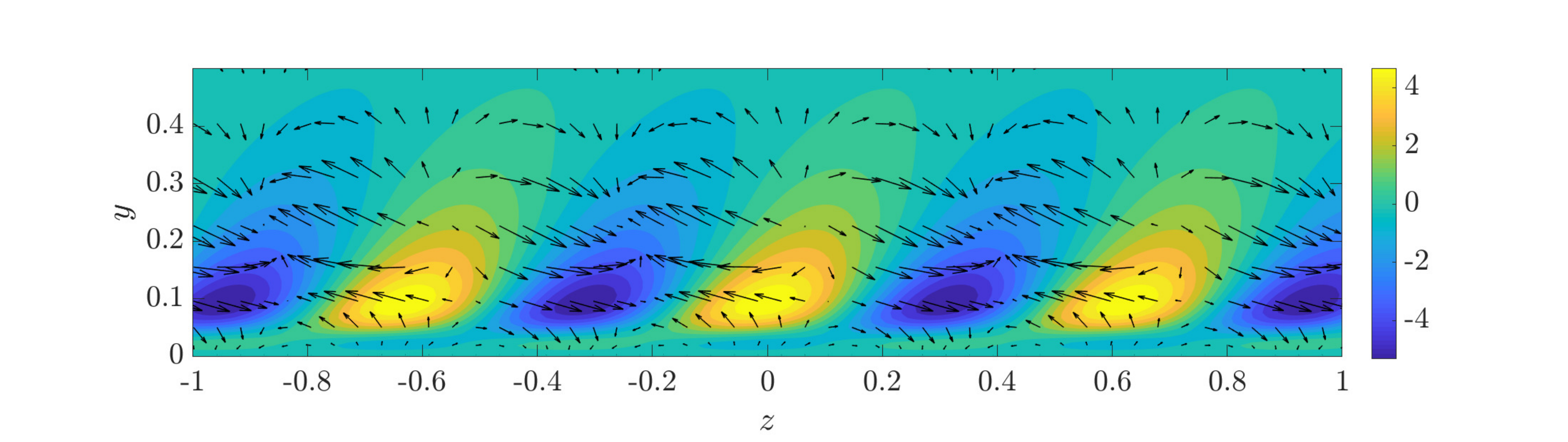}}\put(-305,62){$a)$}\\
{\includegraphics[width=0.79\textwidth,trim={70 50 80 40},clip]{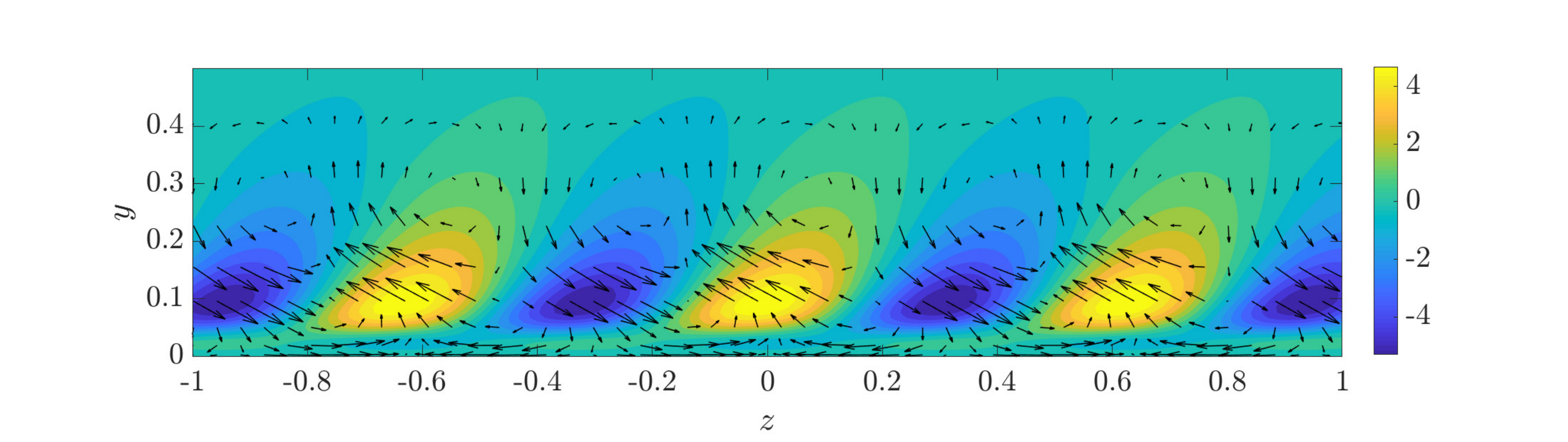}}\put(-305,60){$b)$}\\
{\includegraphics[width=0.79\textwidth,trim={70 50 80 40},clip]{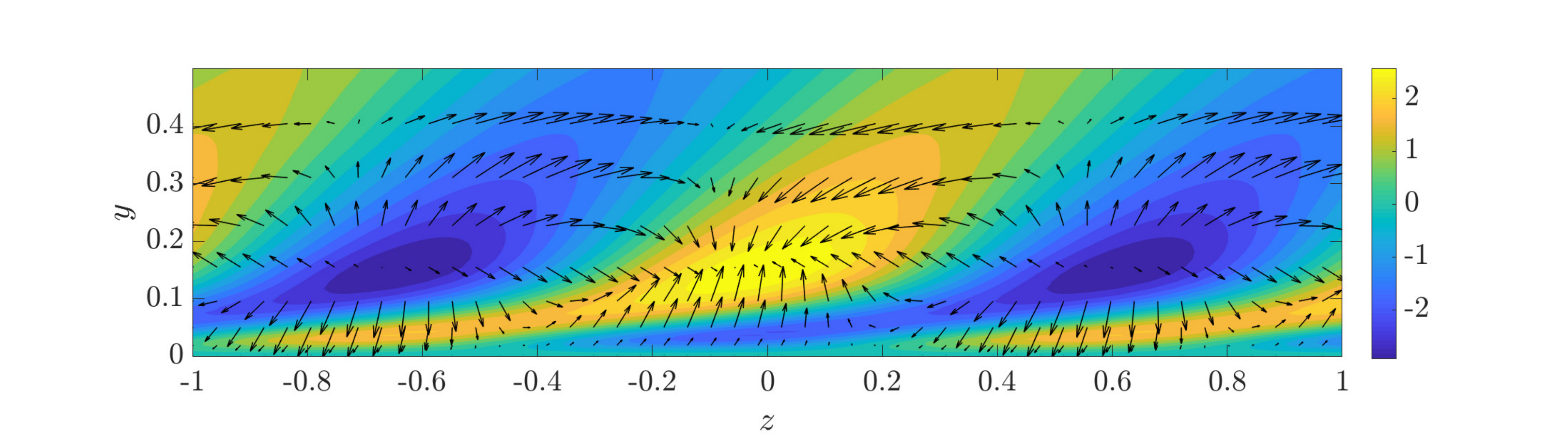}}\put(-305,60){$c)$}\\
{\includegraphics[width=0.79\textwidth,trim={70 50 80 40},clip]{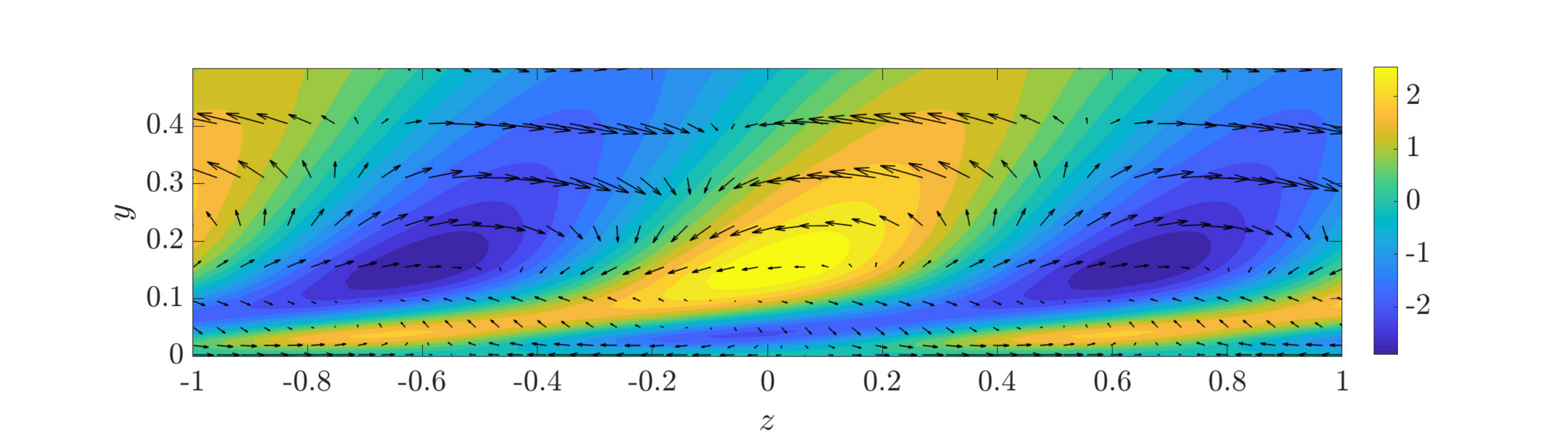}}\put(-305,60){$d)$}\\
{\includegraphics[width=0.79\textwidth,trim={70 52 80 20},clip]{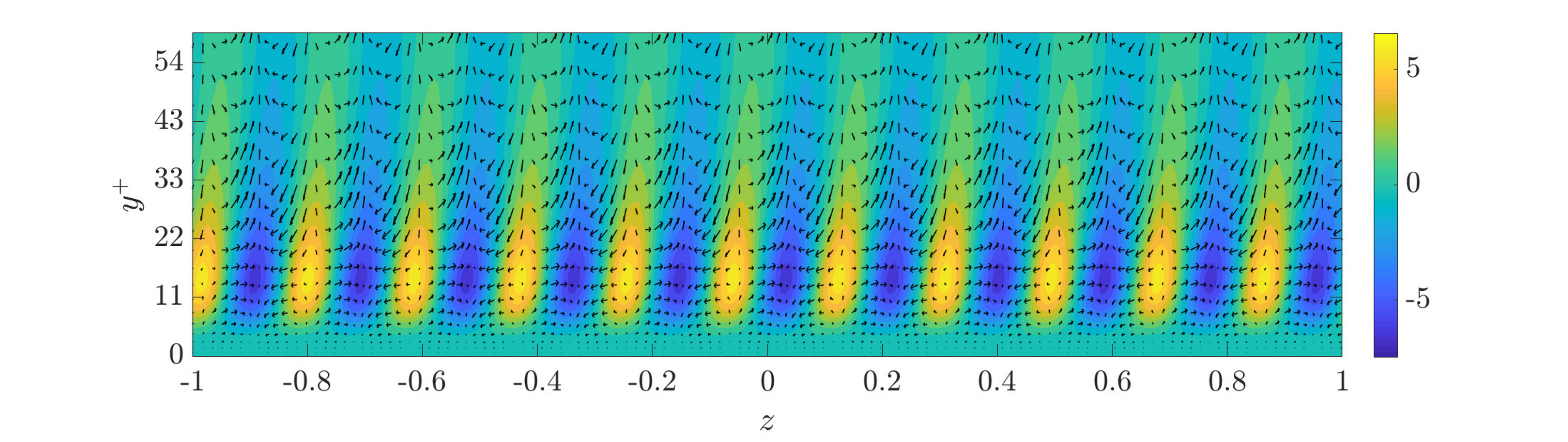}}\put(-305,68){$e)$}\\
{\includegraphics[width=0.79\textwidth,trim={70 52 80 20},clip]{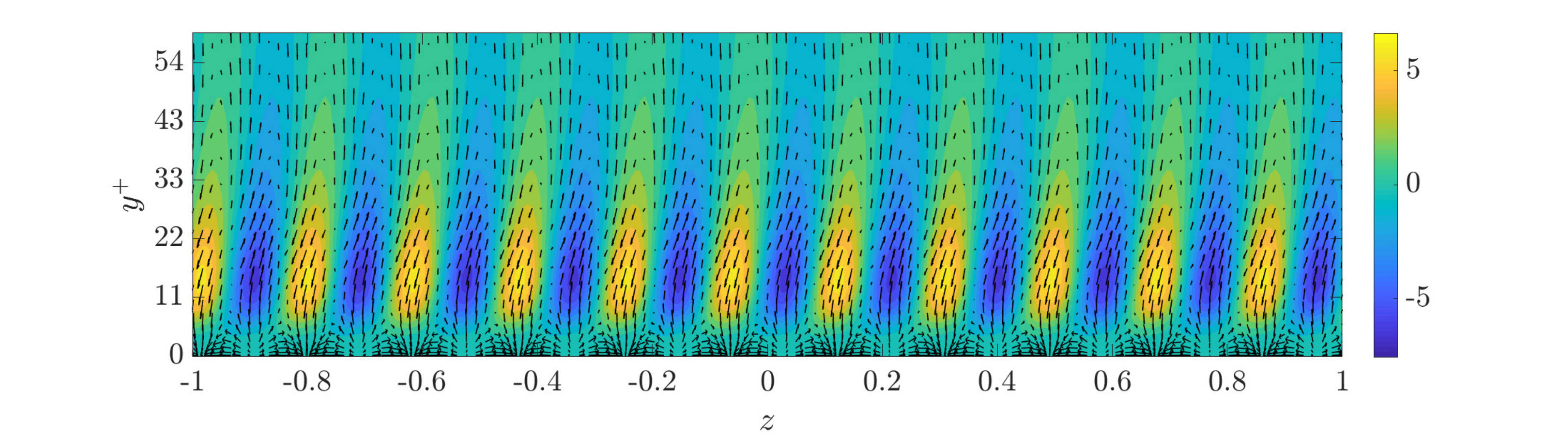}}\put(-305,68){$f)$}\\
{\includegraphics[width=0.79\textwidth,trim={70 52 80 13},clip]{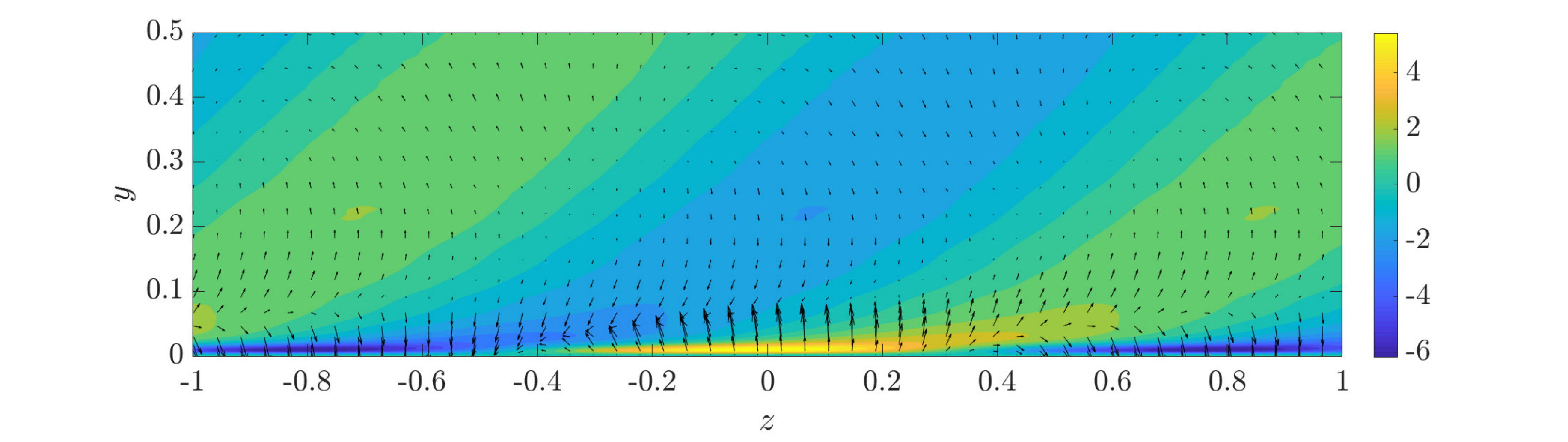}}\put(-305,68){$g)$}\\
{\includegraphics[width=0.79\textwidth,trim={70 05 80 13},clip]{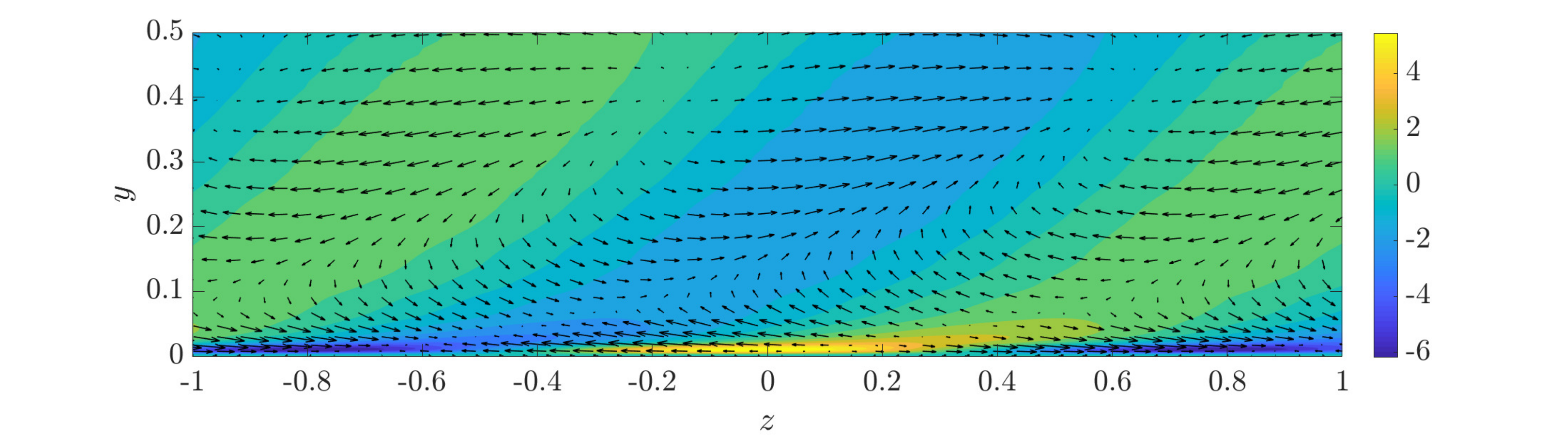}}\put(-305,86){$h)$}
\end{center}
\caption{First SPOD mode of $\bfP$: $\bfzeta_1$. a,b,c,d) $\ret = 179$: e,f,g,h) $\ret = 543$; a,b,e,f) Small scales; c,d,g,h) Large scale; a,c,e,g) Full $\bfzeta_1$; b,d,f,h) Solenoidal part of $\bfzeta_1$. Contours: streamwise component of $\bfzeta_1$. Vectors: spanwise and wall-normal components of $\bfzeta_1$.}
\label{fig:P1m}
\end{figure}
Effort has been made to reconstruct the input $\bfP$ from sensor measurements with system identification techniques \citep{jovanovic2001a,zare2017a,illingworth2018a,towne2020a}. Here, a low-rank approximation of $\bfP$, referred to as $\bfP_{lr}$, is computed a posteriori by retaining the first two SPOD modes of $\bfP$, so that
\begin{equation}
\bfP_{lr} = \eta_1 \bfzeta_{1}^{}\bfzeta^{H}_{1} + \eta_2 \bfzeta_{2}^{}\bfzeta^{H}_{2}.
\end{equation}
The second SPOD mode is also included since in a channel flow the modes are paired because of the symmetry of the flow case. The approximated output $\bfS_{\bfP_{lr}}$ is computed with $\bfR$ and compared to the output $\bfS$ from DNS data in figure~\ref{fig:Slr}. %
It is remarkable that considering only the first two SPOD modes of $\bfP$ leads to a recovery of most of the output in all the cases presented. This shows that the forcing is highly structured and coherent, as already suggested by the results in \S~\ref{sec:DNSresults}, and its dominant structure, which is represented by the first two symmetrical and anti-symmetrical SPOD modes, is responsible for the bulk of the flow response for both near-wall and large-scale motions  in both $\ret = 179$ and $\ret = 543$. This result suggests a direction for the modeling of the structures considered here: the identification of the non-linear processes which lead to the observed coherent forcing may form a foundation for reduced-order models of self-sustaining mechanisms in turbulent channel flows, similarly to the developments of \cite{hamilton1995a} and \cite{farrell2012a} for low Reynolds turbulent Couette flow in minimal boxes.\\ \indent
Assessing the low-rank behavior of the forcing is helpful for those techniques which account for the forcing through system identification methods, in a similar fashion of \cite{jovanovic2001a}, \cite{zare2017a}, or \cite{illingworth2018a}. The low-rank behavior of the forcing can be a benefit because identification techniques may lose accuracy when the number of unknown parameters increases \citep{hjalmarsson2007a}. In this case identifying the first eigenvector of the cross-power spectral density $\bfP$ instead of the whole $\bfP$ reduces the number of unknowns from $3N_y(3N_y+1)/2$ to $3N_y$ ($N_y$ is the number of discrete point in the wall-normal direction). Thus, in this case the number of unknowns would be reduced of two orders of magnitude. The second SPOD mode can be retrieved later by exploiting the symmetry of the flow case.\\ \indent
Since most of the output can be obtained from the presented low-rank approximation of the input $\bfP$, the field given by the first SPOD mode $\bfzeta^{}_{1}$ is of interest, so it is presented in figure~\ref{fig:P1m} together with its solenoidal part. The second SPOD mode $\bfzeta^{}_{2}$ is not presented because the modes are symmetric and anti-symmetric with respect to the center line of the channel and their behavior on one wall is exhaustive. Even though the mode $\bfzeta_1$ does not clearly appear to force the lift-up mechanism, its solenoidal part, which is the only one responsible for the velocity field, is actually accelerating the flow in the directions and areas typical for the occurrence of the lift-up mechanism. In fact, the solenoidal field takes the shape of oblique vortices which appear to push near-wall fluid particles away from the wall and further fluid particles towards the wall. This occurs for both $\ret=179$ and $\ret= 543$, and for both large- and small-scale structures. Thus, it appears that the feedback of the non-linear terms arising from the fluctuations of the velocity field mainly amounts to a vortical structure forcing the lift-up mechanism. It is also noticeable that the magnitude of the streamwise component of the solenoidal part of the forcing is higher than it is for the wall-normal and the spanwise components. This behavior is opposite to the one predicted by the linear optimal forcing, which coincides with the first right-singular vector of the resolvent \citep{hwang2010c}. Figure~\ref{fig:P1m} suggests that the streamwise component counteracts the effect of the wall-normal and spanwise components, as it appears in figures~\ref{fig:SPcRet180} and \ref{fig:SPcRet550} and discussed in \S~\ref{sec:Pij}. In figure~\ref{fig:P1m} it can be observed that in regions where the wall-normal and spanwise components of the forcing would generate a negative velocity fluctuation, the streamwise component pushes the flow in the positive direction, and vice versa, in a destructive interference. This is in accordance with the fact that the energy amplification predicted with an approach that focuses only on the linear operator is much higher than the one observed in DNS of the non-linear Navier--Stokes. In fact, focusing solely on the linear operator implies assuming that the forcing is mainly the linear optimal forcing, or the first right-singular vector of $\bfR$, whose streamwise component has a negligible magnitude with respect to the other two \citep{hwang2010c}. In other terms, focusing the analysis of a turbulent channel flow solely on the non-normal linear operator can be definitely misleading, as pointed out already by \cite{waleffe1995b}. Nevertheless, since expanding the Navier-Stokes system in a reference state and its relative fluctuations is entirely general, as already clarified by \cite{henningson1996a}, rewriting the system in terms of a (non-normal) linear operator without dropping the non-linear forcing terms in the analysis can be helpful to shed light on the `recycling' of the amplified outputs in the input from the non-linear terms.\par
%
%
\subsection{$\ret = 543$. Influence of the near-wall forcing on the large-scale output}
\begin{figure}
\begin{center}
{\includegraphics[width=0.48\textwidth,trim={0 0 0 0},clip]{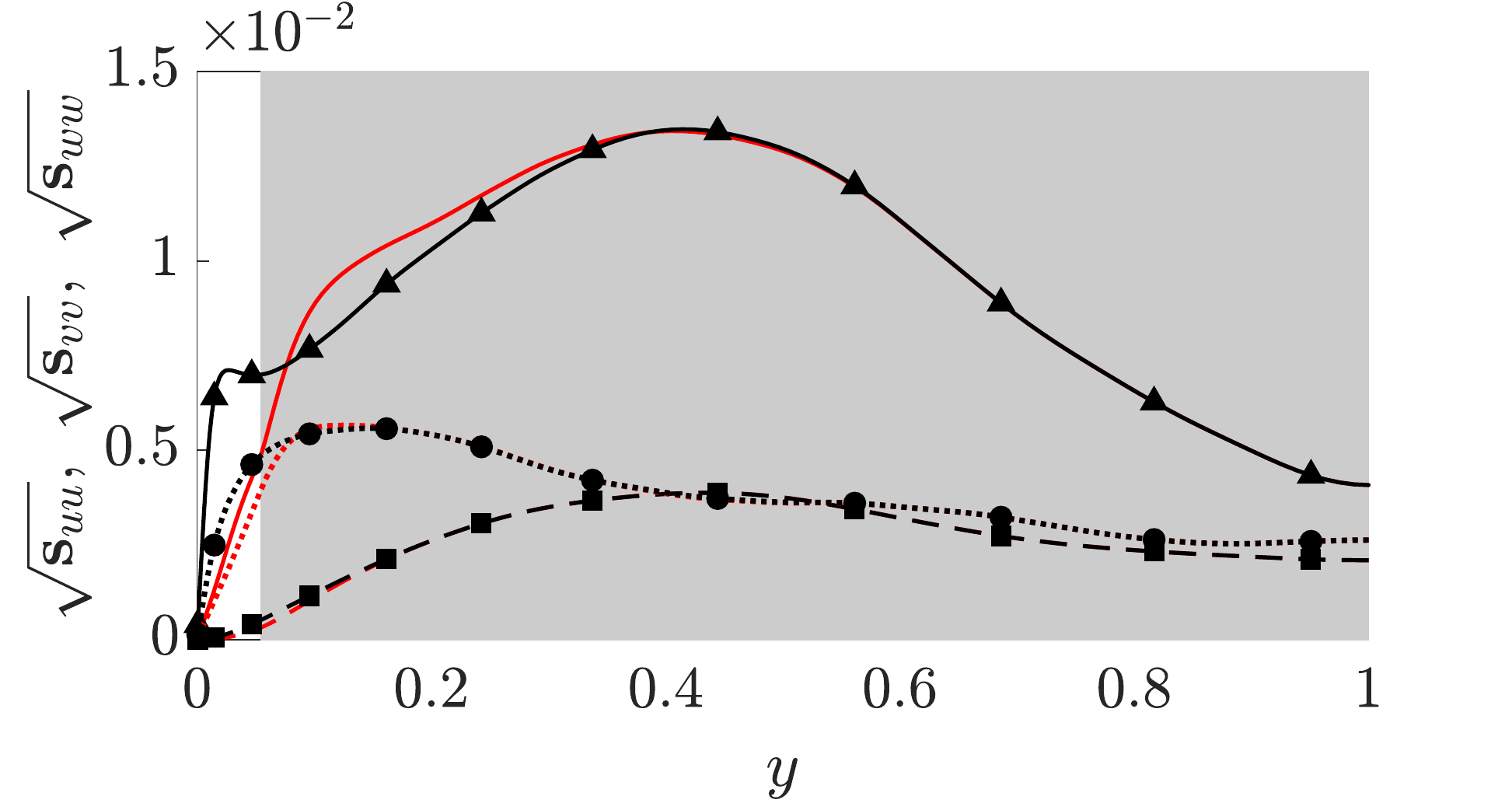}}\put(-190,93){$a)$}\hspace{0.1cm}
{\includegraphics[width=0.48\textwidth,trim={0 0 0 0},clip]{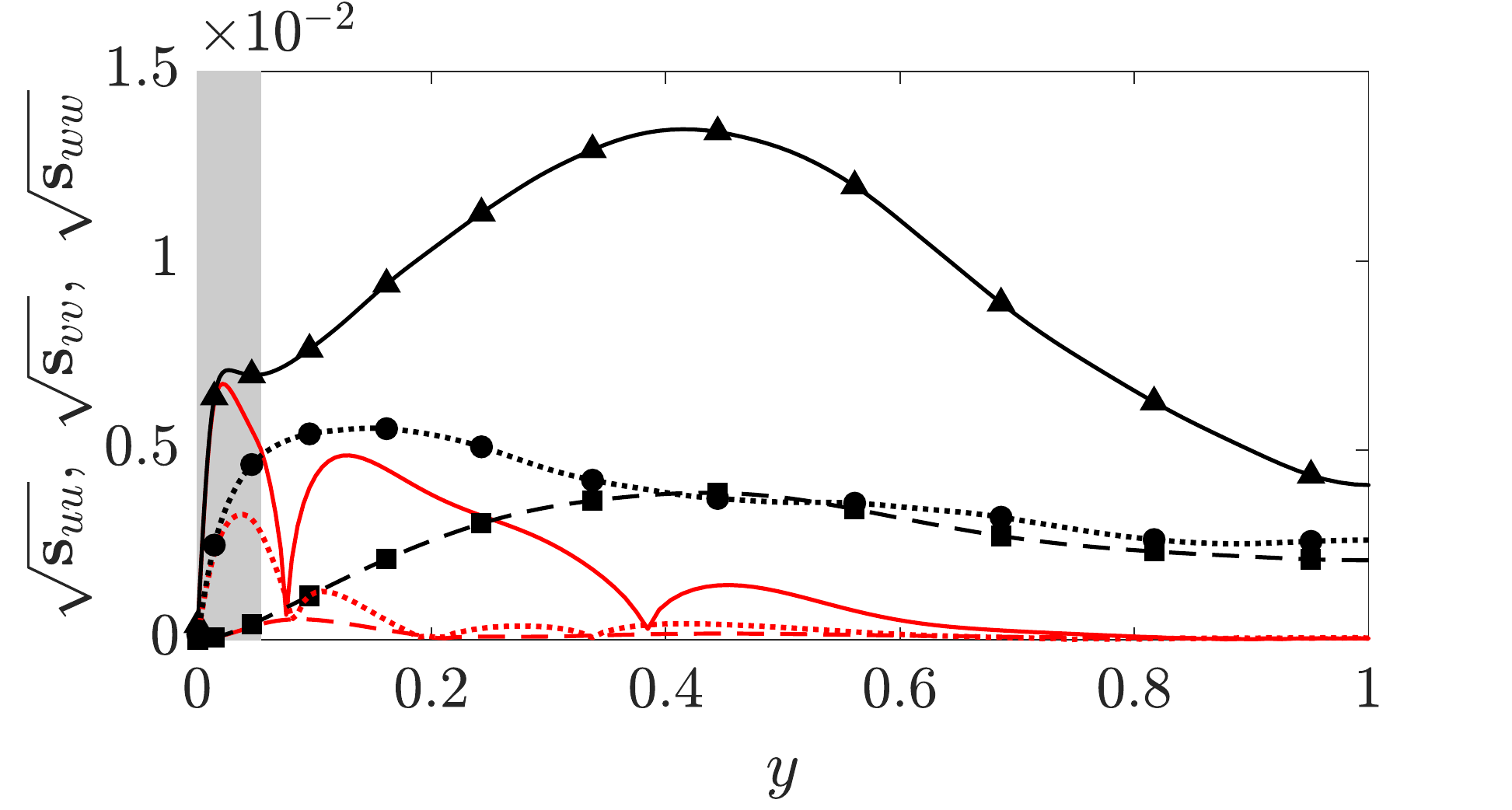}}\put(-190,93){$b)$}
\end{center}
\caption{Effect of the inner layer on the large-scale motion of $\ret=543$. PSD. Symbols: $\bfS = \bfR\bfP\bfR^H$; triangle: streamwise component; circle: spanwise component; square: wall-normal component. Black lines: DNS. Red lines: $\bfS_{h} = \bfR\bfP_{h}\bfR^H$ (left: outer layer), $\bfS_{y^+} = \bfR\bfP_{y^+}\bfR^H$ (right, inner layer). Solid line: streamwise component; dotted line: spanwise component; dashed line: wall-normal component. Grey area: interval of $y$ where the $\bfP$ is not zero.}
\label{fig:Syp}
\end{figure}%
\cite{flores2006a} and \cite{flores2007a} showed with one-point statistics that changes in the wall boundary conditions in the DNS of a turbulent channel flow, which completely alter the near-wall statistics, did not influence the statistics in the outer layer. \cite{hwang2010a} demonstrated with the aid of large-eddy simulations that large-scale motions are still present in a turbulent channel flow even when the smaller scales typical of near-wall motions are filtered out; the filtering was possible by adjusting the constant of the Smagorinsky eddy-viscosity model employed. \cite{flores2010a} and \cite{hwang2011a} further confirmed that the dynamics of large-scale structures, although affected by smaller near-wall eddies, is mostly related to processes with similar length scales. Despite these results, the large-scale structures $(\lambda_x,\lambda_z)=(6.28,1.57)$ at $\ret=543$ presented here show a low-amplitude peak at $y^+=6$ in the streamwise forcing component $\bfp_{uu}$. In \S~\ref{subsec:blocks} it is inferred that this low-amplitude peak is associated to the streamwise component of the forcing $\bfP_{11}$. Thus, this peak must be related to the peak at $y^+=6$ in the streamwise component of the forcing $\bfp_{uu}$.\\ \indent
In order to verify this statement all the points of the input $\bfP$ from DNS data which are close to the wall, with $y^+<30$, are set to zero, a new input $\bfP_{h}$ is computed, and the output $\bfS_{h} = \bfR \bfP_{h} \bfR^{H}$ is evaluated. The PSD of the resulting output $\bfS_{h}$ is presented in figure~\ref{fig:Syp}a, where the symbols represent the original output from the DNS data, the shaded area is the area where the forcing is non-zero, and the red lines are the PSDs of the output $\bfS_{h}$. The curves of the PSDs of $\bfS_{h}$ are almost unaltered when compared to those of $\bfS$ for $y^+\ge30$, where the forcing is not set to zero. In the near-wall region $y^+<30$, where the forcing is set to zero, the PSDs of $\bfS_{h}$ does not match the DNS data, and clearly the peak in the streamwise component is absent. In order to verify the effects of the near-wall portion of the forcing, another forcing $\bfP_{y^+}$ is computed by setting to zero the points in $y^+\ge30$, and the output $\bfS_{y^+}=\bfR \bfP_{y^+}\bfR^H$ is evaluated. The PSD of the output $\bfS_{y^+}$ is presented in figure~\ref{fig:Syp}b, where it appears that the near-wall peak in the velocity is present, but the PSDs are not matching the DNS data for $y^+\ge30$. It follows that the low-amplitude peak at $y^+=12$ in $\bfs_{uu}$ of the DNS data must be caused by the near-wall portion of the streamwise forcing $\bfP_{11}$, and that the dynamics of large-scale motions is mostly related to processes with similar length scales.\par
%
%
\section{Modelling forcing statistics}\label{sec:prediction}
Having access to $\bfP$ allows to quantify its projection onto the right-singular vectors of the resolvent $\bfR$, which are the key quantity to validate linear resolvent analyses \citep{beneddine2016a}. These projections are quantified here for the first time for the turbulent channel flows presented. The resolvent $\bfR$ is decomposed into its left- and right-singular vectors, and the projections of the forcing term onto the right-singular vectors of $\bfR$ are computed. These projections are evaluated for the modeled forcings and for the forcing $\bfP$ from DNS data, and the effects of the modelling choice on the predictions are discussed.
\begin{table}
\centering
\begin{tabular}{llcccc}
\multicolumn{1}{c}{$\ret$ (length scale)}  & \multicolumn{1}{c}{$\gnu$} &\multicolumn{1}{c}{$\gnut$}  &\multicolumn{1}{c}{$\sigma_1$} & \multicolumn{1}{c}{$b_1$}\\
\\
179  (large scale)   & $2.54 \times10^{-9}$ & $8.70 \times  10^{-7}$&$2.14\times10^{2}$&$7.46\times 10^{-6}$\\
179  (near-wall) & $1.83 \times10^{-8}$ & $5.26 \times  10^{-7}$ &$1.10\times10^{2}$&$6.50\times 10^{-5}$\\
543  (large scale)   & $9.61  \times 10^{-10}$ & $2.29 \times 10^{-6}$ &$1.11\times10^{3}$&$2.91\times 10^{-6}$\\
543  (near-wall) &$1.03 \times 10^{-8}$ & $4.65 \times 10^{-7}$ &$3.74\times10^{1}$&$5.27\times 10^{-6}$\\
\end{tabular}
\caption{Normalization factors $\gnu$ and $\gnut$; first singular value $\sigma_1$; first projection coefficient $b_1$. $\ret=179$ and $\ret=543$. Large scales and near-wall structures.}
\label{tab:gammas}
\end{table}
\begin{figure}%
\begin{center}
{\includegraphics[width=0.49\textwidth,trim={0 0 0 0},clip]{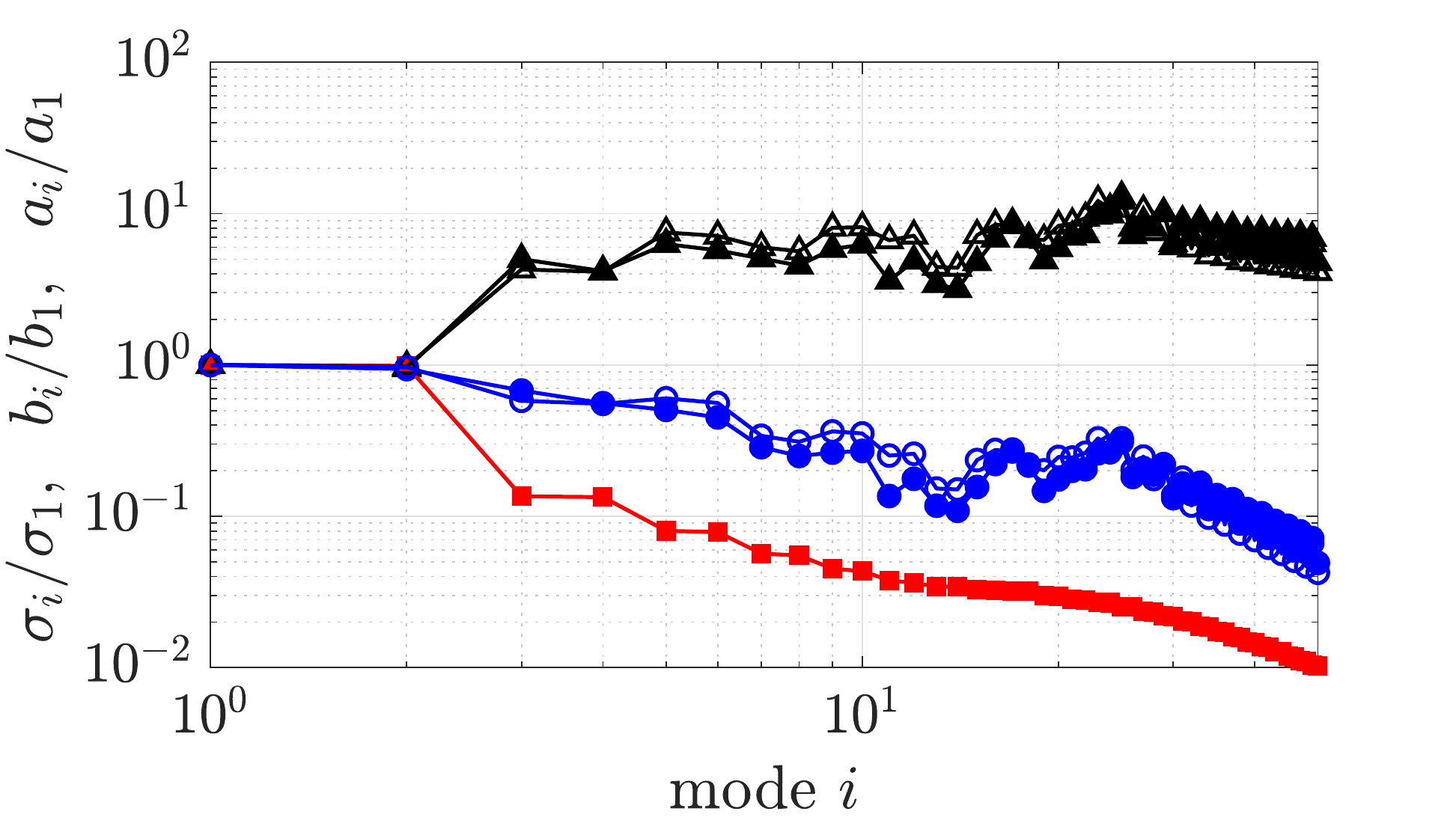}}\put(-187,105){$a)$}\hspace{0.1cm}
{\includegraphics[width=0.49\textwidth,trim={0 0 0 0},clip]{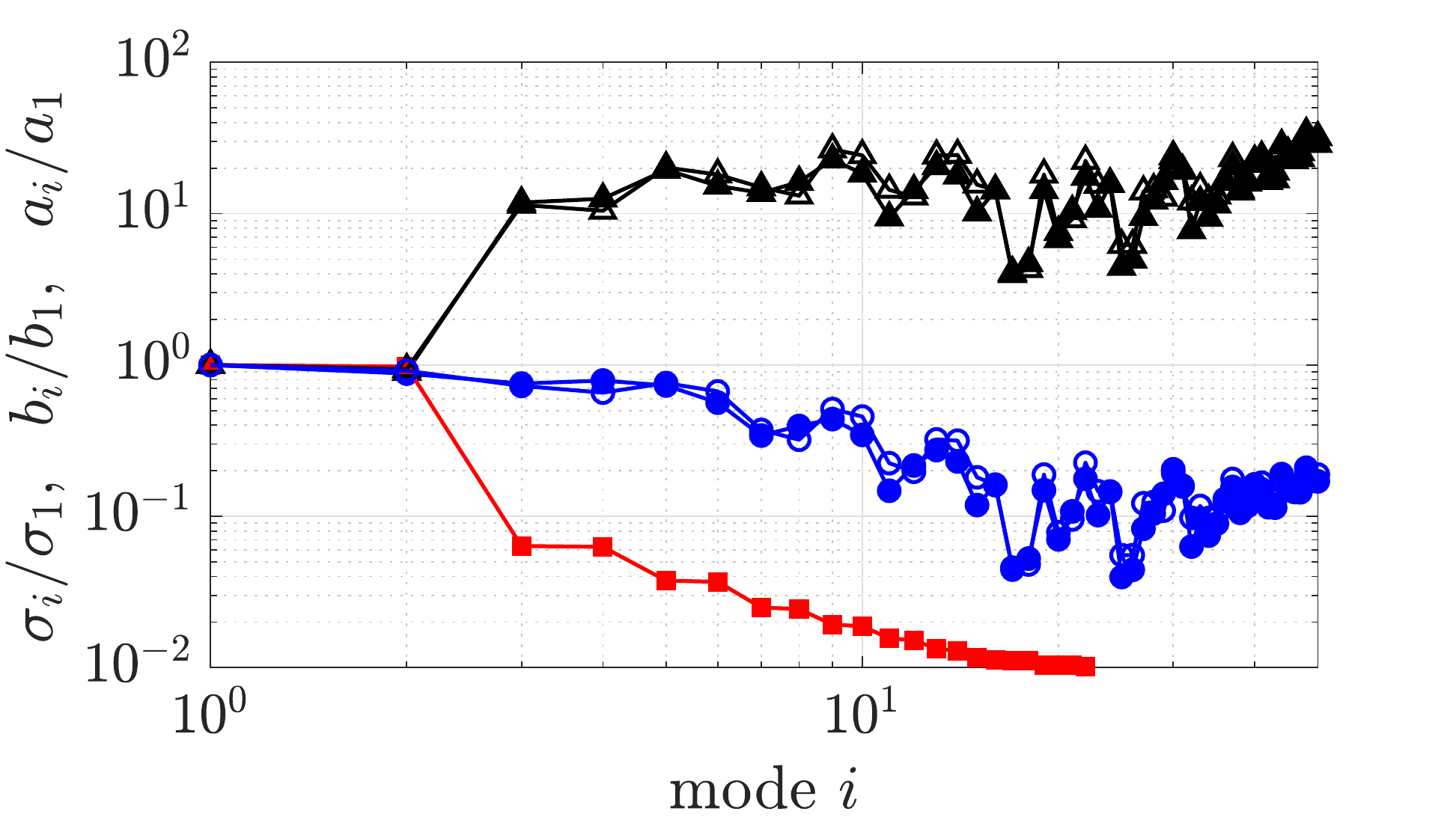}}\put(-187,105){$b)$}\\
{\includegraphics[width=0.49\textwidth,trim={0 0 0 0},clip]{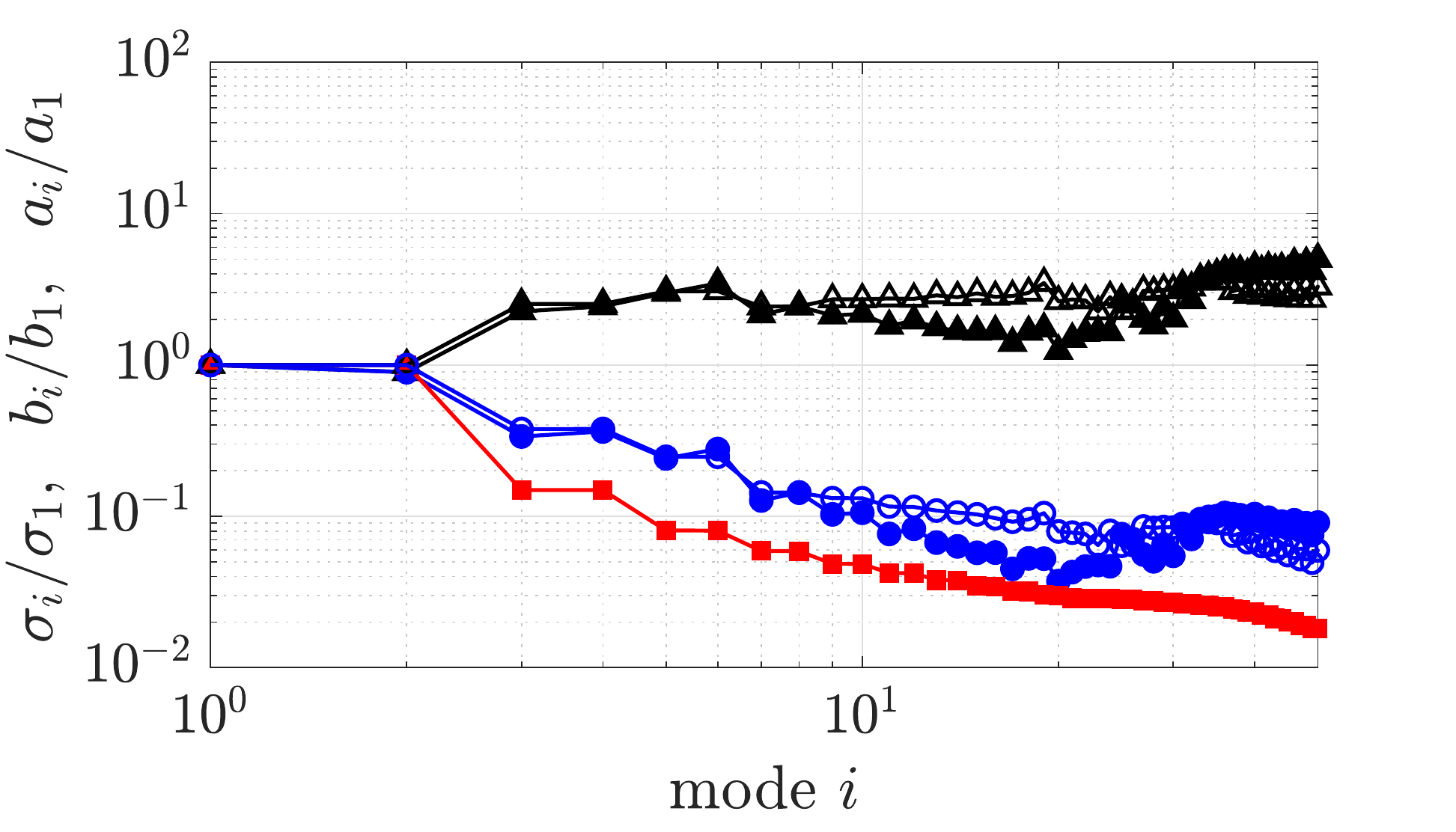}}\put(-187,105){$c)$}\hspace{0.1cm}
{\includegraphics[width=0.49\textwidth,trim={0 0 0 0},clip]{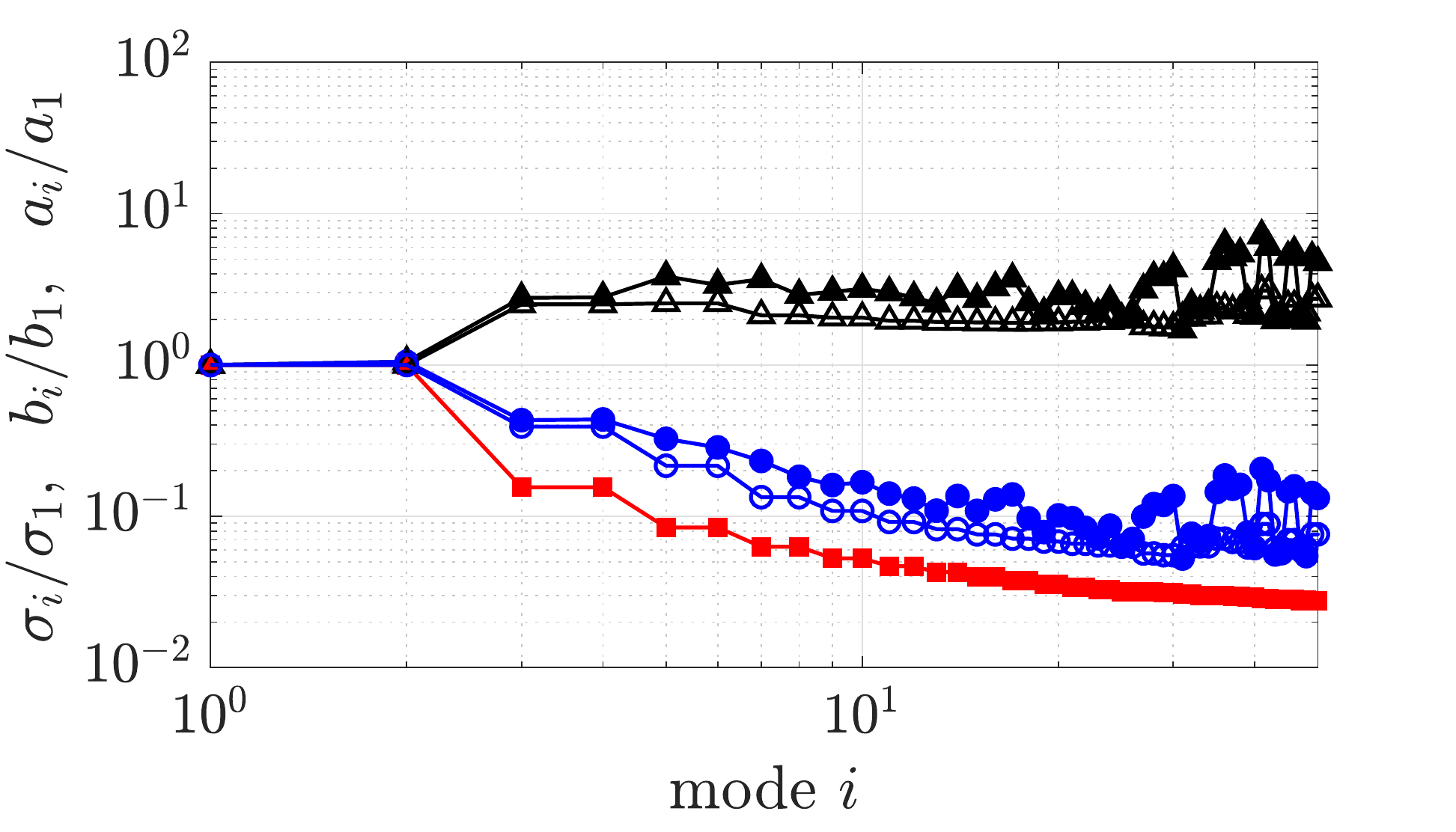}}\put(-187,105){$d)$}\\
\end{center}
\caption{Singular values $\sigma_i$ of $\bfR$, projection coefficients $b_i$, weights $a_i=\sigma_i b_i$. The values are rescaled by their value at $i=1$ (see table~\ref{tab:gammas}). Triangle: projection of $\bfP$ onto the right-singular vectors $\bfpsi_i$ of $\bfR$ ($b_i$ coefficients); filled symbols: DNS data; empty symbols: eddy-viscosity model. Square: singular values $\sigma_i$ of $\bfR$. Circle: $a_i = \sigma_i b_i$; filled symbols: DNS data;  empty symbols: eddy-viscosity model. a,c) $\ret = 179$. b,d) $\ret = 543$. a,b) Large scales. c,d) Small scale.}
\label{fig:RprojP}
\end{figure}%
\begin{figure}
\begin{center}
{\includegraphics[width=0.23\textwidth,trim={0 0 0 0},clip]{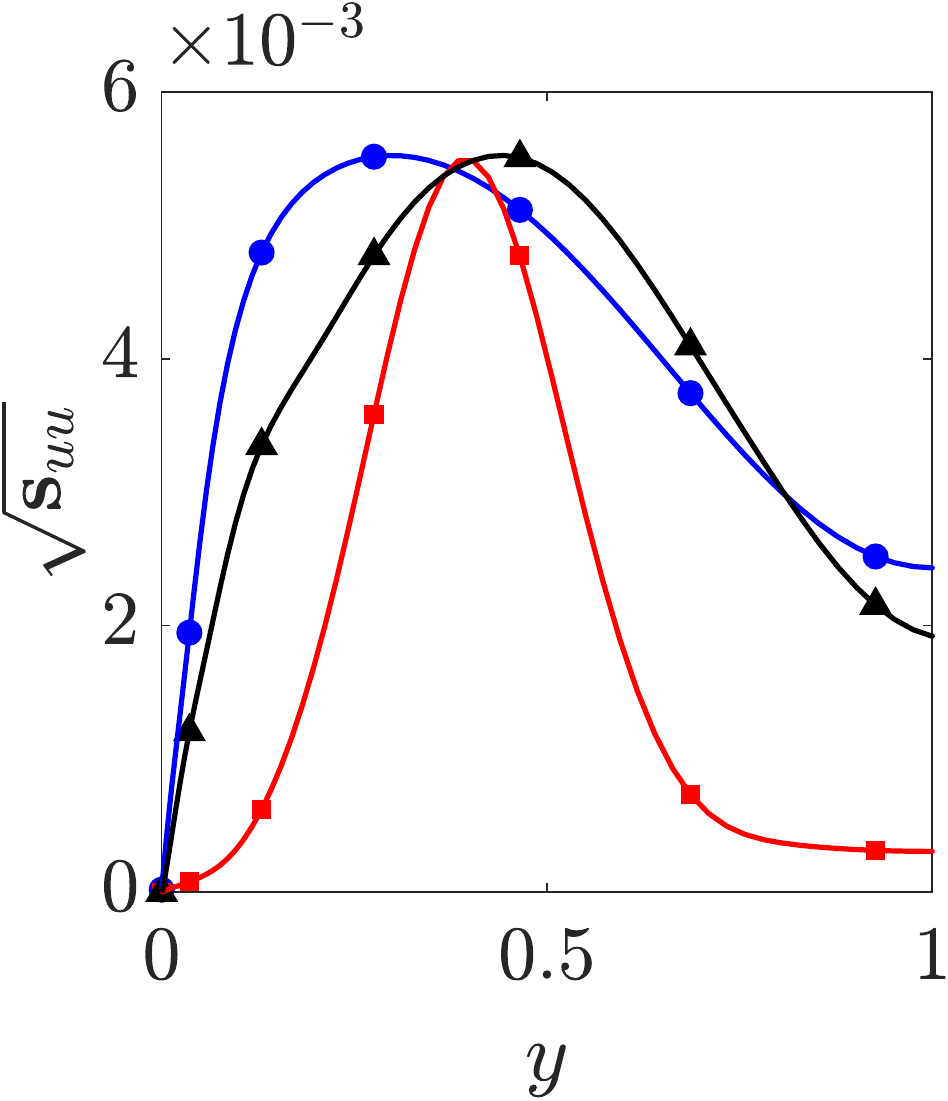}}\put(-88,90){$a)$}\hspace{0.2cm}
{\includegraphics[width=0.23\textwidth,trim={0 0 0 0},clip]{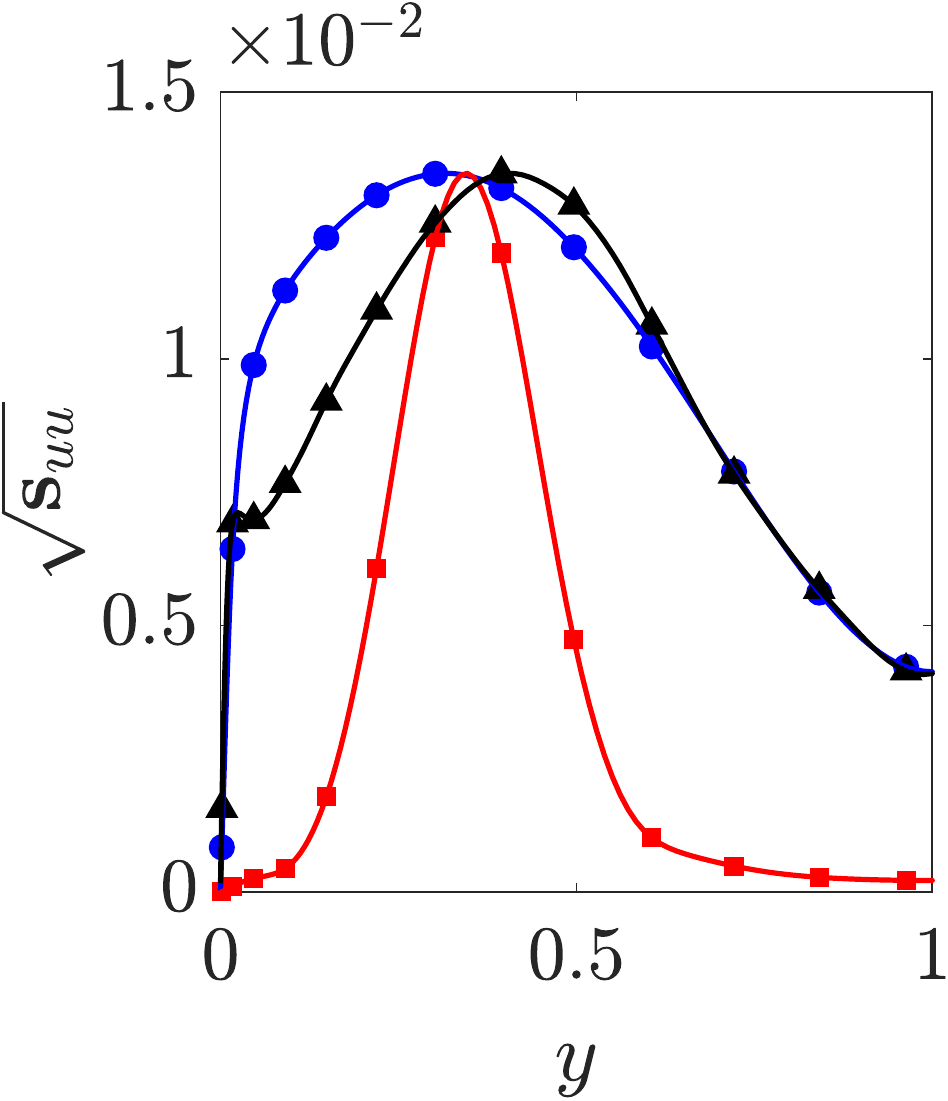}}\put(-88,90){$b)$}\hspace{0.2cm}
{\includegraphics[width=0.23\textwidth,trim={0 0 0 0},clip]{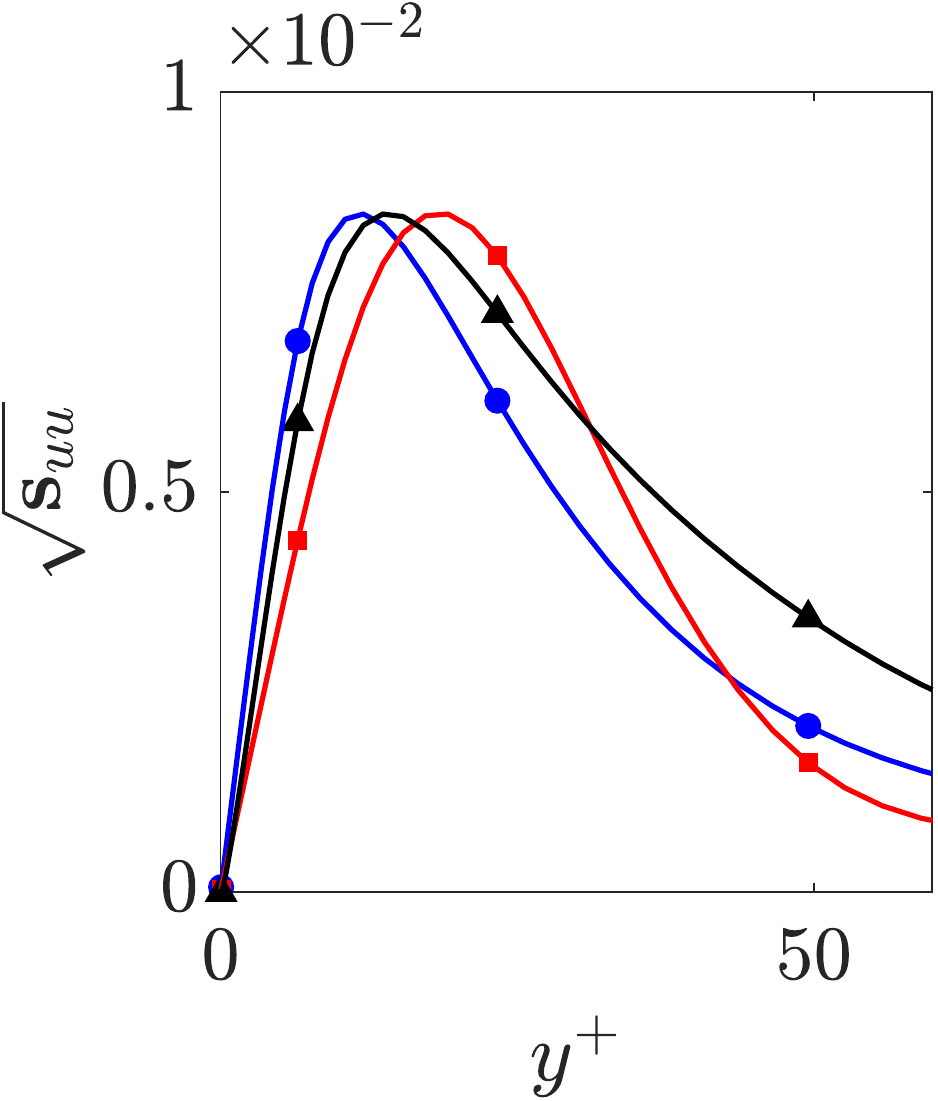}}\put(-88,90){$c)$}\hspace{0.2cm}
{\includegraphics[width=0.23\textwidth,trim={0 0 0 0},clip]{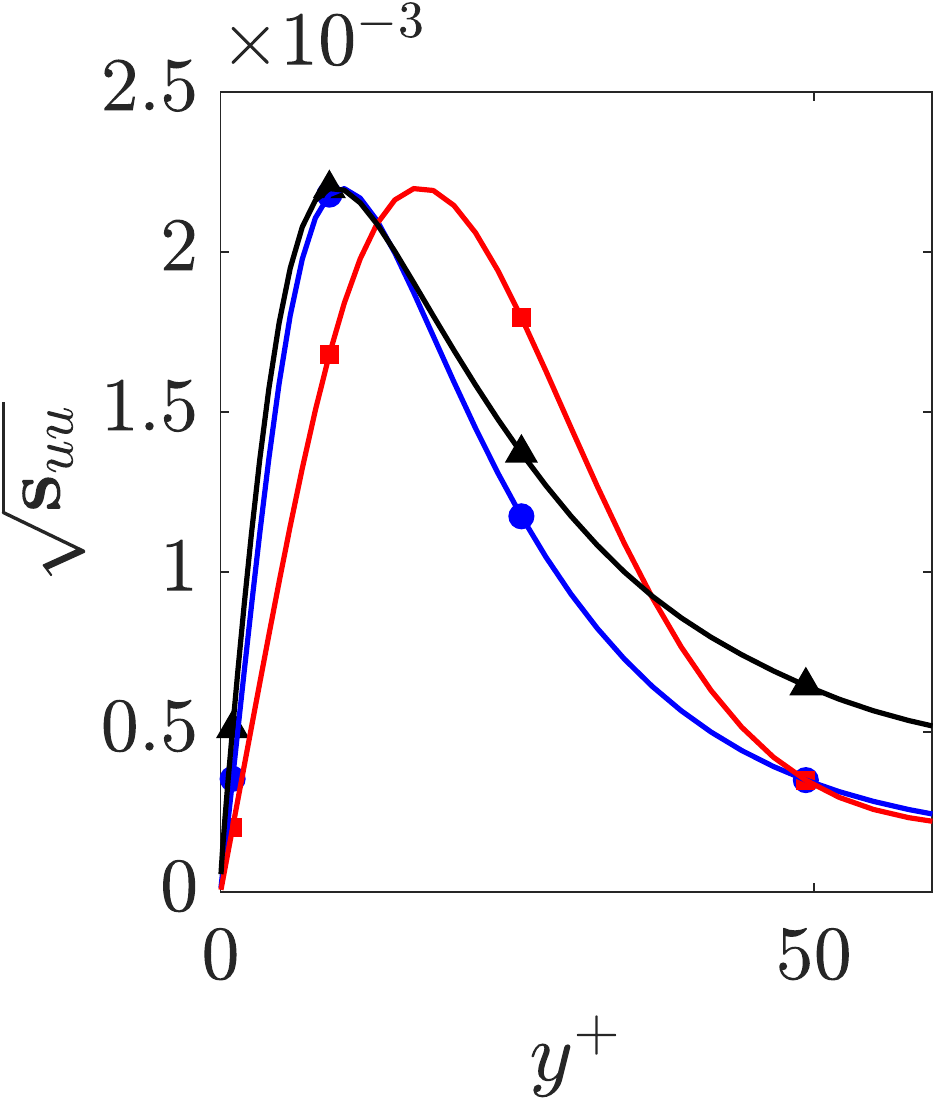}}\put(-88,90){$d)$}\\
\end{center}
\caption{PSD. Streamwise component $\bfs_{uu}$. {\text{\sout{$\ \blacktriangle \ $}}}: DNS data. {\color{red}\text{\sout{\tiny$\ \blacksquare \ $}}}: $\bfS_{\nu} = \gnu\bfR\bfR^H$. {\large \color{blue}\text{\sout{$\ \bullet \ $}} }: $\bfS_{\nu_{t}} = \gamma_{\nu_{t}}\bfR_{\nu_t}\bfR_{\nu_t}^H$. a,c) $\ret = 179$. b,d) $\ret = 543$. a,b) Large scales. c,d) Small scale.}
\label{fig:Spred}
\end{figure}%
\subsection{Forcing uncorrelated in space}
The assumption $\bfP = \gnu \bfI$ leads to the prediction $\bfS_{\bfR_{\nu}} = \gnu \bfR \bfR^H$. In this case the PSD of the output can be written as $\diag(\bfS_{\bfR_{\nu}}) = \gnu \sum_{i} \sigma_i^2| \bfphi_i |^2$, which is the sum of the left singular vectors of $\bfR$ times the square of the corresponding singular values. Since $b_i^2 = \diag(\bfpsi_i^H\mathbf{W}\bfP\mathbf{W}\bfpsi_i)$, $\diag(\bfS_{\bfR_{\nu}}) =  \sum_i |\bfphi_i|^2 a_i^2$, and $a_i = \sigma_i b_i$ from equations~\eqref{eq:bi} and \eqref{eq:bcoeff}, this modeling corresponds to setting $b_i = \gnu$, or $a_i = \sigma_i \gnu$. If there are sufficiently large gains $\sigma_i$ to neglect the others, the shape of $\diag(\bfS_{\bfR_{\nu}})$ is mainly given by the $\bfphi_i$ corresponding to such $\sigma_i$. It is shown in \cite{morra2019a} and \cite{pickering2019a} that for turbulent channel flows and for turbulent jets this sort of prediction does not lead to the best results. In \cite{morra2019a} $\bfP$ is not presented, and the conclusions are drawn by comparing $\bfS_{\bfR_\nu}$ with $\bfS$ from the DNS results. 
The reason for the mismatch between the prediction $\bfS_{\bfR_{\nu}} =  \gnu \bfR \bfR^H$ and the DNS results is clear once the non-linear forcing $\bfP$ from the DNS is projected onto the basis of the right-singular vectors $\bfpsi_i$, as shown in figure~\ref{fig:RprojP}. In figure~\ref{fig:RprojP} the trends are normalized by their value at the first mode $i=1$; the reference value for the normalization and the $\gnu$ used are summarized in table~\ref{tab:gammas}. In figure~\ref{fig:RprojP} the resolvent gains $\sigma_i$ are plotted as red squares for both near-wall and large-scale structures and for both the $\ret=179$ and $\ret = 543$. In turbulent channel flows the resolvent modes appear in pairs with nearly identical gains because of the symmetrical or anti-symmetrical behavior of the flow at the walls. It appears that for all considered cases there is a noticeable gain separation between the leading pair of gains $\sigma_1$ and $\sigma_2$ and the subsequent ones. However, it can be observed that (i) for both $\ret=179$ and $\ret=543$ and for both near-wall and large-scale structures the coefficients $b_i$ are such that the forcing has a non-negligible projection onto the right-singular vectors of $\bfR$, $\bfphi_i$, which correspond to the $\sigma_i$ with lower magnitude (compare $b_i$ with $\sigma_i$ for $i>2$: black-filled triangle symbols in figure~\ref{fig:RprojP}), and that (ii) for both $\ret=179$ and $\ret=543$ the average magnitude of this projection onto higher-order right eigenfunctions is more significant for large-scale structures than for small scale structures. The first observation explains why the assumption $\bfP = \gnu\bfI$ can lead to an erroneous prediction $\diag(\bfS_{\bfR_{\nu}})= \gnu \diag(\bfR \bfR^H) = \gnu \sum_{i} \sigma_i^2| \bfphi_i |^2$: the magnitude of the projections of the forcing $\bfP$ onto the right-singular vector $\bfpsi_i$, which is $b_i$, can compensate a small singular value $\sigma_i$ and increase the relative weight $a_i = \sigma_i b_i$ of a left-singular vector $\bfphi_i$ in the linear combination that leads to the output. The coefficients $a_i$ are shown with blue-filled circle symbols in figure~\ref{fig:RprojP}. The second observation explains why the assumption $\bfP=\gnu \bfI$ gives less erroneous predictions $\bfS_{\bfR_{\nu}} = \gnu \sum_{i} \sigma_i^2| \bfphi_i |^2$ for near-wall structures: the magnitude of the projection $\bfP$ onto $\bfpsi_i$ is such that the trend of the weights $a_i$, which multiply the left-singular vector $\bfphi_i$ in the linear combination, is less modified than it is for large-scale structures. This can be seen by comparing figures~\ref{fig:RprojP}a,b with figures~\ref{fig:RprojP}c,d: for the near-wall structures the trend of the blue-filled circle symbols, which are $a_i$, is more similar to the square red symbols, which represent $\sigma_i$, than it is for the large-scale structures. \\ \indent
The discussion about the coefficients $a_i$, $b_i$, and the singular values $\sigma_i$ and their effect on the output is consistent with the results shown in figure~\ref{fig:Spred}, where the streamwise component of $\bfS_{\bfR_{\nu}}$ is presented. In particular, the assumption $\bfP = \gnu \bfI$ leads to very localized large-scale structures, which do not resemble the DNS data, whereas the near-wall structures approximate better the shape of the DNS data, as presented by \cite{morra2019a} for a turbulent channel at $\ret=1007$. This occurs because of the difference in $\sigma_i$ and $a_i = \sigma_i b_i$: for large-scale structures $a_i$ is farther to $\sigma_i$ than it is for near-wall structures. It is noticeable that for both the $\ret$ presented and for both small-scale and large-scale structures the prediction based on the assumption that the forcing arising from the non-linear terms is uncorrelated in space is erroneous because the assumption is not verified in a turbulent flow, as the shapes of $\bfP$ in figure~\ref{fig:SPRet180} and \ref{fig:SPRet550} show.
%
%
\subsection{Eddy-viscosity modeling}
In eddy-viscosity modeling the non-linear forcing term is modeled by means of the Boussinesq expression, which states that the unknown Reynolds stresses are proportional to the rate of strain tensor given by the known field through a scalar $\nu_t$. $\nu_t$ is the eddy (or turbulent) viscosity, whose structure is prescribed by the chosen modeling approach. In the most general case $\nu_t$ is not constant in space, so its effect on the dynamics are (i) modifying the local dissipation rate by a factor $1 + \nu_t/\nu$ and (ii) introduce some additional terms proportional to the (partial) derivatives of $\nu_t$ and linear in the velocity fluctuations.\\ \indent 
The focus of this work is neither a discussion on the choice of the eddy-viscosity modeling strategy nor a detailed study on the specific effects of a specific eddy-viscosity model. Here, the eddy-viscosity model employed in the resolvent analyses of \cite{hwang2010a}, \cite{hwang2010b} and \cite{morra2019a} is used as an example, and its effects are compared to those produced by the forcing $\bfP$ computed from the DNS data. It is noteworthy that the model for the eddy-viscosity $\nu_t$ adopted here, taken from \cite{cess1958a}, is tuned at $\ret = 2000$, so its performance at lower $\ret$ is not assured. This eddy-viscosity modeling approach was firstly proposed by \cite{reynolds1972a} and is based on the assumption that the fluctuations around the mean flow can be decomposed into a coherent and an incoherent part. The incoherent part is assumed to be unknown, so its contribution to the Reynolds stresses is modeled, whereas the coherent part is assumed to be known and it is used for the modeling (see \cite{reynolds1972a} for more details). The results from the eddy-viscosity modeling are discussed by comparing $\bfP$ from DNS data with $\bfP_{\nu_t}$, where $\bfP_{\nu_t}$ is the equivalent forcing introduced by the eddy-viscosity model such that $\bfS_{\bfR_{\nu_t}} = \bfR \bfP_{\nu_t}\bfR^H=\gnut\bfR\bfR^H$ and is computed as in equation~\eqref{eq:Pnut}. The normalization scalars are presented in table~\ref{tab:gammas}\\ \indent
The square root of the coefficients $b_i^2 = \diag(\bfpsi_i^H\mathbf{W}\bfP_{\nu_t}\mathbf{W}\bfpsi_i)$, which quantify the projection of $\bfP_{\nu_t}$ onto the right-singular vectors $\bfpsi_i$ of $\bfR$, is presented in figure~\ref{fig:RprojP} with black-empty triangles. The coefficients $a_i=\sigma_i b_i$ which define the shape of the prediction $\diag(\bfS_{\bfR_{\nu_t}}) = \sum_i |\bfphi_i|^2 a_i^2$ are also presented in figure~\ref{fig:RprojP} as blue-empty circles. These quantities are presented for both $\ret = 179$ and $\ret = 543$, and for both near-wall and large scale structures. The trends presented in figure~\ref{fig:RprojP} are normalized by their value for the first mode; the reference for the normalization is presented in table~\ref{tab:gammas}. It appears that the equivalent forcing $\bfP_{\nu_t}$, provided by the eddy-viscosity modeling, is able to modify the relative weights $a_i = \sigma_i b_i$ in the linear combination $\diag(\bfS_{\bfR_{\nu_t}}) = \sum_i |\bfphi_i|^2 a_i^2$ of the right-singular vectors $\bfphi_i$ of $\bfR$ similarly to $\bfP$ from DNS data, as shown in figure~\ref{fig:RprojP}. In this figure DNS data are represented by the black-filled triangles and blue-filled circles. This occurs because $\bfP_{\nu_t}$ provides coefficients $b_i$ whose magnitude is comparable to that of the coefficients $b_i$ given by $\bfP$ from DNS data.
\\ \indent
The discussion about the coefficients $a_i$, $b_i$, and the singular values $\sigma_i$ and their effect on the output is consistent with the results shown in figure~\ref{fig:Spred}, where the streamwise component of $\bfS_{\bfR_{\nu_t}}$ is presented. $\bfS_{\bfR_{\nu_t}}$ is in better accordance with $\bfS$ than $\bfS_{\bfR_{\nu}}$, as shown in \cite{morra2019a} for a turbulent channel at $\ret=1007$. In particular, the eddy-viscosity modeling gives an improved approximation of the output also for near-wall structures. This is expected from the trend of the coefficients $a_i$ (and $b_i$) in figure~\ref{fig:RprojP}: the empty circles (and triangles) which represent $a_i$ (and $b_i$) from $\bfP_{\nu_t}$ are closer to the filled circles (and triangles) which represent $a_i$ (and $b_i$) for $\bfP$ from DNS data than they are for $\bfP=\gnu\bfI$. Note that for $\bfP=\gnu\bfI$ it is $a_i = \sigma_i \gnu$ and $a_i/a_1 = \sigma_i/\sigma_1$ in figure~\ref{fig:RprojP} coincide with the red-filled squares.\\ \indent
It is noteworthy that the prediction improves when the relative weight $a_i$ for $i>2$ is $a_i > \sigma_i$ in all the cases presented, which is expected by inspecting the trend of $a_i$ and $b_i$ based on $\bfP$ from DNS data. The right-singular vectors of $\bfR$ with $i>2$ correspond to sub-optimals in linear analyses based on the resolvent $\bfR$. Thus, if a significant part of the forcing is spanned by linear sub-optimals, such linear analyses can be unacceptably inaccurate. It is clearly stated in \cite{beneddine2016a} that the validity of linear analyses based solely on the resolvent $\bfR$ should not be based on the trend of the singular values $\sigma_i$ but on the coefficients $a_i=\sigma_i b_i$, which include the projection of the forcing onto the right-singular vectors of $\bfR$. Here, this statement is quantified for the presented channel flows.\\ \indent
Thus, the results with the eddy-viscosity model suggest that accounting for the non-linear forcing term $\bfP$ improves accuracy in all cases and not only for large-scale structures, as discussed in \cite{pickering2019a} and \cite{morra2019a}. However, it is noticeable that also in this case the prediction is erroneous because the provided forcing is not exactly the $\bfP$ computed from the DNS data. In fact, in figure~\ref{fig:RprojP} the filled symbols and the empty symbols for $a_i$ and $b_i$, which represent the projections $b_i$ and the coefficients $a_i$ of the linear combination of $\phi_i$ for $\bfP$ and for $\bfP_{\nu_t}$ are not superposed. Nevertheless, it is clear that the accuracy of the prediction is improved if the modeled forcing term mimics the coefficients $b_i$ from the projection of $\bfP$ computed from the DNS data. Thus, an eddy-viscosity modeling approach may be tempting for these sort of predictions, as it leads to an effective forcing $\bfP_{\nu_t}$ whose structure, or colour, leads to outputs close to the observations from DNS data.
%
%
\section{Conclusions}\label{sec:concl}
In this work the CSD of the non-linear forcing term associated to the velocity fluctuation, which is the input to the resolvent operator based on the mean flow, is quantified for the first time for turbulent channel flows. The computation is based on snapshots of DNS of turbulent channel flows at $\ret = 179$ and $\ret = 543$. The non-linear forcing is computed at fixed time instants and the realizations are used with the Welch method. The CSD of the velocity fluctuations is computed with the same technique. The CSDs are computed for highly energetic structures typical of buffer-layer motions, $(\lambda_x^+,\lambda_z^+) = (1130,113)$ at $\ret=179$ and $(\lambda_x^+,\lambda_z^+) = (1137,100)$ at $\ret=543$, and large-scale motions: $(\lambda_x,\lambda_z) = (4.19,1.26)$ at $\ret=179$ and $(\lambda_x,\lambda_z) = (6.28,1.57)$ at $\ret = 543$.
The accuracy of the computed non-linear forcing term is assessed by computing its response with the resolvent. This response is then compared to the velocity fluctuations from DNS data. The two PSDs appear to be indistinguishable, which shows the evaluation to be accurate. The computed CSD of the forcing due to the non-linear terms is shown not to be uncorrelated (or white) in space, which implies the forcing is structured. Thus, the recent practice of assuming a colored forcing for resolvent analyses is here explicitly verified for the treated turbulent channel flows.\\ \indent
Since the non-linear forcing is non-solenoidal by construction and the velocity field of the incompressible Navier--Stokes is affected only by the solenoidal part of the forcing \citep{chorin1993a}, the PSD associated to the solenoidal part of the non-linear forcing is evaluated and presented. It is seen that the wall-normal and the spanwise components of the non-linear forcing are very different from their solenoidal counterpart for all the cases presented. In particular, in the solenoidal part of the forcing these two components has the shape of quasi-streamwise vortices, which are typical of the lift-up mechanism. On the other hand, the streamwise component of the forcing is almost unchanged in its solenoidal counterpart. For all the cases presented it is shown that the transverse components of the forcing generate a response which is counteracted by the response of the streamwise component of the forcing, as in a destructive interference. It is also demonstrated that the high-amplitude peak at $y^+=6$ in the streamwise component of the forcing for the large-scale structure $(\lambda_x,\lambda_z) = (6.28,1.57)$ at $\ret = 543$ is not relevant in the response. This explicitly verifies the conclusions of \cite{flores2010a,hwang2011a} that the dynamics of large-scale motions, although affected by smaller near-wall eddies, is mostly related to processes with similar length scales. This last statement is also verified by the fact that a low-rank approximation of the forcing, which includes only the pair of most energetic symmetric and antisymmetric SPOD modes which have the same length scale of the response, leads to the bulk of the response for both near-wall and large-scale structures and for both Reynolds numbers.\\ \indent
The projection of the non-linear forcing term onto the right-singular vectors of the resolvent are evaluated. It appears that the multiplication of the singular values $\sigma_i$ of the resolvent with the projection coefficients $b_i$ modifies the relative importance of the left-singular vectors of the resolvent in the linear combination which defines the shape of the response. It is seen that the left-singular vectors of the resolvent associated with very low-magnitude singular values are non-negligible because the non-linear forcing term has a non-negligible projection onto the linear sub-optimals of the resolvent analysis introduced by \cite{mckeon2010a}. This occurs for both near-wall and large-scale structures at both Reynolds numbers, but the effect is stronger for large-scale structures. Since the response is given by the linear combination of the left-singular vectors of the resolvent weighted with the terms $a_i = \sigma_i b_i$, the evaluation of $b_i$ is an explicit quantification of the validity of resolvent analysis, as discussed in \cite{beneddine2016a}. The same coefficients $b_i$ are computed when the stochastic forcing is modelled with an eddy-viscosity approach. It is here clarified that this modelling leads to an improvement in the accuracy of the prediction of the response \citep{morra2019a} since the resulting coefficients $b_i$ are closer to those associated with the non-linear forcing term evaluated from DNS data.\par \bigskip
%
\section*{Acknowledgements}
The authors would like to thank C. Cossu for fruitful discussions. The authors would like to acknowledge the VINNOVA Projects PreLaFlowDes and SWE-DEMO and the Swedish-Brazilian Research and Innovation Centre CISB for funding. The simulations were performed on resources provided by the Swedish National Infrastructure for Computing (SNIC) at NSC, HPC2N and PDC.
%
%
\appendix
\section{Resolvent operators}\label{app:OSS}
The linear system in equation~\eqref{eq:OSS} is derived from the equation~\eqref{eq:momeq}, as in \cite{schmid2001a}. The matrices $\bfA$ and $\bfB$ are defined as
\gdef\thesubequation{\theequation \textit{a,b}}
\begin{subeqnarray}
\bfA =
\begin{bmatrix}
\Delta^{-1} \mathcal{L}_{\mathcal{OS}} & 0\\
-i\beta U' & \mathcal{L}_{\mathcal{SQ}} \\
\end{bmatrix},
\quad
\bfB = 
\begin{bmatrix}
-i\alpha \Delta^{-1}\mathcal{D} & -k^2\Delta^{-1} & -i\beta \Delta^{-1} \mathcal{D}\\
i\beta & 0 & -i\alpha \\
\end{bmatrix},
\label{eq:ABop}
\end{subeqnarray}
where the discretized version of the generalized Orr-Sommerfeld and Squire operators \citep{cossu2009a,pujals2009a} are
\begin{subequations}
\label{eq:ORR}
\begin{align}
\mathcal{L}_{\mathcal{OS}} &= -i \alpha (U\Delta-U'') + \nu_{T}\Delta^2+2\nu_T' \Delta\mathcal{D} + \nu_T''(\mathcal{D}^2 + k^2),\\
\mathcal{L}_{\mathcal{SQ}}&= -i \alpha U + \nu_T \Delta + \nu_T'\mathcal{D},
\end{align}
\end{subequations}
with $\mathcal{D}$ and $'$ representing $\mathrm{d}/\mathrm{d} y$, $k^2 = \alpha^2+\beta^2$, $\Delta = \mathcal{D}^2-k^2$, and $U(y)$ the reference flow. $\nu_{T} = \nu + \nu_t$ with $\nu$ the molecular viscosity and $\nu_t$ the eddy-viscosity model. The eddy-viscosity used is the one proposed by \cite{cess1958a}, as reported by \cite{reynolds1967a},
\begin{equation}
\label{eq:Cess}
\frac{\nu_t}{\nu} = \frac{1}{2}\left[ 1 + \frac{\kappa^2 \ret^2}{9}(1-y^2)^2(1+2y^2)^2(1-e^{-\ret(1-|y|)/A})^2 \right]^{\frac{1}{2}} - \frac{1}{2},
\end{equation}
with $\ret = u_{\tau} h/\nu$ the Reynolds number based on the friction velocity. The von K\'arman constant is $\kappa=0.426$ and the constant $A=25.4$ as in \cite{pujals2009a}, \cite{hwang2010c}. This model is tuned for $\ret=2000$. Note that setting $\nu_t = 0$ reduces equations~\eqref{eq:OSS} to the standard Orr-Sommerfeld and Squire equations. $\nu_t = 0$ results in equation~\eqref{eq:Rnu}, whereas $\nu_{t}$ as in equation~\eqref{eq:Cess} results in equation~\eqref{eq:Rnut}. Homogeneous boundary conditions are enforced on both walls: $\hat{v}(\pm1) = \partial_y \hat{v}(\pm1) = \hat{\omega}_y(\pm1) = 0$. The matrices relating $\hat{\bfu}$ and $\hat{\mathbf{q}}$ are
\begin{subeqnarray} 
\bfC = \frac{1}{k^2}
\begin{bmatrix}
i\alpha \mathcal{D} &  -i\beta \\
k^2 & 0 \\ 
i\beta\mathcal{D} & i\alpha \\ 
\end{bmatrix},
\quad
\bfD=
\begin{bmatrix}
0 & 1 & 0 \\
i\beta & 0 & -i\alpha \\
\end{bmatrix}.
\label{eq:CDop}
\end{subeqnarray}
\section{Solenoidal part of the forcing with $\bfL$ or $\bfC \bfB$}\label{app:divfree}
Take a general forcing field $\bm{f}$ as the sum of a solenoidal field $\bm{f}^s$ and and irrotational field $\bm{f}^{r}$, such that $\nabla \cdot \bm{f}^{s} = 0$ and $\nabla \times \bm{f}^{r} = 0$. Take $\bm{f}^{r} = \nabla \chi$, with $\chi$ a scalar field. For given wave-numbers $(\alpha,\beta)$ the Fourier modes of $\bm{f}$, $\bm{f}^s$, and $\bm{f}^s$, discretized with $N_y$ points in the wall-normal direction are the $3N_y \times 1$ vectors $\hat{\bff}$, $\hat{\bff}^s$, $\hat{\bff}^r$, and the Fourier mode of $\chi$ discretized with $N_y$ points in the wall-normal direction is the $N_y \times 1$ vector $\hat{\bm{\chi}}$. Applying $\bfB$ to $\hat{\bff}$ corresponds to $\bfB\hat{\bff} = \bfB\hat{\bff}^{s} + \bfB\hat{\bff}^{r}$. It holds
\begin{equation}
\bfB\hat{\bff}^{r} = 
\begin{bmatrix}
k^2\mathcal{D} \hat{\bm{\chi}} - k^2\mathcal{D} \hat{\bm{\chi}}\\
-\alpha \beta \hat{\bm{\chi}} + \beta \alpha \hat{\bm{\chi}}\\
\end{bmatrix} = 0,
\end{equation}
and
\begin{equation}
\bfB\hat{\bff}^{s} = 
\begin{bmatrix}
\hat{\bff}_y^s\\
i \beta \hat{\bff}_x^s - i \alpha \hat{\bff}_z^s\\
\end{bmatrix} \equiv \bfD\hat{\bm{\bff}}^s,
\end{equation}
with $\hat{\bff}_x^s$, $\hat{\bff}_y^s$, and $\hat{\bff}_z^s$ the $N_y\times 1$ vectors of the streamwise, wall-normal and spanwise components of $\hat{\bff}^s$.\\ \indent
Since $\hat{\bfu}=\bfC\hat{\mathbf{q}}$ and $\hat{\mathbf{q}}=\bfD\hat{\bfu}$, it follows $\hat{\bfu}=\bfC \bfD \hat{\bfu}$. Thus, $\hat{\bff}^s=\bfC \bfD \hat{\bff}^s = \bfC \bfB \hat{\bff}$, which provides a first way to isolate the solenoidal part of the forcing.\\ \indent
The Fourier transform in time of equation~\eqref{eq:OSS} leads to $-(i\omega \bfI + \bfA) \tilde{\mathbf{q}} = \bfB\tilde{\bff}$, which is equivalent to $-(i\omega \bfI + \bfA) \bfD\tilde{\bfu} = \bfD\tilde{\bff}^{s}$. Thus, $\bfL\tilde{\bfu} = \bfC\bfD\tilde{\bff}^{s} = \tilde{\bff}^{s}$, with $\bfL=-\bfC(i\omega \bfI + \bfA) \bfD$.\\ \indent
It follows that $\bfL\tilde{\bfu} = \tilde{\bff}^{s} = \bfC \bfB \tilde{\bff}$, which leads to the solenoidal part of the forcing from the velocity field. Thus, using $\bfL$ or $\bfC \bfB$ to compute the solenoidal part of the forcing is equivalent.\par
\section{DNS and data analysis details}\label{app:welch}
The simulations were performed by means of the \texttt{SIMSON} code (see \cite{chevalier2007a} for details) for the simulation of the turbulent channel at $\ret=179$ and by means of  \texttt{Channelflow} code (see \cite{channelflow2018a} for details) for the simulation of the turbulent channel at $\ret=543$.\\ \indent
The initial transient of the simulation is discarded. The Welch's method with Hann windowing and $75$\% overlap is used to compute the cross-spectral densities $\bfS$ and $\bfP$ using a total of $N_s = 10001$ snapshots of the DNS solutions with sampling interval $\Delta t_s = 0.5$ for a total acquisition time $T_{max} = 5000$ for $\ret=179$, and using $N_s = 20000$ snapshots sampled every $\Delta t_s = 0.15$ for a total acquisition time $T_{max} = 299.85$ for $\ret=543$. Data have also been averaged between the two walls. The Welch's method implemented is the one described in \cite{towne2018a} or \cite{pintelon2012a}. Note that the input-output relationship is based on the windowed data. The presence of the window, which is a function of time, needs to be accounted in the time derivative, or the input-output relationship in equation~\eqref{eq:OSS} is not valid \citep{martini2019a}. A compensation term of the form $-\hat{\bm{q}}d{W}(t)/dt$, with $W(t)$ the Hann window, is added to the forcing in order to preserve the identity in equation~\eqref{eq:OSS}. The same procedure presented in \cite{nogueira2020a} is followed here.\par
\bibliographystyle{jfm}
\bibliography{/Users/pmorra/Documents/KTH/Works/Bibliography/Morra_biblio}
\end{document}